\documentclass{article}

\usepackage[]{graphicx} 

\usepackage{enumitem}
\usepackage[margin=1in]{geometry}
\usepackage{float}
\usepackage[dvipsnames]{xcolor}
\definecolor{niceblue}{HTML}{156ABC}

\usepackage{amssymb,amsmath, bm,amsthm}
\usepackage{caption}
\usepackage{subcaption}
\usepackage{url}
\usepackage{setspace}
\usepackage[hidelinks,colorlinks=true,citecolor=niceblue,linkcolor=niceblue]{hyperref}

\usepackage[%
backend=bibtex8,
uniquename=false,
uniquelist=false,
maxcitenames=2,
backref,
style=authoryear,
isbn=false,
dashed=true,
natbib=true
]{biblatex}
\usepackage{array}
\newcolumntype{C}[1]{>{\centering\arraybackslash}p{#1}}

\usepackage{sansmathfonts}
\usepackage[default, scale=.95]{lato}
\usepackage[T1]{fontenc}

\usepackage[hidelinks,colorlinks=true,citecolor=niceblue,linkcolor=niceblue]{hyperref}
\usepackage{xr}

\makeatletter
\newcommand*{\addFileDependency}[1]{
	\typeout{(#1)}
	\@addtofilelist{#1}
	\IfFileExists{#1}{}{\typeout{No file #1.}}
}
\makeatother

\newcommand*{\myexternaldocument}[1]{%
	\externaldocument{#1}%
	\addFileDependency{#1.tex}%
	\addFileDependency{#1.aux}%
}

\myexternaldocument{zz_sup_graphTest_arxiv}

\newtheorem{theorem}{Theorem}

\newtheorem{lemma}{Lemma}

\newcommand{\R}{{}\operatorname{\mathbb{R}}}
\newcommand{\E}{{}\operatorname{\mathbb{E}}}

\newcommand{\Nat}{{}\operatorname{\mathbb{N}}}

\newcommand{\cov}{\operatorname{Cov}}

\newcommand{\var}{\operatorname{Var}}
\newcommand{\mean}[1]{\mkern 1.5mu\overline{\mkern-1.5mu#1\mkern-1.5mu}\mkern 1.5mu}

\newcommand{\p}{{}\operatorname{\mathbb{P}}}

\newcommand{\boldD}{\mathbf{D}}

\newcommand{\boldM}{\mathbf{M}}

\newcommand{\boldS}{\mathbf{S}}

\newcommand{\boldU}{\mathbf{U}}

\newcommand{\boldd}{\mathbf{d}}

\newcommand{\boldp}{\mathbf{p}}

\newcommand{\mcL}{\mathcal{L}}

\newcommand{\given}{ \mathop{\mid}}

\newcommand{\ident}{\mathbf{I}}

\newcommand{\inv}{^{-1}}

\newcommand{\trans}{^\top}
\newcommand{\ones}{\mathbf{1}}
\newcommand{\indicator}{\mathbb{I}}

\newcommand{\iid}{\mathrel{\overset{\makebox[0pt]{\mbox{\normalfont\tiny\sffamily iid}}}{\sim}}}

\newcommand{\dist}{\xrightarrow{\mcL}}

	\newcommand{\N}{\mathcal{N}}

\title{Graph-Based Tests for Multivariate Covariate Balance Under Multi-Valued Treatments}
\author{Eric A. Dunipace \thanks{edunipace@mail.harvard.edu}\\
	David Geffen School of Medicine at UCLA, Los Angeles, CA\\
	Department of Biostatistics, Harvard T.H. Chan School of Public Health, Boston, MA
	}
\date{}

\begin{document}
	\maketitle
	\begin{abstract}
		We propose the use of non-parametric, graph-based tests to assess the distributional balance of covariates in observational studies with multi-valued treatments. Our tests utilize graph structures ranging from Hamiltonian paths that connect all of the data to nearest neighbor graphs that maximally separates data into pairs. We consider algorithms that form minimal distance graphs, such as optimal Hamiltonian paths or non-bipartite matching, or approximate alternatives, such as greedy Hamiltonian paths or greedy nearest neighbor graphs. Extensive simulation studies demonstrate that the proposed tests are able to detect the misspecification of matching models that other methods miss. Contrary to intuition, we also find that tests ran on well-formed approximate graphs do better in most cases than tests run on optimally formed graphs, and that a properly formed test on an approximate nearest neighbor graph performs best, on average. In a multi-valued treatment setting with breast cancer data, these graph-based tests can also detect imbalances otherwise missed by common matching diagnostics. We provide a new \texttt{R} package \texttt{graphTest} to implement these methods and reproduce our results.
	\end{abstract}
	\noindent%
	

	\doublespacing
	
\section{Introduction}\label{sec:intro_graphtest}

In studies of causal effects, balancing the covariate distributions in each treatment group is essential for unbiased estimates.
For example, in a study of the effects of smoking on medical expenditures, multiple smoking behaviors are possible: there may be never smokers, former smokers, and current smokers.
To estimate the relative effects of the different exposures, we would like the groups to be balanced by various demographic characteristics such as age, race, and income.
Distributional balance is expected by design in randomized control trials (RCT) for both measured and unmeasured confounders, but conducting a RCT is often not possible because of either cost or ethical restrictions, as would be the case in such a study of smoking. Instead, researchers are left to rely on observational studies to answer their causal questions. 

If the number of covariates is few and the covariates have a small number of discrete levels, obtaining exact matches between treatment groups may be possible, thereby achieving exact distributional balance. 
However, such simple settings are rarely found in practice. This problem can be further compounded in the multi-valued treatment setting when there are more groups to match.
Achieving distributional balance in such multi-valued treatment settings then relies on adjustments via methods such as propensity score matching \citep{Rosenbaum1983} or cardinality matching  \citep{Zubizarreta2014}. Intuitively, these methods work by matching units that are close in terms of some covariate distance, which will make the distributions more similar between the treatment groups. But how do we know when the distributions between treatment groups are similar enough to proceed?

To answer this question, investigators typically consider univariate assessments of covariate balance. Such assessments can include assessing the mean of each covariate by treatment group before and after matching and a corresponding test for the difference of these means \citep{Stuart2010}.  
Other authors run a Kolmogorov-Smirnov test \citep{Kolmogorov1933, Smirnov1948} for each covariate \citep{Zubizarreta2012, Diamond2013}.
However, these methods will not generally diagnose differences in the joint covariate distribution between treatment groups.

Instead, we seek a way to identify differences in the joint distributions between the multiple groups after matching is completed---in other words, whether the observational study fails to approximate a randomized experiment in terms of observed covariate balance. Several common methods of assessing distributional balance such as likelihood ratio tests or examining the distribution of the propensity scores \citep{Rosenbaum1985} require having correct probability models for the data. Unfortunately, these data models are rarely known in practice. 

Fortunately, there are several non-parametric tests applicable to the multi-valued treatment setting and the ones we consider involve forming a graph-like structure on the data. Some work involves ranking the data in some fashion such that the observations are ordered in a one-dimensional space and hence form a graph \citep{Bhattacharya2019}. \citet{puri1966class}, for example, consider a ranks-based test that uses the data ranks for each individual covariate in a Kruskal-Wallis test \citep{Kruskal1952} while \citet{Chenouri2012} use data-depth---measuring how far each observation is from the overall sample mean in terms of Mahalanobis distance---to calculate another ranks based test. \citet{Marozzi2014} also uses a rank-based test to simultaneously examine location and scale changes. 
Rather than ranks, work by \citet{Mukherjee2020} instead forms a non-bipartite graph and extends the crossmatch test of \citet{Rosenbaum2005} to multi-valued treatments.


In this vein, we consider several graphs structures on the data that range from Hamiltonian paths connecting all of the data into a single graph to kNN graphs.
Hamiltonian paths are graphs that connects all of the observations in a dataset but with the constraint that each observation is only visited once. 
If we imagine the path as a highway, we will see all of the observations along the highway, but there is no way to see the old points we have visited without turning around---in graph terminology, there are no cycles. 
There are many such paths through the data under this broad definition, but we will define the shortest of such possible paths as the optimal Hamiltonian path. Under this graph, if the distributions vary between treatment groups we expect units from the same distribution to be closer to each other along the path. 


For nearest neighbor graphs, the goal is not to find a path but to find matches---\textit{i.e.}, neighbors---that are close to each other in terms of a distance metric. This can be done in a way that allows for overlap between matched sets or in a way that forces those matched sets to be disjoint. In both types of nearest neighbor graphs, if the distributions vary between treatment groups, we expect observations from the same treatment group to be connected more frequently to each other than to observations from different treatmentes.

Several methods use these graphs for  two-sample testing.
In Hamiltonian paths, \citet{Biswas2014} use the Wald-Wolfowitz  runs test \citep{Wald1940} along the path to test for distributional similarity in the two-sample setting.
For nearest neighbor graphs, the crossmatch test of \citet{Rosenbaum2005} creates non-overlapping kNN graphs for $k=1$ and counts the number of matches that occur \textbf{across} samples. Recent work by \citet{Chen2020} develops a similar test on greedy kNN graphs that instead counts the number of times nearest neighbors come from the \textbf{same} sample.

Unfortunately, there has not been much work extending these particular graph-based tests to the multisample setting. As previously mentioned, \citet{Mukherjee2020} extend Rosenbaum's crossmatch test to the multisample setting, but similar examples exist neither for Hamiltonian paths nor for greedy nearest neighbor graphs. In addition, the empirical performance of these methods is unclear in practice, and to our knowledge, there is no readily available computational implementation of these methods.

To bridge these gaps with matched samples, we offer four contributions. First, we develop similar extensions for tests along the Hamiltonian path. Second, we provide a multisample extension to the nearest neighbor test of \citeauthor{Chen2020}. Third, we conduct an extensive simulation study of these tests in a variety of settings, finding that the kNN test outperforms others. Fourth, we implement all of these tests into the new \texttt{R} package \texttt{graphTest}. 

Next, we briefly describe our setting in Section \ref{sec:setting_graphtest}, give a motivating example in Section \ref{sec:motivating_graphtest}, and describe how we construct and perform tests on the graphs in Section \ref{sec:alg_graphtest}. Then we provide results from simulations and a case study in Sections \ref{sec:sims_graphtest} and \ref{sec:case_study_graphtest}, respectively. Finally, Section \ref{sec:summary_graphtest} concludes.

\section{Setting} \label{sec:setting_graphtest}
Suppose we have covariate data from $G$ treatment groups: $\{  X_1^{(1)},\ldots, X_{n_1}^{(1)} \},$ $\ldots,$ $\{X_1^{(G)},$ $\ldots,$ $X_{n_G}^{(G)}\}$  with $n_1 + n_2 + \cdots + n_G = N$. These groups could arise from sampling from different populations or because observations from the same population have chosen different treatments.  As such, our goal is to see if after matching or some other adjustment, we can successfully evaluate if the distribution in each group is the same---\textit{i.e.} $H_0: F_1 = F_2= \cdots = F_G$.

Constructing graph-based tests studied in this work requires a notion of closeness  between units and a way to denote which units are connected. 
We define $\boldD$ as the pairwise distance matrix between all points in the sample with entries $\boldD_{ij} = {\boldd}(X_i, X_j), \,\forall i \neq j, \,i,j \in \{1,...,N\}$, 
for some distance $\boldd: \R^d \times \R^d \mapsto \R_+ $. Further, let $\boldM$ be a matrix whose entries indicate whether unit $i$ is connected to unit $j$. 
If unit $i$ is connected to unit $j$, then $\boldM_{ij} = 1$. Otherwise, $\boldM_{ij} = 0$.

\section{Motivating example}\label{sec:motivating_graphtest}

To better understand this setting and the challenges it presents, we now consider a small example where individuals can choose one of three treatments. Our setting has only two confounders, 	$X_1, X_2 \iid \N(0,1)$ and we then generate a treatment indicator $Z \sim \operatorname{Categorical}(\boldp)$, where $\boldp = ( p_1,  p_2,   p_3  ),\trans \,  p_i = \frac{\exp(\eta_i)}{\sum_{j=1}^3 \exp(\eta_j)} $
and $\eta_1 =  0.1 X_1 - 0.1 X_2 -X_1X_2, 	\eta_2 = - 0.2 X_1 + 0.2 X_2 + 0.5 X_1^2 , \eta_3 = -0.1 X_1 + 0.2 X_2 - 2 X_1 X_2.$
We set the sample size to $N = 150$.

\begin{figure*}[tb]
	\centering
	\includegraphics[width=0.8\linewidth]{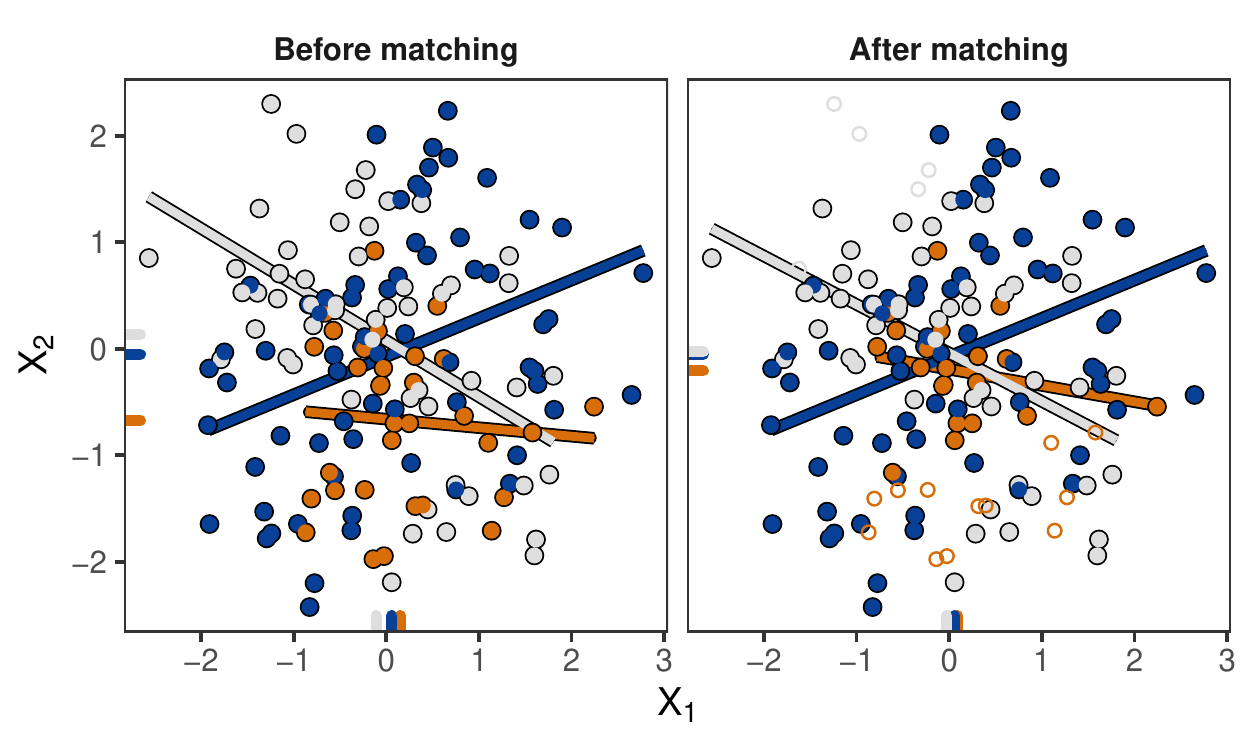}
	\caption{One draw of the covariates before and after matching with cardinality matching targeting the means of each group to the overall sample mean. Open circles in the second plot denote observations that were dropped from the sample because they were unmatched. The lines denote the regression of $X_2$ on $X_1$ which will be equal to the scaled correlation between the variables for centered covariates. Tick marks are the means in each group for the respective covariate. We can see the means are balanced but some imbalance remains in the correlations between the variables. This figure appears in color in the electronic version of this article.}
	\label{fig:match_fig}
\end{figure*}

Typically, researchers will not know the true treatment-assignment mechanism and we emulate this situation in our setup.
We use cardinality matching \citep{Zubizarreta2014} to balance first moments between each group and the overall sample mean but ignore the covariate functions $X_1^2$ and $X_1 X_2$ necessary to capture the true propensity score model. Matching towards the overall sample means will target the average treatment effects for the whole sample \citep{DelosAngelesResa2016, Bennett2020}. We use the \texttt{R} package \texttt{designmatch} \citep{Zubizarreta2018} and set the parameters of the matching such that the group means should differ from the sample mean by no more than 0.1 standard deviations.
The results of such a matching for one draw of the data is shown in Figure \ref{fig:match_fig}. 

The covariate balance before and after matching for this draw in  Table \ref{tab:mot_eg_bal} would indicate that the matching was a success. Further, the absolute standardized mean difference, defined here as $\frac{1}{3} \sum_{g=1}^3 \frac{|\mean{X}_g - \mean{X}|}{\sqrt{\widehat{\var}(X)}}$, also indicates that all groups are within 0.1 standard deviations of the sample mean on average, and $F$ tests for differences in means fail to reject the null hypothesis that the group means are the same.

\begin{figure*}[tb]
	\centering
	\begin{subfigure}[t]{.4\linewidth}
		\includegraphics[width = \textwidth]{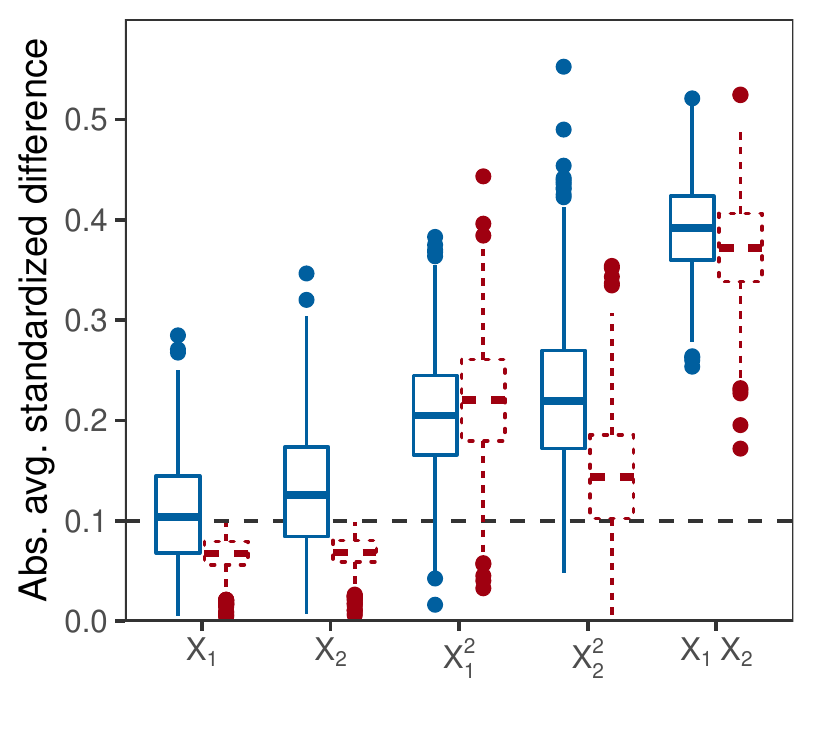}
		\caption{Absolute average standardized difference in means.}
		\label{fig:mot_std_diff}
	\end{subfigure}
	\hspace{0.2em}
	\begin{subfigure}[t]{.52\linewidth}
		\includegraphics[width = \textwidth]{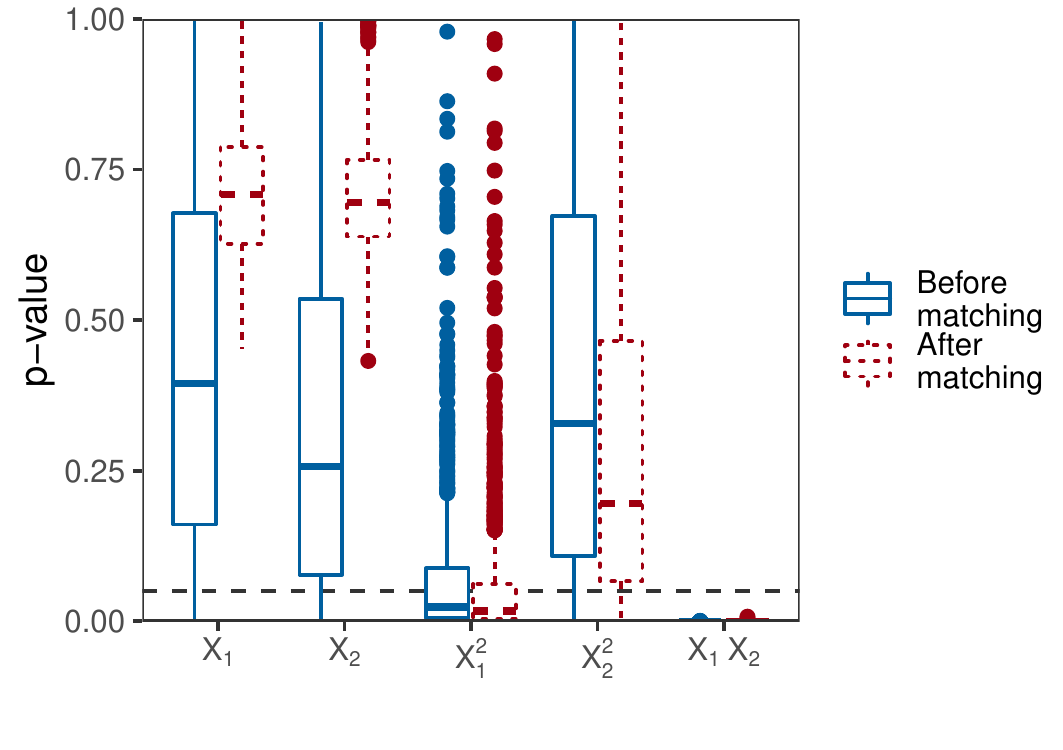}
		\caption{$p$-values.}
		\label{fig:mot_pval}
	\end{subfigure}
	\caption{Univariate evaluations before and after matching evaluated over 1000 ex\-per\-i\-ments. Dotted lines denote 0.1 standard deviations (a) and a $p$-value of 0.05 (b). All tests are $F$ tests of the null hypothesis that the group means do not differ for the listed covariate functions. This figure appears in color in the electronic version of this article.}
\end{figure*}

Unfortunately, the matching model misses two important features of treatment selection: $X_1^2$ and $X_1X_2$. Running this experiment 1000 times, we see that while standardized differences of means $X_1$ and $X_2$ are well balanced on average, the other covariate functions are not (Figure \ref{fig:mot_std_diff}). Similarly, univariate tests of covariate means are insufficient to diagnose the statistically significant deviations in the other covariate functions (Figure \ref{fig:mot_pval}). Clearly, we need an omnibus test to diagnose failure without having to check every possible covariate moment.

\begin{table}[tb]
\centering
\begin{tabular}{l|r|rrrrr|rrrrr}
  \hline \multicolumn{2}{c}{} & \multicolumn{5}{c|}{Before matching} & \multicolumn{5}{c}{After matching} \\ \hline
Cov. & Sample & Group 1 & 2 & 3 & Std. Diff & p & Group 1 & 2 & 3 & Std. Diff & p \\ 
  \hline
$X_1$ & 0.02 & 0.15 & 0.06 & -0.11 & 0.10 & 0.49 & 0.09 & 0.06 & -0.04 & 0.05 & 0.85 \\ 
  $X_2$ & -0.12 & -0.67 & -0.05 & 0.13 & 0.27 & 0.00 & -0.21 & -0.05 & -0.03 & 0.08 & 0.78 \\ 
   \hline
\end{tabular}
\caption{Table of mean balance before and after mixed integer program matching for one sample. Data were matched to obtain group means within 0.1 standard deviations to the mean of the overall sample mean. p-values (p) indicate good balance on the covariate means after matching.} 
\label{tab:mot_eg_bal}
\end{table}

With this goal in mind, we now examine some graph-based tests that may offer a solution.

\section{Structures, algorithms, and tests}\label{sec:alg_graphtest}

This section considers two graph structures: Hamiltonian paths and nearest neighbor graphs. For each, we describe the graph structures, the algorithms to construct them, and the tests that we will run on these structures. Throughout, we will use the motivating example of the previous section for exposition in our figures.

\subsection{Hamiltonian paths}
As stated previously, Hamiltonian paths are graphs through the data where each point is visited once and the starting and ending points are not the same. A Hamiltonian cycle is similar except that the starting and ending points are the same. Optimal Hamiltonian cycles are typically known in the literature as the Traveling Salesman Problem or TSP \citep{Dantzig1954}.

\subsubsection{Algorithms}
\begin{figure*}[tb]
	\begin{subfigure}[t]{\linewidth}
		\includegraphics[width =\textwidth]{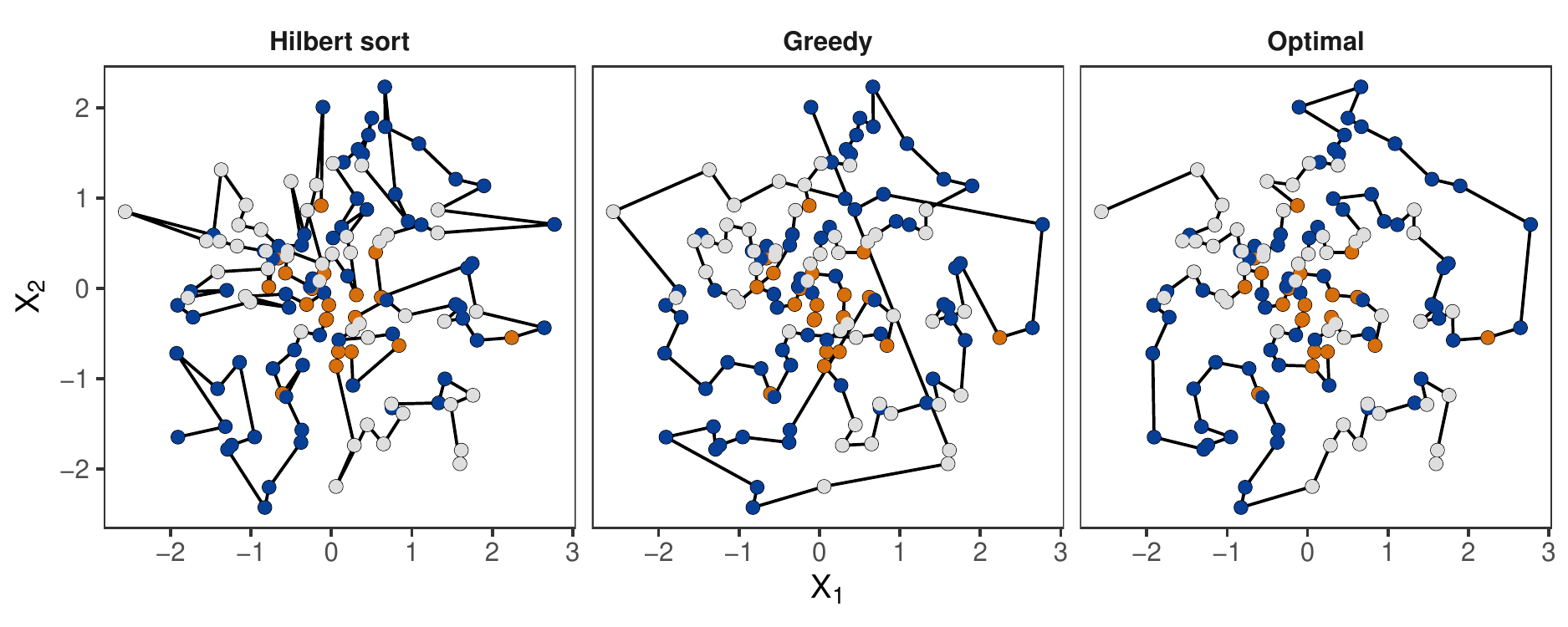}
		\caption{Paths from different algorithms}
	\end{subfigure}
	\begin{subfigure}[b]{\linewidth}
		\includegraphics[width =\textwidth]{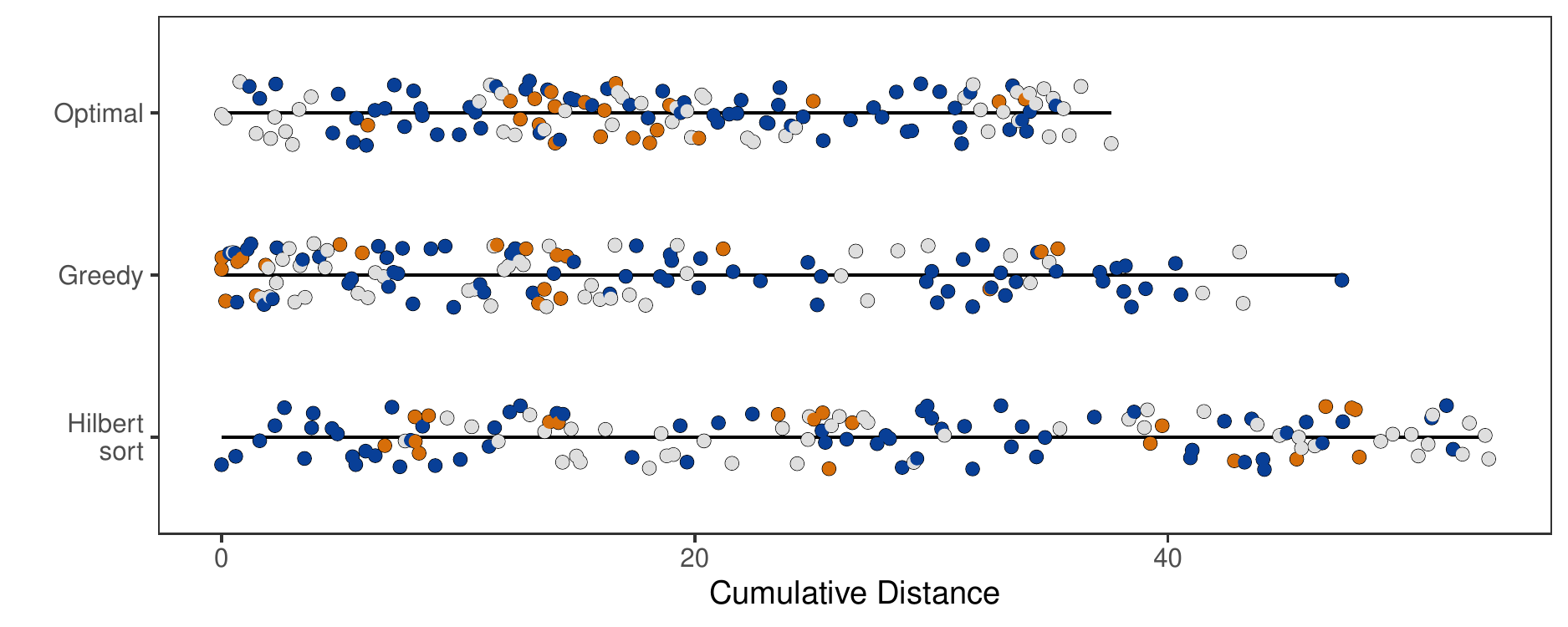}
		\caption{Distances of paths from different algorithms}
		\label{fig:path_length}
	\end{subfigure}
	\caption{Hamiltonian paths constructed by the Concorde solver (optimal), a greedy path constructed by Kruskal's algorithm, and a path constructed via a Hilbert sort. Data is the same as the matched sample from Figure \ref{fig:match_fig}. Distances show the increasing length of the Hamiltonian path for various methods. Points in Figure (b) have been jittered to better visualize their location along the path. This figure appears in color in the electronic version of this article.}
	\label{fig:path_eg}
\end{figure*}

There are a variety of ways to construct Hamiltonian paths. We present three algorithms to do so in order of increasing speed and decreasing optimality (Figure \ref{fig:path_eg}).

Our optimal algorithm is the Concorde solver of of \citet{Applegate2003}. Unfortunately, Concorde requires integer distances, which forces us to transform $\boldD$ and risk either the loss of precision or numeric overflow in order to preserve accuracy.
Despite these concerns, the Concorde solver still finds optimal paths on the moderately sized problems considered in this paper and does so much faster than a brute-force integer programming solution.

The next solution we consider comes from an application of Kruskal's algorithm \citep{Kruskal1956}. Using the pairwise distance matrix $\boldD$,
we add edges in a greedy fashion starting with the two observations with the shortest pairwise distance. Then we sequentially add the next closest unit to the end of the path until we construct a full Hamiltonian path. While not optimal globally, this path will be more optimal locally for some subset of observations.

The final algorithm is a sorting algorithm based on the Hilbert space-filling curve \citep{Hilbert1891}. The idea is to construct a Hilbert space-filling curve until it forms a Hamiltonian path through the data. This method also prescribes an ordering to the data points, and for this reason, the method is called Hilbert sorting.
Since the constructed path is deterministic, Hilbert sorting is incredibly fast---successfully sorting 10 million observations in about 20 seconds on a circa 2018 laptop---but the path is no where close to optimal globally or locally. To perform this sorting method, we use the  implementation freely available from \citet{CGALref2021} with an \texttt{R} interface relying on the \texttt{approxOT} package \citep{Dunipace2021}.

\subsubsection{Test statistics}
After path construction, the Hamiltonian path forms a one-dimensional radon measure on the manifold created by the corresponding graph \citep{Biswas2014}. Due to this one dimensional nature, the Hamiltonian path will admit a variety of one-dimensional statistics and corresponding tests. In this work we will consider two such tests: the ranks of observations along the Hamiltonian path and the runs of observations from the same treatment group \citep{Kolmogorov1933,Smirnov1948, Birnbaum1960, Mood1940}.

For these quantities, we consider two general forms of tests.  The first form is an extremum test---either a minimum or maximum---that looks at the the probability of observing the most extreme group under $H_0$: $\boldS = (S_1, S_2, ..., S_G)\trans$ is the vector of statistics and $T_l = \min(\boldS)$ or $T_u = \max(\boldS)$ are the corresponding test statistics. 
The second test has the form of a Wald test---that is, given $\Sigma = \var(\boldS)$, and $\mu = \E(\boldS)$, then 
$T_w = (\boldS - \mu)\trans \Sigma \inv (\boldS - \mu).$

The probabilities of these test statistics will naturally depend on the distributions of the statistics $\boldS$, which are often asymptotically Gaussian. Under a Gaussian distribution, the extremum can easily be calculated from the \texttt{mvtnorm} package in \texttt{R} \citep{Genz2020}. Meanwhile, $T_w \sim \chi_{\nu}^2,$ where $\nu$ is the degrees of freedom---typically $G$ or $G-1$---with probabilities easily calculated in any statistical software.

Finally, we note that for the ranks-based tests, we will only consider the Wald test since the extremum test is not easily obtained. For the runs-based test, we examine both the Wald and extremum tests.

\subsection{Nearest neighbor graphs}
\begin{figure*}[!tb]
	\includegraphics[width=\textwidth]{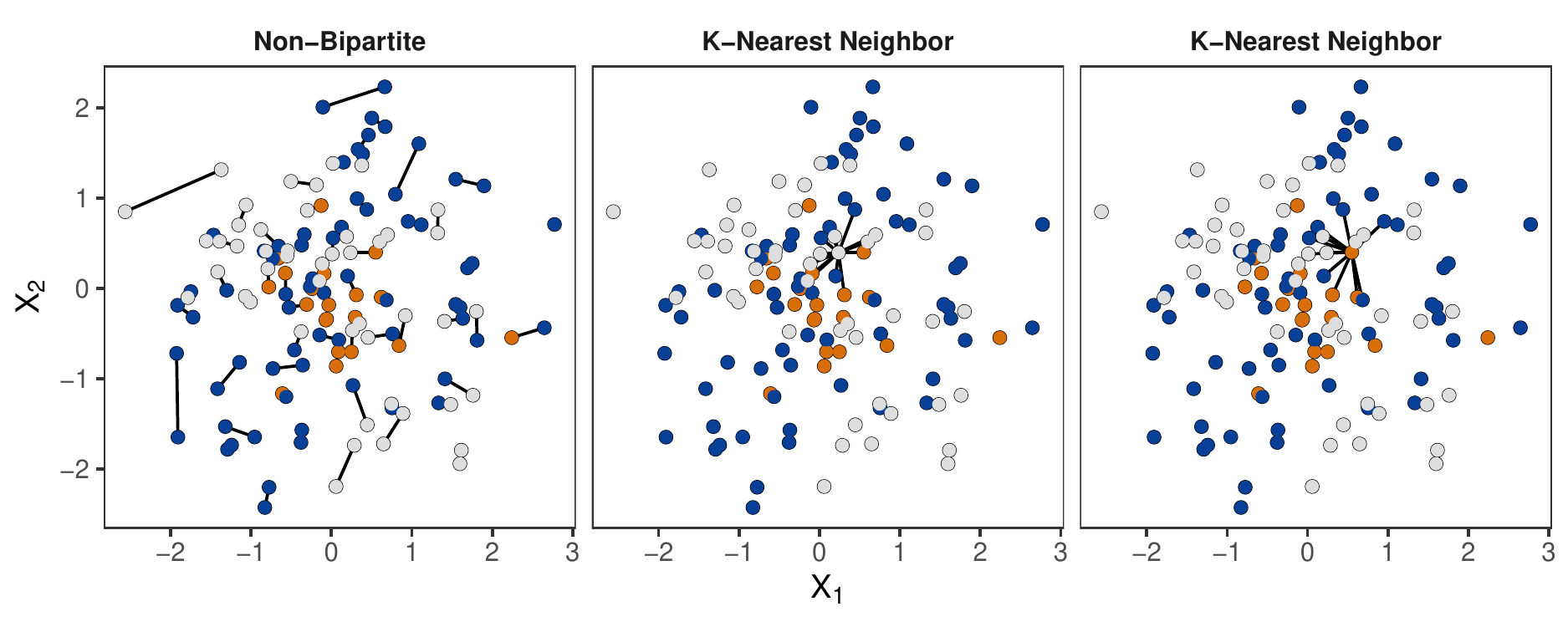}
	\caption{Nearest neighbor graphs constructed by an optimal non-bipartite matching (left) or via a greedy nearest neighbor search with 12 nearest neighbors (middle and right). Data come from the matched sample of Figure \ref{fig:match_fig}. The greedy nearest neighbors only shows the graphs for two separate points for clarity. This figure appears in color in the electronic version of this article.}
	\label{fig:nn_eg}
\end{figure*}

Similar to Hamiltonian paths, nearest neighbor graphs construct groups that are close to each other in terms of their covariate distance, $\boldD$. Also, nearest neighbor graphs can be constructed in an optimal or greedy way, as can be seen in Figure \ref{fig:nn_eg}.

\subsubsection{Algorithms}\label{sec:nbp_graphtest}
Optimal nearest neighbor graphs, also known as non-bipartite matching (NBM), minimizes the total distance between matches, which may lead to units that are not matched to their closest possible match, while the constraints force each unit to only be matched to one individual. Unlike the TSP, this problem is quick to solve for pairwise matching $(k = 1)$, but for larger numbers of matches, the problem is NP-hard \citep{Karp1972}.

In contrast, the greedy formulation is quick to solve for any number of matches, $k \in \Nat, k > 0$. One simply takes the pairwise distance matrix $\boldD$ and finds the $k$ nearest matches to the observation of interest. This is done with replacement so that matches are allowed to be shared across observations, unlike in NBM. More formally, the kNN problem is 
\begin{alignat*}{2}
	\min_{\boldM \in \{0,1\}} & \sum_{i=1} \sum_{j \neq i} \boldD_{ij} \boldM_{ij} \\ 
	\text{such that} & \sum_{j=1, j \neq i} \boldM_{ij} = k, \quad & \forall i \in \{1,...,N\} \\
	& \boldM_{ii} = 0, \quad & \forall i \in \{1,...,N\},
\end{alignat*}
Fortunately, there are algorithms that efficiently compute the nearest neighbor graph without having to construct the full distance matrix. Several of these, such as the k-d tree \citep{Bentley1975}, are implemented in the \texttt{R} package \texttt{FNN} \citep{Beygelzimer2019}.

\subsubsection{Test statistics}\label{sec:nn_stat}

Unlike our tests along Hamiltonian paths, the tests on nearest neighbor graphs respect how the graphs are built. As such, we describe the corresponding statistics for NBM and kNN graphs separately. Also, we consider both extremum and Wald tests as above.

\paragraph{Crossmatch statistic} The crossmatch statistic uses the structure of NBM to see how many observations are matched across treatment groups. Intuitively, if the null is true that $F_1 = \cdots = F_G$, then we would expect there to be a high number of matches across groups \citep{Rosenbaum2005, Mukherjee2020}.
For the crossmatch test, the extremum test is a minimum test since we expect individuals will be matched across groups \textit{less} often than we would expect if $H_0$ is false.

\paragraph{$k$-Nearest neighbors statistic} We now present our multisample version of the statistic developed by \citet{Chen2020}. For  a greedy nearest neighbor graph, we look at each observation $i$ and count how many of  $i$'s nearest neighbors come from the same treatment group. Under the alternative, we would expect this number to be high since it would indicate observations from the same treatment group are more likely to be close to one another, which would indicate a difference between the distributions.
If this happens more than we would expect under the null, we conclude the distributions are different.

More formally, define $C_g = \sum_{i , j} \boldM_{ij} \indicator(Z_i = Z_j)\indicator(Z_i = g), \, \forall g \in \{1,...,G\}$, or the number of times units from group $g$ are matched to other units from group $g$. We then have the following lemma for $C_g$.

\begin{lemma} \label{lemma:knn_moment}
	The expectation and variance of $C_g$ are 
	\begin{equation}
		\E(C_g) =  k \frac{n_g (n_g - 1)}{N-1}
		\label{eq:expect_knn}
	\end{equation}
	and 
	\begin{align}
		\var(C_g) &= \frac{n_g(n_g - 1)(N - n_g)(N - n_g -1)}{N(N-1)(N-2)(N-3)} \, \,\times \label{eq:var_knn}\\
		& \quad \quad \left(kN + 2J + \frac{n_g - 2}{N-n_g  - 1}\left(2 S + k N - k^2 N \right) \right. \nonumber \\ 
		& \quad \quad \quad \left. - \frac{2}{N-1} k^2 N  \right)\nonumber .
	\end{align}
	The covariance of $C_g$ and $C_h$ for $g \neq h$ is 
	\begin{align}
		\cov(C_g, C_h)  &= \frac{n_g(n_g - 1)n_h (n_h - 1)}{N(N-1)(N-2)(N-3)} \times	\label{eq:cov_knn}\\
		 & \quad \quad \left(2J -2S + \frac{k^2 N (N-3)}{N-1} \right) , \nonumber
	\end{align}
	where $J = \sum_{i=1}^{N-1} \sum_{j=1}^N \boldM_{ij} \boldM_{ji}$ is the number of pairs of observations that are mutual nearest neighbors, and $S = \sum_{i=1}^N \sum_{j=1}^{N-1}\sum_{l=j}^N \boldM_{ji} \boldM_{li}$ is the number of pairs of observations that share a nearest neighbor.
\end{lemma}
\noindent For a proof, see \ref*{sec:proof}. 
The basic idea is to sum over $\boldM$ and multiply by the probability that a given set of nodes come from group $g$.

We now turn to the asymptotic distribution of the counts $C_g$. In principle, we can use the exact permutation distribution to calculate the $p$-values of tests using this statistic; however, this is computationally very expensive so we do not do so in practice. Before proceeding, we note that Propositions 3.1 and 3.2 of \citet{Henze1988} show that $2J/N$ and $2S/N$ converge to constants $c_1$ and $c_2$ that only depend on two things: the distance metric used to calculate the nearest neighbors and the dimension of the covariate space. Then we define $U_g =\frac{C_g  - \E(C_g)}{\sqrt{\var(C_g)}}$ as the centered and scaled version of $C_g$. 
	This brings us to the asymptotic distribution of the $U_g$.

\begin{theorem} \label{thm:knn_normality}
	Under the permutation null distribution as $N \to \infty$,
	\[(U_1,...,U_G)\trans \dist \N \left(\mathbf{0}, \Omega \right), \]
	with $\Omega_{gg} = 1$. For $g \neq h$,
	
	\[  \Omega_{gh} =  \frac{p_gp_h}{(1-p_g)(1-p_h)} \frac{c_1 - c_2 + k^2}{\sqrt{f_g f_h}}, \]
	where 
	\begin{align*}
		p_g &= \lim_{n_g, N \to \infty} \frac{n_g}{N},\\
		f_g &=  k + c_1 + \frac{p_g}{1-p_g} (c_2+ k - k^2) .
	\end{align*}
\end{theorem}
\noindent We provide a proof in \ref*{sec:proof}.

In practice, we utilize the finite sample correlations between counts 
\[\hat{\Omega}_{gh} =  R_{gh} \frac{2J - 2S + k^2 \frac{N (N-3)}{N-1}}{\sqrt{\hat{f}(n_g)\hat{f}(n_h)}} ,\]
with
\[R_{gh} = \frac{\sqrt{n_g (n_g-1)n_h(n_h-1)}}{\sqrt{(N-n_g)(N-n_g-1)(N-n_h)(N-n_h - 1)}}\]
and
\[\hat{f}(n) =  kN + 2 J + \frac{n-2}{N-n-1} (2S + kN - k^2N) - \frac{2}{N-1} k^2N.\]
This will have $\Omega_{gh}$ as its limit.
Further, given the discrete nature of the counts, we utilize a continuity correction such that $U_g = \frac{C_g - 0.5 - \E(C_g)}{\sqrt{\var(C_g)}}$.

For the kNN statistics, the extremum test is a maximum based test since we expect that individuals will be matched to their own group \emph{more} often if $H_0$ is false.
The Wald test using the kNN statistic can be calculated by $T_{\text{wald-knn}} = \boldU \trans \Omega \inv \boldU$ and $T_{\text{wald-knn}} \sim \chi_G^2.$

Finally, we note that the power of the tests based on the kNN statistic can be sensitive to the choice of $k$. \citet{Bhattacharya2019} finds that $k$ should increase with the sample size in order for the test to be well-powered. 

\section{Simulation studies}\label{sec:sims_graphtest}
We now present simulation results in a variety of settings. Note that we set $k = \lfloor 0.1 N \rfloor$ in each case where we use the kNN test. 

In the first simulation study, we return to the motivating example from Section 3 to see if the new tests can correctly diagnose deviations from the null hypothesis that $F_1 = \cdots = F_G$ missed by univariate tests. Then in the second simulation study, we utilize the settings of \citet{Mukherjee2020} to evaluate the performance of the graph-based tests. 
Overall, we find greedy paths and greedy nearest neighbor constructions perform better than their optimal counterparts while the kNN test performs best on average.

\subsection{Motivating example, continued}\label{sec:motiv_eg_res}

Returning to the motivating example from Section \ref{sec:motivating_graphtest}, we again use cardinality matching to target distributional balance. As before, we use a misspecified setting that matches on $X_1$ and $X_2$ but misses the important covariate functions  $X_1^2$ and $X_1 X_2$, but this time we also add a well-specified setting that includes all relevant covariate moments. We again match treatment groups so that the group means on the selected covariate functions are within 0.1 standard deviations, and we still utilize 1000 experiments with a sample size of 150. For each experiment, we run the kNN and crossmatch tests on their corresponding graphs and the runs and ranks test on a greedy Hamiltonian path built via Kruskal's algorithm. Our implementation of Kruskal's algorithm relies on the \texttt{TSP} package developed by \citet{Hahsler2007} in \texttt{R}.

\begin{figure*}[!tb]
	\centering
	\includegraphics[width= \textwidth]{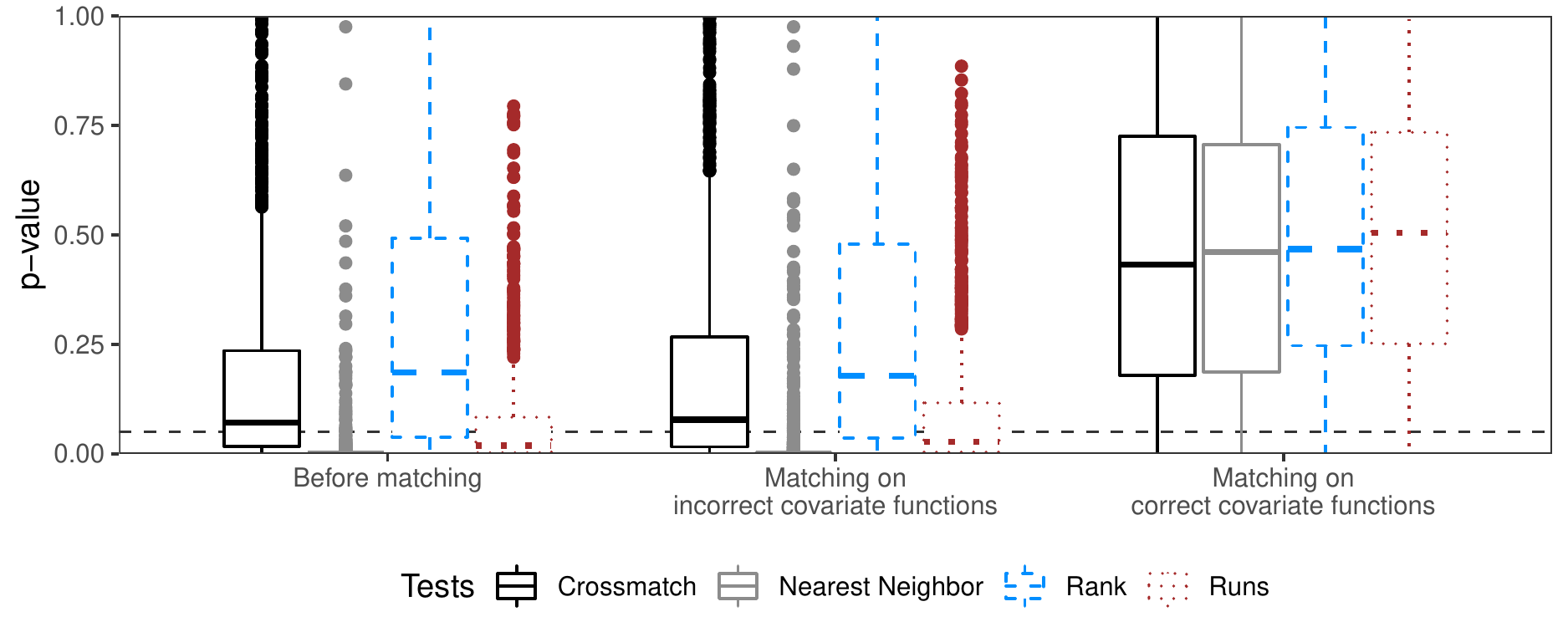}
	\caption{Graphical tests run before matching (left), after matching with a misspecified model (middle), and after matching with a well-specified model (right). This figure appears in color in the electronic version of this article.}
	\label{fig:graph_motiv_eg}
\end{figure*}

Figure \ref{fig:graph_motiv_eg} reveals that kNN and the runs test successfully detect distributional imbalance both initially and in the misspecified models. Of concern is that the crossmatch and ranks test also fail to correctly reject the null even at baseline in many cases.

\subsection{Setting of \citet{Mukherjee2020}}\label{sec:agarwal_desc}
For our second simulation study, we use the setting of \citet{Mukherjee2020}.
In their experiments, the authors use a multivariate Gaussian data and either change (1) the parameters of the distribution or (2) the structure of the problem. The distributions are always drawn from 
\[X_1^{(g)},...,X_{n_g}^{(g)} \iid \N(\mu_g, \Sigma_g),\] where $\mu_g \in \R^d$, $\Sigma_k \in \R^{d \times d}$, and $g \in \{1,...,G\}$. We set $n_g = 50 g $ and run 1000 experiments in each setting. 

\paragraph{(1) Parameter values} We change the parameter values in one of three ways. The first is location changes $\mu_g = (g-1) \delta \ones_d,$ with $\Sigma_g = \ident_d;$ the second is scale changes with $\mu_g = \mathbf{0}$ and $\Sigma_g = \{1 + (g-1)\delta\} \, \ident_d$; and the third is correlation changes with $\mu_g = \mathbf{0}$ and $\Sigma_g = (1- \rho_g)\ident_d  + \rho_g \ones_{d \times d}, \quad \rho_g = \frac{(g-1)\delta}{g-1},$
where $\ones_{d \times d}$ is a $d \times d$ is a matrix of ones, $\ones_{d}$ is a $d$-length vector of ones, and $\ident_d$ is a $d \times d$ identity matrix.
In location change, we set $\delta \in \{0.04, 0.06, 0.08, 0.10, 0.12\}$; in scale changes and correlation changes, we set $\delta \in \{0.0, 0.05, 0.10, 0.15, 0.20, 0.25, 0.30\}$.

\paragraph{(2) Structure changes} For changes to the problem structure, we either change the dimension of the covariates---$d \in \{5,10,25,50,100,250\}$---or the number of treatment groups---$G \in \{3,4,5,6\}$.
Also note that as we change the dimension of the covariates, we keep $G$ fixed at five and as we change the number of treatment groups, we keep $d$ fixed at 150.

Finally, we present results averaged across simulation settings in this section to help simplify results. To see a detailed presentation of the results for each setting, please see Web Figures \ref*{fig:web_1}--\ref*{fig:knn_k_power}. 

\subsubsection{Extremum versus Wald tests}\label{sec:test_extrem_wald}

As previously discussed, there are two ways we can construct our tests: either an extremum (max or min) or a Wald test. Ultimately, we find that extremum tests perform better in tests along Hamiltonian paths while the Wald tests perform better for the nearest neighbor graphs.

To simplify exposition, we average across parameter change settings and present the relative power of Wald tests as compared to extremum tests, that is 
$\frac{\p(\text{reject }H_0  \given \text{Wald test})}{\p(\text{reject }H_0 \given  \text{ extremum test})}.$  Numbers greater than one indicate that the Wald test has higher power on average while numbers less than one indicate the extremum test has higher power for the given algorithm and structure. 

\begin{table}[tb]
\centering
\begin{tabular}{l|rrrrrr}
  \hline & \multicolumn{6}{c}{Dimension}\\ \hline
Algorithm & 5 & 10 & 25 & 50 & 100 & 250 \\ 
  \hline
Optimal & 0.69 & 0.72 & 0.71 & 0.75 & 0.80 & 0.87 \\ 
  Greedy & 0.68 & 0.71 & 0.77 & 0.83 & 0.90 & 0.92 \\ 
  Hilbert sort & 0.68 & 0.71 & 0.76 & 0.72 & 0.68 & 0.72 \\ 
  NBP & 1.44 & 1.54 & 1.50 & 1.52 & 1.42 & 1.38 \\ 
  kNN & 0.93 & 1.06 & 1.16 & 1.17 & 1.15 & 1.09 \\ 
   \hline
\end{tabular}
\caption{The relative power of the Wald-like test to the extremum test for the listed algorithms. The results are with an increasing dimension of the covariate space and averaged across the location, scale, and correlation changes. The algorithms above the double line are ones used to construct Hamiltonian paths while those below the double line, non-bipartite matching (NBP) and k-nearest neighbors (kNN), are used to construct nearest neighbor graphs. The number of groups $G$ is held fixed at 5.} 
\label{tab:extrem_dim}
\end{table}

\begin{table}[htb]
\centering
\begin{tabular}{l|rrrr}
  \hline & \multicolumn{4}{c}{Treatments}\\ \hline
Algorithm & 3 & 4 & 5 & 6 \\ 
  \hline
Optimal & 1.03 & 0.87 & 0.82 & 0.77 \\ 
  Greedy & 1.09 & 0.95 & 0.89 & 0.88 \\ 
  Hilbert sort & 1.59 & 1.06 & 0.71 & 0.56 \\ 
  NBP & 0.92 & 1.19 & 1.43 & 1.80 \\ 
  kNN & 1.30 & 1.21 & 1.15 & 1.06 \\ 
   \hline
\end{tabular}
\caption{The relative power of the Wald-like test to the extremum test for the listed algorithms. The results are with an increasing number of treatments and averaged across the location, scale, and correlation changes. The algorithms above the double line are ones used to construct Hamiltonian paths while those below the double line, non-bipartite matching (NBP) and k-nearest neighbors (kNN), are used to construct nearest neighbor graphs. The number of covariates $d$ is held fixed at 150.} 
\label{tab:extrem_group}
\end{table}

Table \ref{tab:extrem_dim} presents the average results for each graph structure as the covariate dimension is increased while holding the number of treatment groups fixed at $G = 5$. For the Hamiltonian path algorithms, the Wald tests have lower relative power for smaller covariate dimensions, with decreasing differences as the dimension of the problem increases. In contrast, for the nearest neighbor graphs, the Wald-like tests have higher power on average as the covariate dimension increases. Only kNN with 5 covariates bucks this trend.

Similarly, Table \ref{tab:extrem_group} shows the results as we change the number of groups. We note that the Wald tests have higher power for the Hamiltonian paths with three groups. However, this reverses as the number of groups increases and there seems to be a general trend that the extremum tests have higher relative power as we consider a larger numbers of groups. 

In contrast, the nearest neighbor graph algorithms display higher power for the Wald tests in most cases; however, kNN and NBP display opposite trends in relative power as the number of groups increase. The kNN algorithm has a \emph{decreasing} difference between the Wald and extremum tests as the number of groups increases; the NBP algorithm displays an \emph{increasing} difference between the two tests as the number of groups increases with the Wald test having higher relative power.

\subsubsection{Greedy paths perform better than optimal ones}\label{sec:test_opt}
We might assume \textit{a priori} that the optimal Hamiltonian path will perform better than its greedy version because it is the shortest path through the data. Unfortunately, our experiments do not support this idea.

In Table \ref{tab:path_dim}, we present the results from changing the the dimension of the covariate space for our algorithms: optimal, greedy, and Hilbert sorting. 
For comparison purposes, we present results relative to the power of the tests on an optimal Hamiltonian path: $\frac{\p(\text{reject }H_0 )}{\p(\text{reject }H_0 \given  \text{ optimal path})}.$
As before, we average the results across parameter changes and present the average relative power for each path as we change the problem dimension, $d$, and number of treatment groups, $G$.
On average, the greedy path does better than the optimal path, with increasing power as the dimension increases. In contrast, the Hilbert sort path displays increasing relative  power as the dimension of the problem increases. 

Similarly, in Table \ref{tab:path_group}, we see that the relative power of tests on the greedy Hamiltonian path is consistently four times higher than tests on the optimal path found by Concorde. As before, the Hilbert sort path still has the lowest relative power, but this effect decreases as the number of groups increases.

\begin{table}[tb]
\centering
\begin{tabular}{l|rrrrrr}
  \hline & \multicolumn{6}{c}{Dimension}\\ \hline
Path & 5 & 10 & 25 & 50 & 100 & 250 \\ 
  \hline
Optimal & 1.00 & 1.00 & 1.00 & 1.00 & 1.00 & 1.00 \\ 
  Greedy & 2.27 & 3.41 & 3.99 & 4.36 & 4.71 & 4.35 \\ 
  Hilbert sort & 1.02 & 1.03 & 0.84 & 0.74 & 0.65 & 0.51 \\ 
   \hline
\end{tabular}
\caption{The relative power of the listed paths compared to the optimal path found by Concorde. The results are with an increasing dimension of the covariate space and averaged across the location, scale, and correlation changes as well as the results from the runs test and rank test. The number of groups $G$ is held fixed at 5.} 
\label{tab:path_dim}
\end{table}

\begin{table}[tb]
\centering
\begin{tabular}{l|rrrr}
  \hline & \multicolumn{4}{c}{Treatments}\\ \hline
Path & 3 & 4 & 5 & 6 \\ 
  \hline
Optimal & 1.00 & 1.00 & 1.00 & 1.00 \\ 
  Greedy & 4.73 & 4.64 & 4.35 & 4.09 \\ 
  Hilbert sort & 0.50 & 0.56 & 0.57 & 0.60 \\ 
   \hline
\end{tabular}
\caption{The relative power of the listed paths compared to the optimal path found by Concorde. The results are with an increasing number of treatments and averaged across the location, scale, and correlation changes as well as the results from the runs test and rank test. The number of covariates $d$ is held fixed at 150.} 
\label{tab:path_group}
\end{table}

\subsubsection{The $k$-nearest neighbor test outperforms others}\label{sec:test_sim}
We now compare the two other graph-based tests---the multisample crossmatch and the kNN test---to the Kruskal-Wallis ranks and runs test ran on the greedy Hamiltonian paths. All power calculations are presented as relative to the power of the kNN test.

First, we examine the effect of increasing covariate dimension, $d$, while holding the number of treatment groups fixed at $G=5$. We can see in Table \ref{tab:agarwal_dim} that the kNN test outperforms the other tests, on average.
There are a few cases where some of the other tests perform better---for example, in location changes with a small covariate dimension, the crossmatch and runs test can sometimes perform better. 
However, such effects are not observed as dimension continues to increase and are also not observed in settings where the covariance matrix varies by group.

Next, we examine the effect of changing the number of treatment groups, $G$, while holding the covariate dimension fixed at $d=150$. In Table \ref{tab:agarwal_group}, we see that again the kNN test has the highest relative power compared to any other test on average. For smaller numbers of groups, the runs test performs as well as the kNN test, but this effect diminishes as the number of groups continues to increase. 

\begin{table}[htb]
\centering
\begin{tabular}{l|rrrrrr}
  \hline & \multicolumn{6}{c}{Dimension}\\ \hline
Test & 5 & 10 & 25 & 50 & 100 & 250 \\ 
  \hline
Rank & 0.61 & 0.64 & 0.68 & 0.69 & 0.72 & 0.76 \\ 
  Runs & 0.34 & 0.39 & 0.53 & 0.68 & 0.74 & 0.87 \\ 
  Crossmatch & 0.37 & 0.39 & 0.41 & 0.52 & 0.57 & 0.71 \\ 
  Nearest Neighbor & 1.00 & 1.00 & 1.00 & 1.00 & 1.00 & 1.00 \\ 
   \hline
\end{tabular}
\caption{The relative power of the listed tests compared to the nearest neighbor test. The results are with an increasing dimension of the covariate space and averaged across the location, scale, and correlation changes. Nearest neighbor test uses 10\% of the sample size as the number of nearest neighbors and the rank and runs test use a greedily constructed Hamiltonian path. The number of groups $G$ is held fixed at 5.} 
\label{tab:agarwal_dim}
\end{table}

\begin{table}[htb]
\centering
\begin{tabular}{l|rrrr}
  \hline & \multicolumn{4}{c}{Treatments}\\ \hline
Test & 3 & 4 & 5 & 6 \\ 
  \hline
Rank & 0.75 & 0.72 & 0.73 & 0.77 \\ 
  Runs & 0.72 & 0.76 & 0.81 & 0.84 \\ 
  Crossmatch & 0.49 & 0.55 & 0.63 & 0.70 \\ 
  Nearest Neighbor & 1.00 & 1.00 & 1.00 & 1.00 \\ 
   \hline
\end{tabular}
\caption{The relative power of the listed tests compared to the nearest neighbor test. The results are with an increasing number of treatments and averaged across the location, scale, and correlation changes. Nearest neighbor test uses 10\% of the sample size as the number of nearest neighbors and the rank and runs test use a greedily constructed Hamiltonian path. The number of covariates $d$ is held fixed at 150.} 
\label{tab:agarwal_group}
\end{table}

\section{Case study}\label{sec:case_study_graphtest}
Our goal is to use matching to make the distributions in the respective treatment groups more similar to the overall sample distribution. We will then use our graph-based tests to see if univariate tests miss any distributional differences. To do so, we use data from the 1987 National Medical Expenditures Survey data as collected by the \texttt{TriMatch} package in \texttt{R} \citep{breyer_jason_trimatch_2017}.

\subsection{Data}
The original survey included detail information about participants quantity and duration of smoking as pack-years, with one pack-year being defined as smoking one pack of cigarettes per day for one year. In our data, we include the following covariates: the individual's age at the last survey, their gender, whether the individual identified as white, whether the individual attended college, whether the individual had ever been married, whether the individual fell into the low income category, and whether the individual always wore a seat belt. For the binary variables, we add a very small jitter to make matching easier for the software. Finally, we constructed a composite treatment variable based on smoking history as: 1 = never smoker, 2 = former smoker with less than 15 pack-years of smoking, 3 = former smoker with more than 15 pack-years of smoking, 4 = current smoker with less than 15 pack-years of smoking, and 5 = current smoker with more than 15 pack-years of smoking. 
The number of individuals in each treatment group is presented in Table \ref{tab:nmes_n}.
\begin{table}[htb]
\centering
\begin{tabular}{rrrrrr}
  \hline
 & Treatment 1 & Treatment 2 & Treatment 3 & Treatment 4 & Treatment 5 \\ 
  \hline
Baseline & 9804 & 2073 & 2003 & 2326 & 3146 \\ 
  (A) & 8456 & 1512 & 502 & 696 & 1365 \\ 
  (B) & 8065 & 1381 & 458 & 637 & 1247 \\ 
  (C) & 8018 & 1367 & 453 & 629 & 1234 \\ 
  (D) & 6460 & 1070 &   0 &   0 & 755 \\ 
   \hline
\end{tabular}
\caption{Sample size in each treatment group. Baseline is the data before any adjustment, and cardinality matching is done matching individuals in each treatment group to achieve means within the desired calipers of the overall sample means. (A) seeks matches on the first moments such that the means are within 0.05 standard deviations (S.D.), (B) matches on the first moments such that means are within 0.01 S.D., (C) matches on the first moments such that means are within 0.005 S.D., and (D) matches first moments within 0.005 S.D. and matches second to fifth moments within 0.025 S.D.} 
\label{tab:nmes_n}
\end{table}

\subsection{Methods}
We consider four forms of cardinality matching to achieve distributional balance.
The first three forms of cardinality matching we use seeks to match individuals in each treatment group so that the group mean of each covariate is within either 0.05, 0.01, or 0.005 standard deviations of the sample mean. We denote these as Card. Matching (A), (B), and (C), respectively. The fourth and final form targets first through fifth moments of the covariate distributions and we denote this form as Card. Matching (D).  For all matching tasks, we set the maximum time at one hour.

To diagnose failures of distributional balance we utilize a couple of methods. First, we
examine single covariate measures of distributional balance such as the standardized difference of means and the relevant tests for differences in means. We present this in table form for the first moments (see Table \ref{tab:applied_nmes_uni_test}).
Second, we utilize the graphical tests described in this work including the kNN test, the runs and ranks tests on greedy Hamiltonian paths, and the crossmatch test.

\subsection{Results}
We now look to see how various balance assessment methods diagnose a lack of distributional overlap. 
Table \ref{tab:applied_nmes_uni_test} displays the average of the absolute standardized differences in means between treatment group and the sample means for all covariate first moments, and in column 2 we see that there are some covariates with substantial differences between the sample mean and the group means, on average. After cardinality matching on the first moments with a 0.05 standard deviation caliper (columns 4 and 5 of Table \ref{tab:applied_nmes_uni_test}), we see that the average absolute standardized difference in means between groups is within 0.1 standard deviations but all of the $p$-values are below 0.05. For a caliper of 0.01 (columns 6 and 7), the $p$-values are now above 0.05, which again holds true for a caliper of 0.005 standard deviations. After cardinality matching on the first through fifth moments (columns 10 and 11 of Table \ref{tab:applied_nmes_uni_test}), the picture looks similar, though the $p$-values are slightly lower.
From these metrics, we would conclude that sufficient balance has been achieved in most settings.
\begin{table}[htb]
\centering
\resizebox{\columnwidth}{!}{%
\begin{tabular}{|l|lr|lr|lr|lr|lr|}
  \hline & \multicolumn{2}{C{2.7cm}|}{Baseline} & \multicolumn{2}{C{2.7cm}|}{Card. Matching (A)} & \multicolumn{2}{C{2.7cm}|}{Card. Matching (B)} & \multicolumn{2}{C{2.7cm}|}{Card. Matching (C)} & \multicolumn{2}{C{2.7cm}|}{Card. Matching (D)} \\ \hline
Variable & Std. diff & p-value & Std. diff & p-value & Std. diff & p-value & Std. diff & p-value & Std. diff & p-value \\ 
  \hline
Age & 0.26 & $<0.01$ & 0.03 & 0.02 & $<0.01$ & 0.99 & $<0.01$ & 1.00 & $<0.01$ & 0.94 \\ 
  Any college & 0.05 & $<0.01$ & 0.04 & $<0.01$ & $<0.01$ & 0.95 & $<0.01$ & 1.00 & $<0.01$ & 1.00 \\ 
  Ever married & 0.21 & $<0.01$ & 0.05 & $<0.01$ & $<0.01$ & 0.92 & $<0.01$ & 0.99 & $<0.01$ & 0.95 \\ 
  Low Income & 0.08 & $<0.01$ & 0.03 & 0.04 & $<0.01$ & 0.98 & $<0.01$ & 1.00 & $<0.01$ & 0.99 \\ 
  Male & 0.19 & $<0.01$ & 0.05 & $<0.01$ & $<0.01$ & 0.94 & $<0.01$ & 1.00 & $<0.01$ & 0.99 \\ 
  Seatbelt always & 0.11 & $<0.01$ & 0.05 & $<0.01$ & $<0.01$ & 0.97 & $<0.01$ & 1.00 & $<0.01$ & 0.98 \\ 
  White & 0.09 & $<0.01$ & 0.05 & $<0.01$ & $<0.01$ & 0.97 & $<0.01$ & 1.00 & $<0.01$ & 0.98 \\ 
   \hline
\end{tabular}
}
\caption{Univariate tests and absolute standardized mean differences applied to the matches of various methods. Baseline is the data before any adjustment, and cardinality matching is done matching individuals in each treatment group to achieve means within the desired calipers of the overall sample means. (A) seeks matches on the first moments such that the means are within 0.05 standard deviations (S.D.), (B) matches on the first moments such that means are within 0.01 S.D., (C) matches on the first moments such that means are within 0.005 S.D., and (D) matches first moments within 0.005 S.D. and matches second to fifth moments within 0.025 S.D.} 
\label{tab:applied_nmes_uni_test}
\end{table}

%

However, all is not well with the joint distributions. While most of the univariate tests looked fine after adjustment, almost none of the multivariate tests in Table \ref{tab:applied_nmes_mvt_test} indicate that distributional balance has been achieved. For example, the kNN test (column 2) has $p$-values that are below numeric tolerance for tests of distributional balance at baseline and in cardinality matching on the first moments (Card. Matching A--C). Only matching on the first through the fifth moments achieves a non-significant $p$-value (Card. Matching D). A similar case holds for both the runs test and the crossmatch test of \citet{Mukherjee2020}. The ranks tests on the Hamiltonian path also achieves a non-significant $p$-value in Card. Matching (C) as well.

\begin{table}[tb]
\centering
\begin{tabular}{lrrrr}
  \hline
 & k-Nearest Neighbor & Runs & Ranks & Crossmatch \\ 
  \hline
Baseline & $1.11 \times 10^{-16}$ & $1.11 \times 10^{-16}$ & $1.11 \times 10^{-16}$ & $1.11 \times 10^{-16}$ \\ 
  Card. matching (A) & $1.11 \times 10^{-16}$ & $1.50 \times 10^{-10}$ & $1.90 \times 10^{-03}$ & $1.11 \times 10^{-16}$ \\ 
  Card. matching (B) & $1.11 \times 10^{-16}$ & $1.89 \times 10^{-15}$ & $2.12 \times 10^{-14}$ & $1.84 \times 10^{-10}$ \\ 
  Card. matching (C) & $1.11 \times 10^{-16}$ & $1.15 \times 10^{-09}$ & 0.07 & $8.39 \times 10^{-10}$ \\ 
  Card. matching (D) & 0.17 & 0.13 & 0.47 & 0.51 \\ 
   \hline
\end{tabular}
\caption{Multisample tests applied to the matches from the cardinality matching methods. Baseline is the data before any adjustment and cardinality matching is done matching individuals in each treatment group to achieve group means within the desired calipers of the overall sample mean for the selected covariate moments. (A) seeks matches on the first moments such that the means are within 0.05 standard deviations (S.D.), (B) matches on the first moments such that means are within 0.01 S.D., (C) matches on the first moments such that means are within 0.005 S.D., and (D) matches first moments within 0.005 S.D. and matches second to fifth moments within 0.025 S.D.} 
\label{tab:applied_nmes_mvt_test}
\end{table}

\section{Summary and remarks}\label{sec:summary_graphtest}
In this paper, we proposed multisample extensions to the Hamiltonian path testing of \citet{Biswas2014} and the nearest neighbor testing of \citet{Chen2020} and we evaluated these tests in a variety of settings.
We considered various graph structures, algorithms to form these structures, and tests of distributional balance. 

\paragraph{Findings}  In our simulation studies, we found that the extremum test performs best for the runs statistic, while for the statistics on nearest neighbor graphs, the Wald tests perform best.

For both Hamiltonian paths and nearest neighbor graphs, we found in our simulation studies that approximate solutions---that is greedy Hamiltonian paths and the greedy nearest neighbor graphs---perform better than their optimal counterparts. We suspect that the approximate solutions do better locally and may have higher power in certain regions of the data to detect differences in distribution.

Overall, we found the kNN test on the approximate nearest neighbor graph to be the highest power test compared to various alternatives. The Hamiltonian path based tests performed slightly worse that the kNN test in most settings and the crossmatch test fared the worst. Of some concern, both the crossmatch test and the ranks test on a greedy Hamiltonian path missed the model misspecification in the motivating example, while the results in the other simulation studies in Section \ref{sec:test_sim} are more mixed. Based on these results, we would recommend utilizing the kNN test to evaluate multivariate balance in settings with multi-valued treatments.

Our case study revealed how the univariate metrics typically used to assess balance fail to diagnose inadequate overlap in the joint distribution. Of some concern, we found that an exceedingly high number of moments were required to obtain non-significant tests for the overlap of the joint distribution.

\paragraph{Further research} One area that requires further exploration is that some settings may not require the strict distributional balance we are able to measure with the kNN test. Ultimately, balancing the conditional mean functions between treatment groups is sufficient since this will remove the bias in estimating treatment effects. For example, if the conditional mean function of the outcome $Y_i$ is $\E(Y_i \given X_i, Z_i) =  X_i \trans \beta+ \alpha Z_i$, then ensuring balance on the first moments of the covariates would ensure unbiased estimation of the treatment effects determined by $\alpha$. However, since this conditional mean function is not known in practice, determining when sufficient balance has been achieved on the correct functions of the covariates remains a challenge.

Finally, there may be room to invert the kNN test as a method for achieving distributional balance. One could construct an optimization problem that seeks to minimize the test statistic of the kNN test in order to achieve distributional balance. This would be useful in observational settings where balance cannot be achieved through randomization.

\section*{Acknowledgments}
The author would like to thank Claire Chaumont, Gang Liu, Aar\'on Sonabend, Lorenzo Trippa, and Jos\'e Zubizarreta for helpful comments and feedback on an earlier version of this manuscript.
This research was funded by generous support from NIH grant 5T32CA009337-40, the Department of Biostatistics at the Harvard T.H. Chan School of Public Health, and the David Geffen Scholarship from the David Geffen School of Medicine at UCLA. 
	
	\clearpage
	\printbibliography

@article{Biswas2014,
	author = {Biswas, Munmun and Mukhopadhyay, Minerva and Ghosh, Anil K.},
	doi = {10.1093/biomet/asu045},
	issn = {14643510},
	journal = {Biometrika},
	number = {4},
	pages = {913--926},
	title = {{A distribution-free two-sample run test applicable to high-dimensional data}},
	volume = {101},
	year = {2014}
}

@article{Rosenbaum1985,
	author = {Rosenbaum, Paul R. and Rubin, Donald B.},
	doi = {10.1080/00031305.1985.10479383},
	issn = {15372731},
	journal = {American Statistician},
	number = {1},
	pages = {33--38},
	title = {{Constructing a control group using multivariate matched sampling methods that incorporate the propensity score}},
	volume = {39},
	year = {1985}
}

@article{Kruskal1956,
	author = {Kruskal, Joseph B.},
	doi = {10.2307/2033241},
	issn = {00029939},
	journal = {Proceedings of the American Mathematical Society},
	month = {feb},
	number = {1},
	pages = {48},
	title = {{On the Shortest Spanning Subtree of a Graph and the Traveling Salesman Problem}},
	url = {https://www.jstor.org/stable/2033241?origin=crossref},
	volume = {7},
	year = {1956}
}

@article{Dantzig1954,
	author = {Dantzig, G. and Fulkerson, R. and Johnson, S.},
	doi = {10.1287/opre.2.4.393},
	issn = {0096-3984},
	journal = {Journal of the Operations Research Society of America},
	month = {nov},
	number = {4},
	pages = {393--410},
	title = {{Solution of a Large-Scale Traveling-Salesman Problem}},
	url = {http://pubsonline.informs.org/doi/abs/10.1287/opre.2.4.393},
	volume = {2},
	year = {1954}
}

@article{Rosenbaum1983,
	author = {Rosenbaum, Paul R. and Rubin, Donald B.},
	journal = {Biometrika},
	number = {1},
	pages = {41--55},
	title = {{The Central Role of the Propensity Score in Observational Studies for Causal Effects}},
	volume = {70},
	year = {1983}
}

@article{Hilbert1891,
	author = {Hilbert, David},
	doi = {10.1007/BF01199431},
	issn = {0025-5831},
	journal = {Mathematische Annalen},
	month = {sep},
	number = {3},
	pages = {459--460},
	title = {{Ueber die stetige Abbildung einer Line auf ein Fl\"achenst\"uck}},
	url = {http://link.springer.com/10.1007/BF01199431},
	volume = {38},
	year = {1891}
}

@article{Birnbaum1960,
	author = {Birnbaum, Z. W. and Hall, R. A.},
	doi = {10.1214/aoms/1177705797},
	issn = {0003-4851},
	journal = {The Annals of Mathematical Statistics},
	month = {sep},
	number = {3},
	pages = {710--720},
	title = {{Small Sample Distributions for Multi-Sample Statistics of the Smirnov Type}},
	volume = {31},
	year = {1960}
}

@article{Rosenbaum2005,
	author = {Rosenbaum, Paul R.},
	doi = {10.1111/j.1467-9868.2005.00513.x},
	issn = {13697412},
	journal = {Journal of the Royal Statistical Society. Series B: Statistical Methodology},
	number = {4},
	pages = {515--530},
	title = {{An exact distribution-free test comparing two multivariate distributions based on adjacency}},
	volume = {67},
	year = {2005}
}

@article{Bentley1975,
	author = {Bentley, Jon Louis},
	doi = {10.1145/361002.361007},
	isbn = {0001-0782},
	issn = {0001-0782},
	journal = {Communications of the ACM},
	month = {sep},
	number = {9},
	pages = {509--517},
	title = {{Multidimensional binary search trees used for associative searching}},
	url = {https://dl.acm.org/doi/10.1145/361002.361007},
	volume = {18},
	year = {1975}
}

@article{Stuart2010,
	archivePrefix = {arXiv},
	arxivId = {1010.5586v1},
	author = {Stuart, Elizabeth A.},
	doi = {10.1214/09-STS313},
	eprint = {1010.5586v1},
	isbn = {0883-4237 (Print)\r0883-4237 (Linking)},
	issn = {0883-4237},
	journal = {Statistical Science},
	number = {1},
	pages = {1--21},
	pmid = {20871802},
	title = {{Matching Methods for Causal Inference: A Review and a Look Forward}},
	url = {http://projecteuclid.org/euclid.ss/1280841730},
	volume = {25},
	year = {2010}
}

@incollection{Karp1972,
	address = {New York},
	author = {Karp, Richard M.},
	booktitle = {Complexity of Computer Computations},
	doi = {10.1007/978-3-540-68279-0_8},
	editor = {Miller, R.E. and Thatcher, J.W. and Bohlinger, J.D.},
	isbn = {9783540682745},
	pages = {219--241},
	publisher = {Plenum},
	title = {{Reducibility among combinatorial problems}},
	year = {1972}
}

@manual{Dunipace2021,
	author = {Dunipace, Eric A.},
	title = {{{approxOT}: approximate optimal transport}},
	url = {https://CRAN.R-project.org/package=approxOT},
	year = {2022}
}

@manual{Zubizarreta2018,
	author = {Zubizarreta, Jos{\'{e}} R and Kilcioglu, Cinar and Vielma, Juan P},
	title = {{designmatch: Matched Samples that are Balanced and Representative by Design}},
	url = {https://cran.r-project.org/package=designmatch},
	year = {2018}
}

@article{DelosAngelesResa2016,
	author = {{de los Angeles Resa}, Mar{\'{i}}a and Zubizarreta, Jos{\'{e}} R.},
	doi = {10.1002/sim.7036},
	issn = {10970258},
	journal = {Statistics in Medicine},
	number = {27},
	pages = {4961--4979},
	pmid = {27442072},
	title = {{Evaluation of subset matching methods and forms of covariate balance}},
	volume = {35},
	year = {2016}
}

@manual{Genz2020,
	author = {Genz, Alan and Bretz, Frank and Miwa, Tetsuhisa and Mi, Xuefei and Leisch, Friedrich and Scheipl, Fabian and Hothorn, Torsten},
	title = {{{mvtnorm}: Multivariate Normal and t Distributions}},
	url = {https://cran.r-project.org/package=mvtnorm},
	year = {2020}
}

@manual{Beygelzimer2019,
	author = {Beygelzimer, Alina and Kakadet, Sham and Langford, John and Arya, Sunil and Mount, David and Li, Shengqiao},
	title = {{FNN: Fast Nearest Neighbor Search Algorithms and Applications}},
	url = {https://cran.r-project.org/package=FNN},
	year = {2019}
}

@article{Bhattacharya2019,
	archivePrefix = {arXiv},
	arxivId = {1508.07530},
	author = {Bhattacharya, Bhaswar},
	doi = {10.1111/rssb.12319},
	eprint = {1508.07530},
	issn = {14679868},
	journal = {Journal of the Royal Statistical Society. Series B: Statistical Methodology},
	number = {3},
	pages = {575--602},
	title = {{A general asymptotic framework for dist\-ri\-bu\-tion-free graph-based two-sam\-ple tests}},
	volume = {81},
	year = {2019}
}

@article{Chenouri2012,
	author = {Chenouri, Shojaeddin and Small, Christopher G.},
	doi = {10.1214/12-EJS692},
	issn = {19357524},
	journal = {Electronic Journal of Statistics},
	pages = {760--782},
	title = {{A nonparametric multivariate multisample test based on data depth}},
	volume = {6},
	year = {2012}
}

@book{CGALref2021,
	author = {{The CGAL Project}},
	edition = {5.2.1},
	publisher = {CGAL Editorial Board},
	title = {{{CGAL} User and Reference Manual}},
	url = {https://doc.cgal.org/5.2.1/Manual/packages.html},
	year = {2021}
}

@article{Applegate2003,
	author = {Applegate, David and Bixby, Robert and Chv{\'{a}}tal, Va{\v{s}}ek and Cook, William},
	doi = {10.1007/s10107-003-0440-4},
	issn = {0025-5610},
	journal = {Mathematical Programming},
	number = {1},
	pages = {91--153},
	title = {{Implementing the Dantzig-Fulkerson-Johnson algorithm for large traveling salesman problems}},
	volume = {97},
	year = {2003}
}

@article{Henze1988,
	author = {Henze, Norbert},
	doi = {10.1214/aos/1176350835},
	issn = {0090-5364},
	journal = {The Annals of Statistics},
	month = {jun},
	number = {2},
	pages = {1403--1433},
	title = {{A Multivariate Two-Sample Test Based on the Number of Nearest Neighbor Type Coincidences}},
	url = {https://projecteuclid.org/journals/annals-of-statistics/volume-16/issue-2/A-Multivariate-Two-Sample-Test-Based-on-the-Number-of/10.1214/aos/1176350835.full},
	volume = {16},
	year = {1988}
}

@article{puri1966class,
	author = {Puri, Madan Lal and Sen, Pranab Kumar},
	journal = {Sankhy{\=a}: The Indian Journal of Statistics, Series A},
	pages = {353--376},
	publisher = {JSTOR},
	title = {{On a class of multivariate multisample rank-order tests}},
	year = {1966}
}

@article{Chen2017a,
	archivePrefix = {arXiv},
	arxivId = {1307.6294},
	author = {Chen, Hao and Friedman, Jerome H.},
	doi = {10.1080/01621459.2016.1147356},
	eprint = {1307.6294},
	issn = {1537274X},
	journal = {Journal of the American Statistical Association},
	number = {517},
	pages = {397--409},
	title = {{A New Graph-Based Two-Sample Test for Multivariate and Object Data}},
	volume = {112},
	year = {2017}
}

@article{Bennett2020,
	archivePrefix = {arXiv},
	arxivId = {1810.06707},
	author = {Bennett, Magdalena and Vielma, Juan Pablo and Zubizarreta, Jos{\'{e}} R.},
	doi = {10.1080/10618600.2020.1753532},
	eprint = {1810.06707},
	issn = {15372715},
	journal = {Journal of Computational and Graphical Statistics},
	number = {4},
	pages = {744--757},
	publisher = {Taylor & Francis},
	title = {{Building Representative Matched Samples With Multi-Valued Treatments in Large Observational Studies}},
	url = {https://doi.org/10.1080/10618600.2020.1753532},
	volume = {29},
	year = {2020}
}

@article{Wald1940,
	author = {Wald, A. and Wolfowitz, J.},
	doi = {10.1214/aoms/1177731909},
	issn = {0003-4851},
	journal = {The Annals of Mathematical Statistics},
	number = {2},
	pages = {147--162},
	title = {{On a Test Whether Two Samples are from the Same Population}},
	volume = {11},
	year = {1940}
}

@article{Hahsler2007,
	author = {Hahsler, Michael and Hornik, Kurt},
	doi = {10.18637/jss.v023.i02},
	issn = {1548-7660},
	journal = {Journal of Statistical Software},
	month = {dec},
	number = {2},
	pages = {1--21},
	title = {{TSP -- {I}nfrastructure for the traveling salesperson problem}},
	url = {http://www.jstatsoft.org/v23/i02/},
	volume = {23},
	year = {2007}
}

@article{Marozzi2014,
	author = {Marozzi, Marco},
	doi = {10.1007/s10260-014-0255-x},
	issn = {1613981X},
	journal = {Statistical Methods and Applications},
	number = {2},
	pages = {209--227},
	title = {{The multisample Cucconi test}},
	volume = {23},
	year = {2014}
}

@article{Kolmogorov1933,
	author = {Kolmogorov, A L},
	journal = {G. Ist. Ital. Attuari},
	pages = {83--91},
	title = {{Sulla determinazione empirica di una legge di distribuzione}},
	url = {https://ci.nii.ac.jp/naid/10030673552/en/},
	volume = {4},
	year = {1933}
}

@inproceedings{Stein1972,
	address = {Berkeley, CA},
	author = {Stein, Charles},
	booktitle = {Proc. Sixth Berkeley Symp. on Math. Statist. and Prob},
	editor = {{Le Cam}, Lucien M. and Neyman, Jerzy and Scott, Elizabeth L.},
	issn = {0097-0433},
	pages = {583--602},
	title = {{A Bound for the Error in the Normal Approximation to the Distribution of a Sum of Dependent Random Variables}},
	url = {https://projecteuclid.org/euclid.bsmsp/1200514239},
	volume = {6},
	year = {1972}
}

@article{Zubizarreta2014,
	author = {Zubizarreta, Jos{\'{e}} R. and Paredes, Ricardo D. and Rosenbaum, Paul R.},
	doi = {10.1214/13-AOAS713},
	issn = {19417330},
	journal = {Annals of Applied Statistics},
	number = {1},
	pages = {204--231},
	title = {{Matching for balance, pairing for heterogeneity in an observational study of the effectiveness of for-profit and not-for-profit high schools in Chile}},
	volume = {8},
	year = {2014}
}

@article{Smirnov1948,
	author = {Smirnov, N.},
	doi = {10.1214/aoms/1177730256},
	issn = {0003-4851},
	journal = {The Annals of Mathematical Statistics},
	month = {jun},
	number = {2},
	pages = {279--281},
	title = {{Table for Estimating the Goodness of Fit of Empirical Distributions}},
	url = {http://projecteuclid.org/euclid.aoms/1177730256},
	volume = {19},
	year = {1948}
}

@article{Bickel1983,
	author = {Bickel, Peter J. and Breiman, Leo},
	doi = {10.1214/aop/1176993668},
	issn = {0091-1798},
	journal = {The Annals of Probability},
	month = {feb},
	number = {1},
	pages = {347--370},
	title = {{Sums of Functions of Nearest Neighbor Distances, Moment Bounds, Limit Theorems and a Goodness of Fit Test}},
	volume = {11},
	year = {1983}
}

@article{Chen2014a,
	archivePrefix = {arXiv},
	arxivId = {1208.5755},
	author = {Chen, Hao and Zhang, Nancy R.},
	doi = {10.5705/ss.2012.125s},
	eprint = {1208.5755},
	issn = {10170405},
	journal = {Statistica Sinica},
	number = {4},
	pages = {1479--1503},
	title = {{Graph-based Tests for two-sample comparisons of categorical data}},
	url = {http://www3.stat.sinica.edu.tw/statistica/J23N4/J23N43/J23N43.html},
	volume = {23},
	year = {2013}
}

@article{Diamond2013,
	author = {Diamond, Alexis and Sekhon, Jasjeet S.},
	doi = {10.1162/REST_a_00318},
	issn = {00346535},
	journal = {Review of Economics and Statistics},
	number = {3},
	pages = {932--945},
	title = {{Genetic matching for estimating causal effects: A general multivariate matching method for achieving balance in observational studies}},
	volume = {95},
	year = {2013}
}

@article{Zubizarreta2012,
	author = {Zubizarreta, Jos{\'{e}} R.},
	doi = {10.1080/01621459.2012.703874},
	issn = {01621459},
	journal = {Journal of the American Statistical Association},
	number = {500},
	pages = {1360--1371},
	title = {{Using mixed integer programming for matching in an observational study of kidney failure after surgery}},
	volume = {107},
	year = {2012}
}

@article{Mood1940,
	author = {Mood, A. M.},
	journal = {The Annals of Mathematical Statistics},
	number = {4},
	pages = {367--392},
	title = {{The distribution theory of runs}},
	url = {http://pubsonline.informs.org/doi/abs/10.1287/opre.7.1.58},
	volume = {11},
	year = {1940}
}

@article{Kruskal1952,
	author = {Kruskal, William H. and Wallis, W. Allen},
	doi = {10.1080/01621459.1952.10483441},
	issn = {1537274X},
	journal = {Journal of the American Statistical Association},
	number = {260},
	pages = {583--621},
	title = {{Use of Ranks in One-Criterion Variance Analysis}},
	volume = {47},
	year = {1952}
}

@article{Mukherjee2020,
	author = {Mukherjee, Somabha and Agarwal, Divyansh and Zhang, Nancy R. and Bhattacharya, Bhaswar B.},
	doi = {10.1080/01621459.2020.1791131},
	issn = {1537274X},
	journal = {Journal of the American Statistical Association},
	number = {0},
	pages = {1--31},
	publisher = {Taylor & Francis},
	title = {{Distribution-Free Multisample Tests Based on Optimal Matchings With Applications to Single Cell Genomics}},
	url = {https://doi.org/10.1080/01621459.2020.1791131},
	volume = {0},
	year = {2020}
}

@article{Chen2020,
	author = {Chen, Hao and Small, Dylan S},
	doi = {10.1111/biom.13395},
	issn = {0006-341X},
	journal = {Biometrics},
	month = {nov},
	title = {{New multivariate tests for assessing covariate balance in matched observational studies}},
	url = {https://onlinelibrary.wiley.com/doi/10.1111/biom.13395},
	year = {2020}
}

@book{Chen2011,
	address = {Berlin, Heidelberg},
	author = {Chen, Louis H.Y. and Goldstein, Larry and Shao, Qi-Man},
	doi = {10.1007/978-3-642-15007-4},
	editor = {Asmussen, S. and Gani, J. and Jagers, P. and Kurtz, T.G.},
	isbn = {978-3-642-15006-7},
	issn = {1593-8883},
	month = {may},
	publisher = {Springer Berlin Heidelberg},
	series = {Probability and Its Applications},
	title = {{Normal Approximation by Stein's Method}},
	url = {http://link.springer.com/10.1007/s102030050003 http://link.springer.com/10.1007/978-3-642-15007-4},
	year = {2011}
}

@article{Cramer1936,
	author = {Cram{\'{e}}r, H. and Wold, H.},
	doi = {10.1112/jlms/s1-11.4.290},
	issn = {00246107},
	journal = {Journal of the London Mathematical Society},
	month = {oct},
	number = {4},
	pages = {290--294},
	title = {{Some Theorems on Distribution Functions}},
	url = {http://doi.wiley.com/10.1112/jlms/s1-11.4.290},
	volume = {s1-11},
	year = {1936}
}

@Manual{breyer_jason_trimatch_2017,
	title = {{TriMatch}: Propensity Score Matching of Non-Binary Treatments},
	url = {https://CRAN.R-project.org/package=TriMatch},
	version = {0.9.9},
	author = {Breyer, Jason},
	year = {2017},
}

\end{document}


\maketitle
		
		
\doublespacing
		\section{Proofs} \label{sec:proof}
		\subsection{Proof of Lemma \ref{lemma:knn_moment}} \label{sec:knn_moment_proof}
		
		We define a function $\sigma: \{1,...,N\} \mapsto \{1,...,N\}$  that permutes the ordering of the numbers.  Then given a $k$-nearest neighbor graph, we permute the group labels  $Z$ keeping the matrix of $k$-nearest neighbors $\boldM$ fixed. Without loss of generality we focus first on group $g=1$ and counts of nearest neighbors in the same group $C_1$. With all summations taken from $1$ to $N$, the expected number of counts under this distribution is
		
		\begin{align*}
			\E(C_1) &= \E \left[ \sum_{i,j} \boldM_{ij}  \indicator(Z_{\sigma(i)} = 1, Z_{\sigma(j)} = 1) \right] \\
			&=  \sum_{i,j} \boldM_{ij}  \E \left [ \indicator(Z_{\sigma(i)} = 1, Z_{\sigma(j)} = 1) \right]\\
			&=  \sum_{i,j} \boldM_{ij}  \p \left(Z_{\sigma(i)} = 1, Z_{\sigma(j)} = 1 \right)\\
			&=  \sum_{i,j} \boldM_{ij}  \p \left(Z_{\sigma(j)} = 1 \given Z_{\sigma(i)} = 1\right) \p \left( Z_{\sigma(i)} = 1\right)\\
			&=  \sum_{i,j} \boldM_{ij}  \frac{n_1 - 1}{N-1} \frac{n_1}{N}\\
			&=  \frac{n_1}{N} \frac{n_1 - 1}{N-1} \sum_{i} k\\
			&=  k n_1 \frac{n_1 - 1}{N-1}
		\end{align*}
		
		To calculate the variance, we first need the second moment of the distribution. As a reminder, $J$ is the number of pairs of observations that are mutual nearest neighbors and $S$ is the number of pairs of observations that share nearest neighbors. The expectation of the second moment is
		\begin{align*}
			\E(C_1^2) &= \E \left[ \left( \sum_{i,j} \boldM_{ij}  \indicator(Z_{\sigma(i)} = 1, Z_{\sigma(j)} = 1) \right)^2\right]\\
			&= \E \left[ \sum_{i,j,l,r} \boldM_{ij} \boldM_{lr}  \indicator(Z_{\sigma(i)} = 1, Z_{\sigma(j)} = 1,Z_{\sigma(l)} = 1,Z_{\sigma(r)} = 1)\right],
			\intertext{which can be split into a few cases,}
			&= \sum_{i,j; \, \, i \neq j } \left[\boldM_{ij}^2 +  \boldM_{ij}\boldM_{ji} \right]  \p(Z_{\sigma(i)} = 1, Z_{\sigma(j)} = 1,Z_{\sigma(l)} = 1,Z_{\sigma(r)} = 1) \\
			&\quad \quad \sum_{i,j,l; \,\, i \neq j \neq l} \left[\boldM_{ij} \boldM_{il}  + \boldM_{ij} \boldM_{li} + \boldM_{ij} \boldM_{jl}  + \boldM_{ij} \boldM_{lj} \right]  \p(Z_{\sigma(i)} = 1, Z_{\sigma(j)} = 1,Z_{\sigma(l)} = 1) \\
			& \quad \quad  \sum_{i,j,l,r; \,\, i \neq j \neq l \neq r} \boldM_{ij} \boldM_{lr}  \p(Z_{\sigma(i)} = 1, Z_{\sigma(j)} = 1,Z_{\sigma(l)} = 1,Z_{\sigma(r)} = 1) \\
			&= \frac{n_1}{N} \frac{n_1-1}{N-1}\sum_{i,j; \, \, i \neq j } \left[\boldM_{ij}^2 +  \boldM_{ij}\boldM_{ji} \right] \\
			&\quad \quad \frac{n_1}{N} \frac{n_1-1}{N-1} \frac{n_1-2}{N-2} \sum_{i,j,l; \,\, i \neq j \neq l} \left[\boldM_{ij} \boldM_{il}  + \boldM_{ij} \boldM_{li} + \boldM_{ij} \boldM_{jl}  + \boldM_{ij} \boldM_{lj} \right]   \\
			& \quad \quad  \frac{n_1}{N} \frac{n_1-1}{N-1} \frac{n_1-2}{N-2} \frac{n_1-3}{N-3} \sum_{i,j,l,r; \,\, i \neq j \neq l \neq r} \boldM_{ij} \boldM_{lr}  \\
			&= \frac{n_1}{N} \frac{n_1-1}{N-1} (kN + 2 J)\\
			&\quad \quad \frac{n_1}{N} \frac{n_1-1}{N-1} \frac{n_1-2}{N-2}  (N k^2 - Nk + 2 N k^2 -4 J + 2 S)  \\
			& \quad \quad  \frac{n_1}{N} \frac{n_1-1}{N-1} \frac{n_1-2}{N-2} \frac{n_1-3}{N-3} (N^2 k^2  - 3 N k^2 + 2 J - 2 S).
		\end{align*}
		Then the variance is simply $\var(C_1) = \E(C_1^2) - \E^2(C_1)$. The same arguments can be applied to any group $g \in \{1,...,G\}$.
		
		Without loss of generality, now we take $C_1$ and $C_2$. The covariance between these two counts relies on the cross moment
		
		\begin{align*}
			\E(C_1 C_2) &= \E\left[ \left( \sum_{i,j} \boldM_{ij} \indicator \left(Z_{\sigma(i)} = 1, Z_{\sigma(j)} = 1\right)\right) \left( \sum_{l,r} \boldM_{lr} \indicator \left(Z_{\sigma(l)} = 2, Z_{\sigma(r)} = 2\right)\right) \right] \\
			&= \E\left[ \sum_{i,j,l,r; \,\, i \neq j \neq l \neq r} \boldM_{ij} \boldM_{lr} \indicator \left(Z_{\sigma(i)} = 1, Z_{\sigma(j)} = 1, Z_{\sigma(l)} = 2, Z_{\sigma(r)} = 2\right) \right] \\
			&=\sum_{i,j,l,r; \,\, i \neq j \neq l \neq r} \boldM_{ij} \boldM_{lr}  \E\left[ \indicator \left(Z_{\sigma(i)} = 1, Z_{\sigma(j)} = 1, Z_{\sigma(l)} = 2, Z_{\sigma(r)} = 2\right) \right] \\
			&=\sum_{i,j,l,r; \,\, i \neq j \neq l \neq r} \boldM_{ij} \boldM_{lr}  \p\left(Z_{\sigma(i)} = 1, Z_{\sigma(j)} = 1, Z_{\sigma(l)} = 2, Z_{\sigma(r)} = 2\right) \\
			&=\frac{n_1}{N} \frac{n_1 - 1}{N-1}\frac{n_2}{N-2}\frac{n_2}{N-3}\sum_{i,j,l,r; \,\, i \neq j \neq l \neq r} \boldM_{ij} \boldM_{lr}  \\
			&=\frac{n_1}{N} \frac{n_1 - 1}{N-1}\frac{n_2-1}{N-2}\frac{n_2}{N-3}(N^2k^2 - 3Nk^2 + 2 J - 2 S) ,
		\end{align*}
		which is then used to calculate the covariance, $\cov(C_1,C_2) = \E(C_1 C_2) - \E(C_1)\E(C_2)$. This applies to any $C_g, C_h$ for all $g,h \in \{1,...,G\}$ where $g \neq h$.
		
		\subsection{Proof of Theorem \ref{thm:knn_normality}} \label{sec:graphtest_knn_normal_proof}
		
		We now provide a proof of the asymptotic normality of the scaled and centered version of the nearest neighbor counts. Our proof relies on the techniques of \citet{Chen2020}, which in turn relies on the proof techniques of \citet{Chen2014a} and \citet{Chen2017a}. 
		
		Before proceeding, we offer a plan for the proof. Our goal is to prove that the asymptotic distribution permutation null for nearest neighbor counts is Gaussian with the desired expectation and variance. To do this, we utilize the bootstrap distribution of treatment assignments and then show that the counts go to a Gaussian distribution in the limit through a Cram\'er-Wold device \citep{Cramer1936}. This Cram\'er-Wold device in turn relies on Stein's Method \citep{Stein1972} to show that the linear combination of variables from the bootstrap null converges to a Gaussian. Finally, we will show that there is an equivalence between the bootstrap and permutation random variables, such that the asymptotic distribution of the permutation null will also be Gaussian.
		
		Define the bootstrap null distribution such that each observation has probability $p_g = n_g/N$ of being assigned to group $g$ for all $g \in \{1,...,G\}$, which will give $n_g^B$ observations in treatment group $g$. The permutation null will be equivalent to the bootstrap null when $n_g^B = n_g$ for each group.
		The expectation and variance under the unconditional bootstrap distribution are
		\begin{align*}
			\E_B(C_g) &= kN  p_g ^2 \\
			\intertext{and}
			\var_B(C_g) &= p_g^2 (1-p_g)^2\left[ kN + 2J + \frac{p_g}{1-p_g} \left( kN + 2 S + 3 N k^2 \right)  \right] ,
		\end{align*}
		Define $\mu_{g}^{B} = \E_B(C_g)$ and $(\sigma_{g}^B)^2 = \var_B(C_g)$. Then the standardized counts under the bootstrap are 
		\[W_g = \frac{C_g - \mu_{g}^B}{\sigma_{g}^{B}}\] and the standardized number of treated individuals in each group are 
		\[W_{G+g} = \frac{n_g^B - n_g }{\sqrt{N p_g(1-p_g)}}\] for $g \in \{1,...,G-1\}$,
		where $W_{2G}$ is not included since it is linearly dependent on the other group sizes.
		Finally, we define the vector
		\[\boldW = \begin{pmatrix} W_1 & W_2 & \cdots& W_G & W_{G+1} & \cdots& W_{2G - 1} \end{pmatrix} \trans.\] 
		
		We now show that $\boldW$ converges to a Gaussian.
		To do so, we use the Cram\'er-Wold device, which says that $\boldW$ has a multivariate Gaussian distribution if $\mean{W} = \sum_g a_g W_g$ is distributed as a univariate Gaussian for any combination of $a_1, ..., a_g$ such that $\var_B(\mean{W})> 0$. To show this convergence in distribution, we use Stein's method which says that if $\sup_{h \in \operatorname{Lip}(1)} | \E_B h(\mean{W}) - \E h (V) | =0$, then $\mean{W}$ and $V$ have the same distribution. 
		
		\begin{lemma}
			As $N \to \infty$, $\boldW$ converges to a multivariate Gaussian  distribution.
			\label{lemma:boot_conv}
		\end{lemma}
		
		\begin{proof}[Proof of Lemma \ref{lemma:boot_conv}]
			First, we need to define some extra notation to use relevant theoretical results. Let $\mcG$ be the graph $\{e = (i,j): X_j \text{ is the nearest neighbor of } X_i\}$ and 
			\begin{align*}
				F_e &= \{e\} \cup \{e' \in \mcG: e' \text{ and } e  \text{ share a node}\},\\
				H_e &= F_e \cup \{e'' \in \mcG: \exists e' \in F_e,\text{ such that  } e'' \text{ and} e' \text{ share a node}\}.
			\end{align*}
			$F_e$ is the subgraph in $\mcG$ that is connected to edge $e$ while $H_e$ is the subgraph in $G$ connecting to any edge in $A_e$. Write $|\mcG|$ to denote the total number of edges in $\mcG$.
			
			For an observation $i \in \{1,...,N\}$, let $T_i = 1$ if it is assigned to treatment group 1 under the bootstrap distribution, $T_i=2$ if it is assigned to treatment group 2, and so forth up to treatment group $G$. For an edge $e = (e^+, e^-)$, where $e^-$ is the nearest neighbor of $e^+$, write $T_e = G+g$ if both $e^-$ and $e^+$ are both assigned to the same treatment group $g \in \{1,..,G\}.$
			
			Define
			\begin{align*}
				\xi_e &= a_1 \frac{\indicator(T_e = G+1)-p_1^2}{\sigma_1^B} + a_2 \frac{\indicator(T_e = G+2) - p_2^2}{\sigma_2^B} +\cdots + a_G \frac{\indicator(T_e = G+G) - p_G^2}{\sigma_G^B}\\
				\xi_i  &= a_{G+1} \frac{\indicator(T_i = 1) - p_1}{\sqrt{N p_1(1-p_1)}} + \cdots + a_{2G-1} \frac{\indicator(T_i = G-1) - p_{G-1}}{\sqrt{N p_{G-1}(1-p_{G-2})}}
			\end{align*}
			Then $\mean{W} = \sum_{e \in \mcG} \xi_e +  \sum_{i =1}^N \xi_i$.  Let $a = \max(|a_1|, |a_2|,...,|a_{2G-1}|)$ and let $p = \min(p_1,...,p_g)$, this means that  
			$\xi_i \leq a/\sqrt{Np(1-p)}$ and $|\xi_e| \leq  a/ (p(1-p)\sqrt{kN})$. Write
			$\mathbb{A} = \frac{a}{\sqrt{Np(1-p)}}$ and $\mathbb{B} = \frac{a}{p(1-p)\sqrt{kN}}$.
			Define $\mcI = \{1,...,N\}
			\cup \{e \in \mcG\}$ as the index of all edges and nodes.

			Further, let
			\begin{align*}
				A_i &= \{e \in \mcG_i\} \cup \{i\},\\
				B_i &= \{e \in \mcG_{i,2}\} \cup \{\text{nodes in } \mcG_i\},
			\end{align*}
			where $\mcG_i$ is the subgraph of $\mcG$ that contains edges associated with $X_i$ and $\mcG_{i,2}$ is the subgraph that contains edges associated with at least one node in $\mcG_i$.
			Also define
			\[A_i\comp = \{j \in \mcI: j \notin A_i \} \text{  and  } \xi_{A_i} = \{\xi_j: j \in A_i\}.\]
			Finally, for $e = (e^+, e^-)$, let
			\begin{align*}
				A_e&= F_e\cup \{e^+, e^-\},\\
				B_e &= H_e \cup \{\text{nodes in } A_e\}.
			\end{align*}
			Then $A_i$ and $B_i$ satisfy Assumption \ref{ass:ld2}, as do $A_e$ and $B_e$.
			\begin{assumption}[LD2, p. 133, \citealp{Chen2011}]
				For each $i \in \mcI$, there exists $A_i \subset B_i \subset \mcI$ such that $\xi_i$ is independent of $\xi_{A_i\comp}$ and $\xi_{A_i}$ is independent of $\xi_{B_i \comp}$.
				\label{ass:ld2}
			\end{assumption}
			We write for $i \in \mcI$ 
			\[\eta_i = \sum_{j\in A_i} \xi_j  \text{  and  } \tau_i = \sum_{j \in B_i} \xi_j.\]
			Since Assumption \ref{ass:ld2} holds, by Theorem 4.13 of \citet{Chen2011}  for $V \sim \N(0,1)$,
			\[\sup_{h \in \operatorname{Lip}(1)} \left| \E_B h(\mean{W}) - \E h (V) \right| \leq \delta,\]
			for
			\begin{align*}
				\delta & =  \frac{1}{\sqrt{\var_B(\mean{W})}} \left (2\sum_{i \in \mcI} (\E|\xi_i \eta_i \tau_i| + \left| \E(\xi_i \eta_i) \right| \E|\tau_i|) + \sum_{i \in \mcI} \E \left| \xi_i \eta_i^2 \right| \right )\\
				&\leq \frac{1}{\sqrt{\var_B(\mean{W})}} \left (5 \sum_{i \in \mcI}\mathbb{A}^3 (|\mcG_i| + 1)(|\mcG_{i,2}| + |\mcG_i| + 1)  + 5 \sum_{e \in \mcG} \mathbb{B}^3 (|F_e| + 2)(|H_e| + |F_e| + 1)  \right ) \\
				&\leq \frac{1}{\sqrt{\var_B(\mean{W})}} \left (10 \mathbb{A}^3 \sum_{i \in \mcI}  (|\mcG_i| + 1)(|\mcG_{i,2}|+ 1)  + 45\mathbb{B}^3  \sum_{e \in \mcG} |F_e||H_e|  \right )
			\end{align*}
			Note that for $e = (i,j) $,  we have $\mcG_i, \mcG_j \subseteq F_e$ and $\mcG_{i,2}, \mcG_{j,2} \subseteq H_e$, which means $(|\mcG_i |+ 1)(|\mcG_{i,2 }|+ 1) \leq (|F_e|+1)(|H_e| + 1)$.
			
			For each node $i$ we can randomly pick an edge that has node $i$ as one of the end points, and thus each edge can be picked at most twice in the graph since each edge has only two end points.
			Thus,
			\[\sum_{i=1}^N (|\mcG_i |+ 1)(|\mcG_{i,2 }|+ 1) \leq 2 \sum_{e \in \mcG} (|F_e|+1)(|H_e| + 1) \leq 8 \sum_{e \in \mcG} |F_e||H_e|.\]
			Similarly,
			\[\frac{\mathbb{A}}{\mathbb{B}} = \sqrt{p(1-p) k} > 0.\]
			Therefore,
			\[\delta \leq \frac{85a^3}{\sqrt{\var_B(\mean{W})}} \frac{1}{(Np(1-p))^{3/2}} \sum_{e \in \mcG} |F_e||H_e|.\]
			By Corollary S1 of  \citet{Bickel1983}, $\sup_{1 \leq j \leq N} \sum_{i = 1}^N \boldM_{ij} \leq \mathbb{C}$, for a positive constant $\mathbb{C}.$ This means $|F_e| \leq \mathbb{C} + 1$ and $|H_e| \leq (\mathbb{C} + 1)^2$. This means that 
			$\sum_{e \in \mcG} |F_e||H_e| = \mcO(|\mcG|) = \mcO(N)$ and $\delta \to 0$ as $N \to \infty$. Thus, $\sup_{h \in \operatorname{Lip}(1)} \left| \E_B h(\mean{W}) - \E h (V) \right|  \to 0$ and $\boldW$ converges to a multivariate Gaussian.
		\end{proof}
		
		Now we are ready to prove Theorem \ref{thm:knn_normality}.
		\begin{proof}[Proof of Theorem \ref{thm:knn_normality}]
			Let $\E(C_g) = \mu_g^P$ be the expectation of $C_g$ under the permutation null and similarly let $\var(C_g) = (\sigma_g^P)^2$ be the variance under the permutation null for $g \in \{1,...,G\}$. As a reminder,
			\[U_g = \frac{C_g - \mu_g^P}{\sigma_g^P},\]
			and note that 
			\[U_g = \frac{\sigma_g^B}{\sigma_g^P}\left(W_g + \frac{\mu_g^B - \mu_g^P}{\sigma_g^B} \right).\]
			
			By Lemma \ref{lemma:boot_conv}, $\boldW$ converges to a multivariate Gaussian distribution. 
			This means that $W_1, ...,W_G \given W_{G +1}=0,...,W_{2G-1} = 0$ will also have a multivariate Gaussian as its limiting distribution. 
			Therefore,  $(U_1,...,U_G )\trans$ also converges to a multivariate Gaussian since it is a linear transformation of $(W_1,...,W_G )\trans$ when we condition on the event  $n_g^B = n_g.$
			
			The asymptotic expectation of  $(U_1,...,U_G )\trans$ is 0 since
			\begin{align*}
				\lim_{N \to \infty}\frac{\mu_g^B - \mu_g^P}{\sigma_g^B} &=\lim_{N \to \infty}  \frac{k p_g^2 N - k N \frac{n_g(n_g - 1)}{N(N-1)}}{\sigma_g^B } \\
				&= \lim_{N \to \infty}  \frac{k p_g^2  - k  p_g  \frac{n_g - 1}{N-1}}{\sigma_g^B/N}= 0 .
			\end{align*}
			This implies that $U_g$ has the same asymptotic expectation as the $W_g$  which is also 0.
			The variance of the $U_g$ is 1 since we divide by the standard deviation of the counts. Finally,
			the empirical correlation is
			\[\hat{\Omega}_{gh} =  \frac{\sqrt{n_g (n_g-1)n_h(n_h-1)}}{\sqrt{(N-n_g)(N-n_g-1)(N-n_h)(N-n_h - 1)}} \frac{2J - 2S + k^2 \frac{N (N-3)}{N-1}}{\sqrt{\hat{f}(n_g)\hat{f}(n_h)}} ,\]
			with
			\[\hat{f}(n) =  kN + 2 J + \frac{n-2}{N-n-1} (2S + kN - k^2N) - \frac{2}{N-1} k^2N.\]
			If we let $N \to \infty$, then the asymptotic correlation is
			\[  \Omega_{gh} =  \frac{p_gp_h}{(1-p_g)(1-p_h)} \frac{c_1 - c_2 + k^2}{\sqrt{f_g f_h}}, \]
			where 
			\begin{align*}
				c_1 &= \lim_{N \to \infty} \frac{2J}{N}\\
				c_2 &= \lim_{N \to \infty} \frac{2S}{N}
				\intertext{and } 
				f_g &=  k + c_1 + \frac{p_g}{1-p_g} (c_2+ k - k^2)
			\end{align*}
			
		\end{proof}
		\section{ \citeauthor{Mukherjee2020} simulations}\label{sec:agarwal_graph}
		This section presents the detailed results for the \citet{Mukherjee2020} simulation settings. As described in Section \ref{sec:agarwal_desc}, we either change (a) location parameters, (b) scale parameters, or (c) correlation parameters while also varying (i) covariate dimension or (ii) number of treatment groups. The first set of tests examine the power of optimal paths while the second set of tests compare path tests to the two other graph based tests. All data are drawn from multivariate Gaussian distributions with possibly different mean vectors and different covariance matrices.
		
		\subsection{Extremum versus Wald tests}\label{sec:extrem_wald_graph}
		Here we present detailed results comparing a extremum (max or min) tests to Wald-like tests of $F_1 = \cdots = F_G$. Overall, the extremum tests seem to have higher power for the tests along Hamiltonian paths, while the Wald tests have higher power for the tests on nearest neighbor graphs.
		
		\subsubsection{Increasing covariate dimension}\label{sec:extrem_graph_dim}
		We increase covariate dimension, $d$ to see how this affects the power of extremum tests (max or min) as compared to Wald-like tests. First, we examine the path-based tests using the Mood runs test.
		
		\begin{figure*}[h]
			\includegraphics[width=\textwidth]{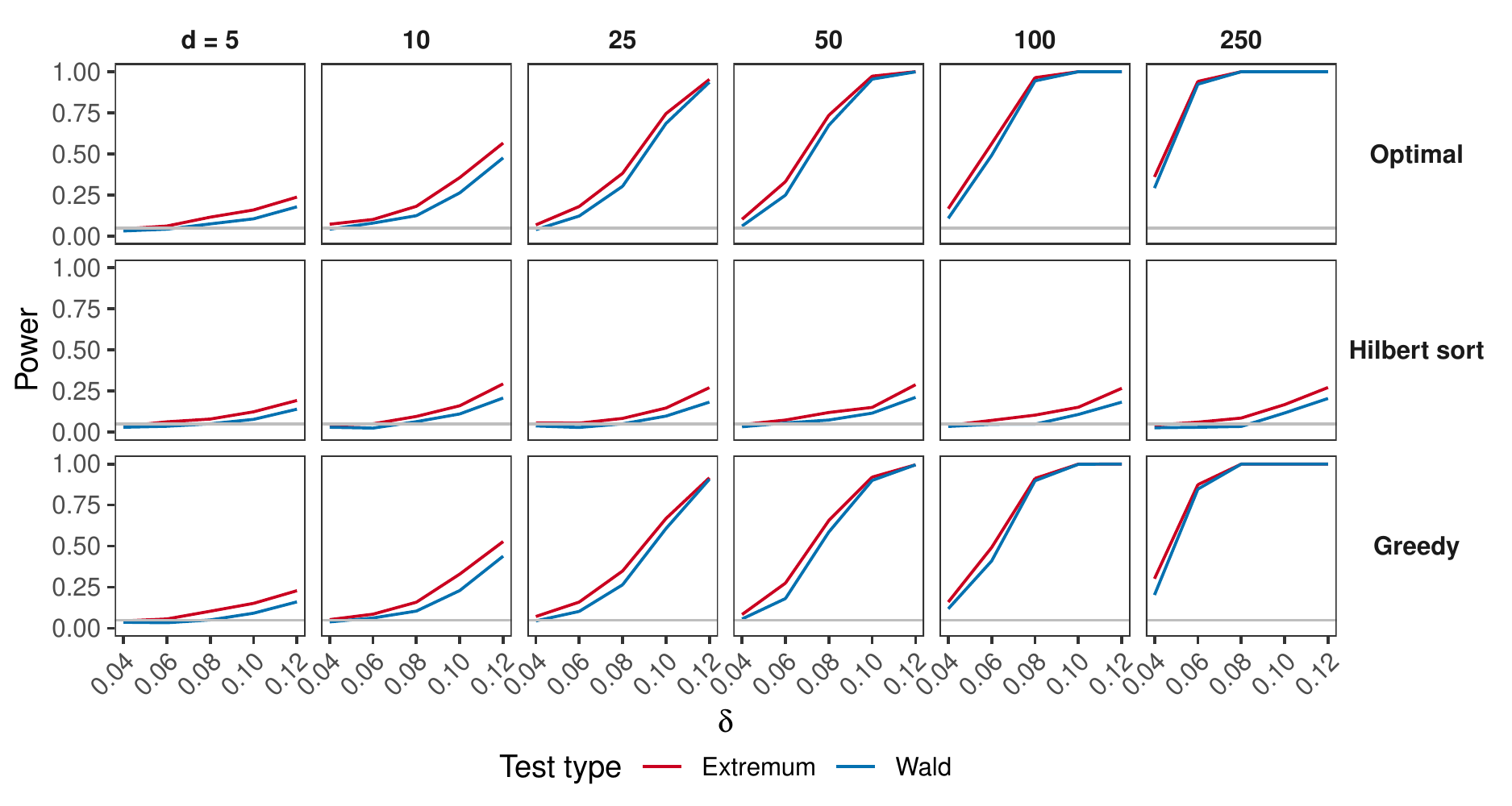}
			\caption{Location change with changing covariate dimension: $\mu_g = (g-1)\delta$ where $g$ is the group number. Comparison of minimum statistic tests (extremum) to Wald-like tests for various paths using the Mood runs test.}
			\label{fig:web_1}
		\end{figure*}
		
		\begin{figure*}[h]
			\includegraphics[width=\textwidth]{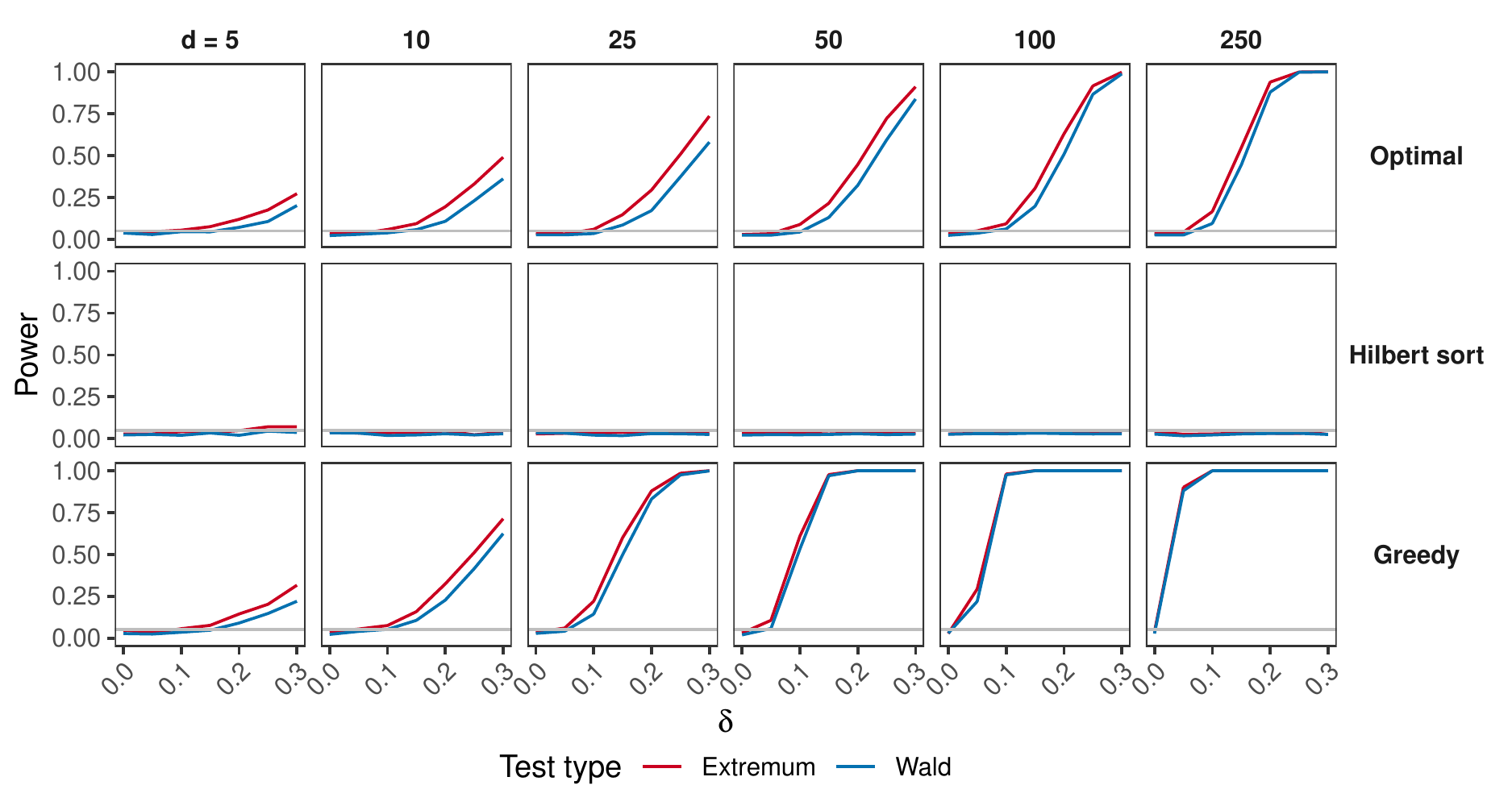}
			\caption{Scale change with changing covariate dimension: $\sigma_g ^2= 1 + (g-1)\delta$ where $g$ is the group number.  Comparison of minimum statistic tests (extremum) to Wald-like tests for various paths using the Mood runs test.}
		\end{figure*}
		
		\begin{figure*}[h]
			\includegraphics[width=\textwidth]{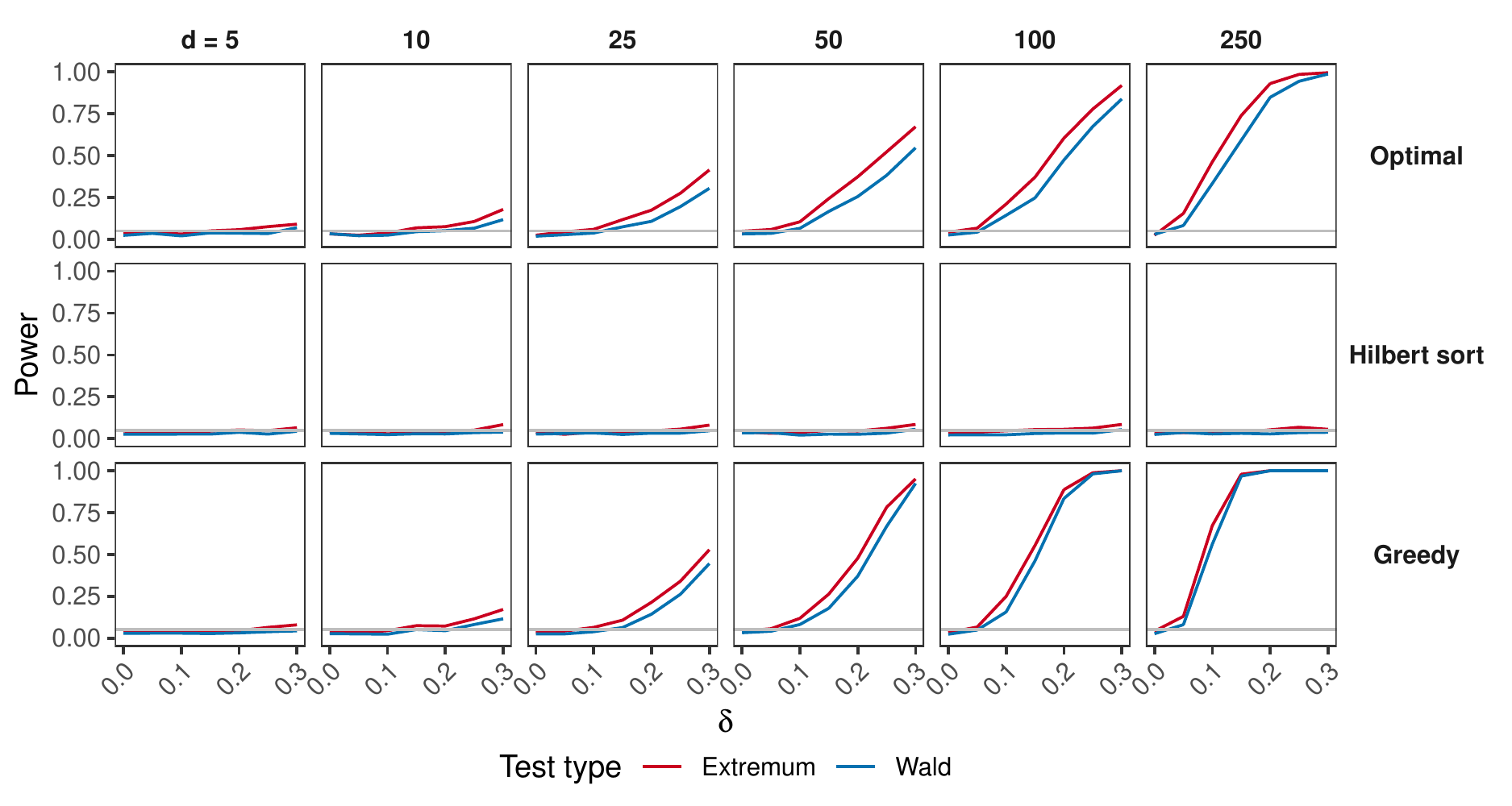}
			\caption{Correlation change with changing covariate dimension: $\rho_g = (g-1)\delta/(G-1)$ where $g$ is the group number.  Comparison of minimum statistic tests (extremum) to Wald-like tests for various paths using the Mood runs test.}
		\end{figure*}
		\clearpage 
		Then we examine the effect of using the extremum tests on the nearest neighbor graphs. For the crossmatch test, the comparison is between a minimum-based extremum test and a Wald-like test while for the kNN test the comparison is between a maximum-based extremum test and a Wald-like test.
		
		\begin{figure*}[h]
			\includegraphics[width=\textwidth]{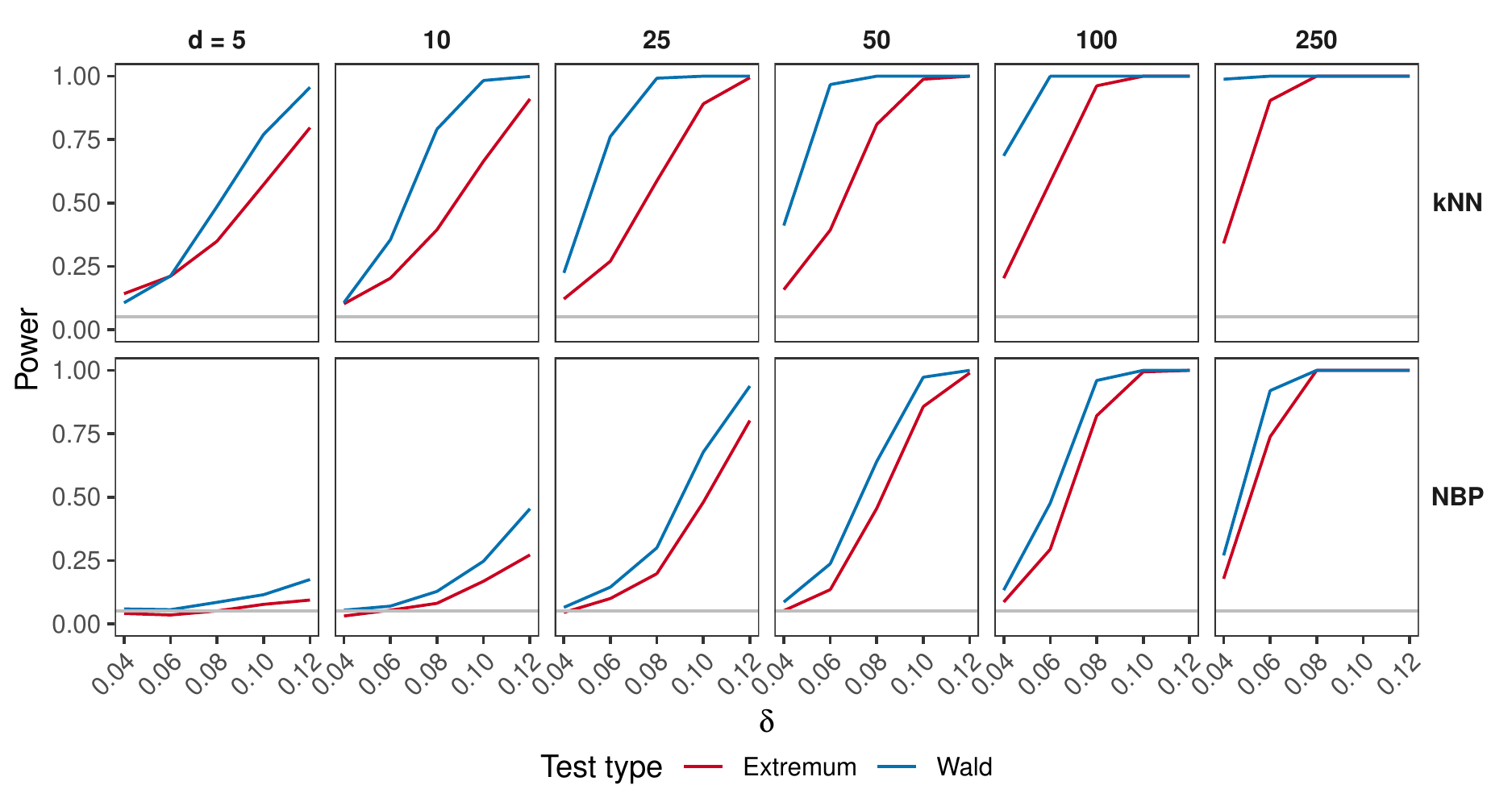}
			\caption{Location change with changing covariate dimension: $\mu_g = (g-1)\delta$ where $g$ is the group number. Comparison of extremum tests to Wald-like tests for the non-bipartite matching (NBP) and k-nearest neighbor (kNN) graphs.}
		\end{figure*}
		
		\begin{figure*}[h]
			\includegraphics[width=\textwidth]{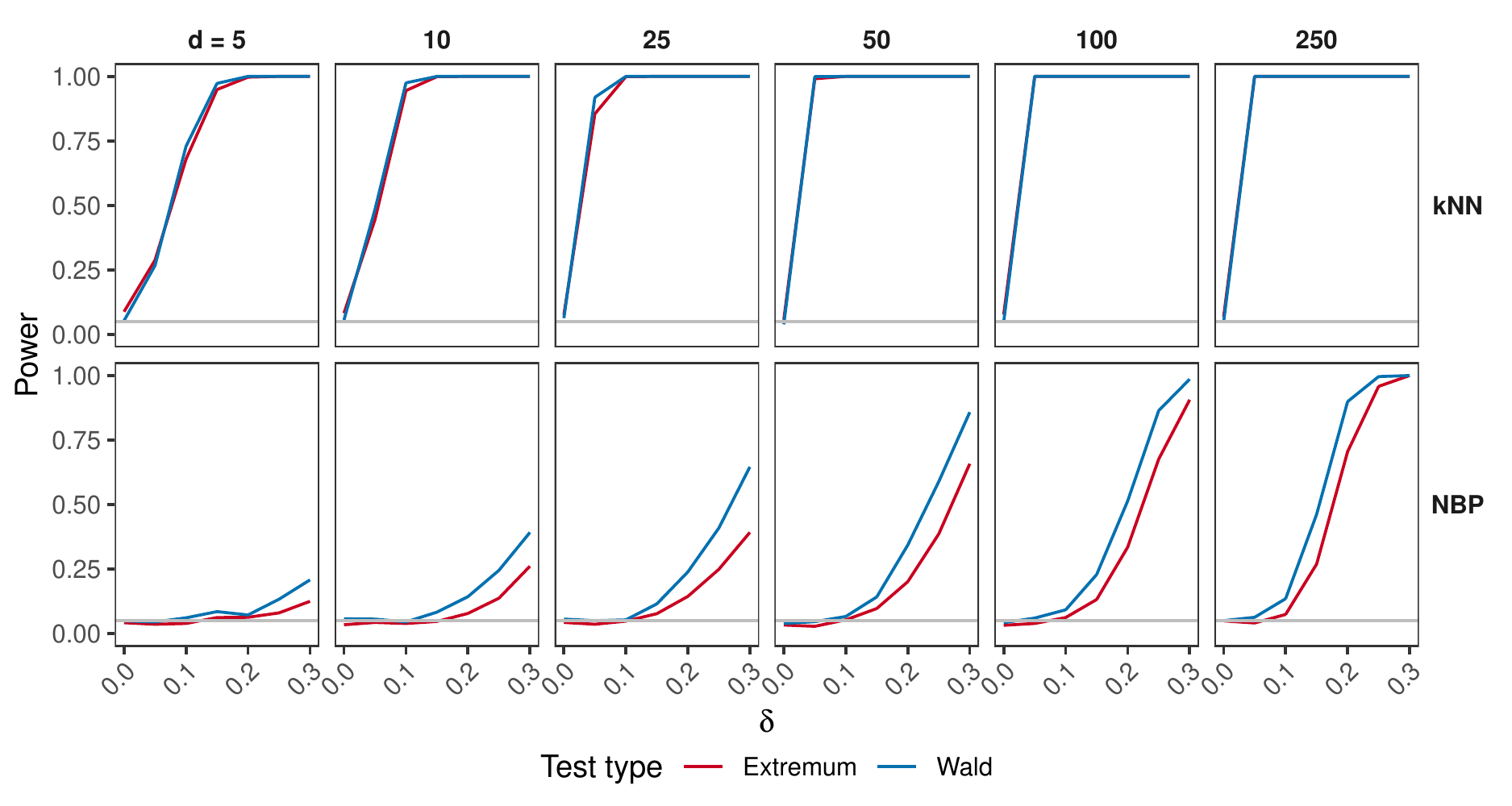}
			\caption{Scale change with changing covariate dimension: $\sigma_g ^2= 1 + (g-1)\delta$ where $g$ is the group number. Comparison of extremum tests to Wald-like tests for the non-bipartite matching (NBP) and k-nearest neighbor (kNN) graphs.}
		\end{figure*}
		
		\begin{figure*}[h]
			\includegraphics[width=\textwidth]{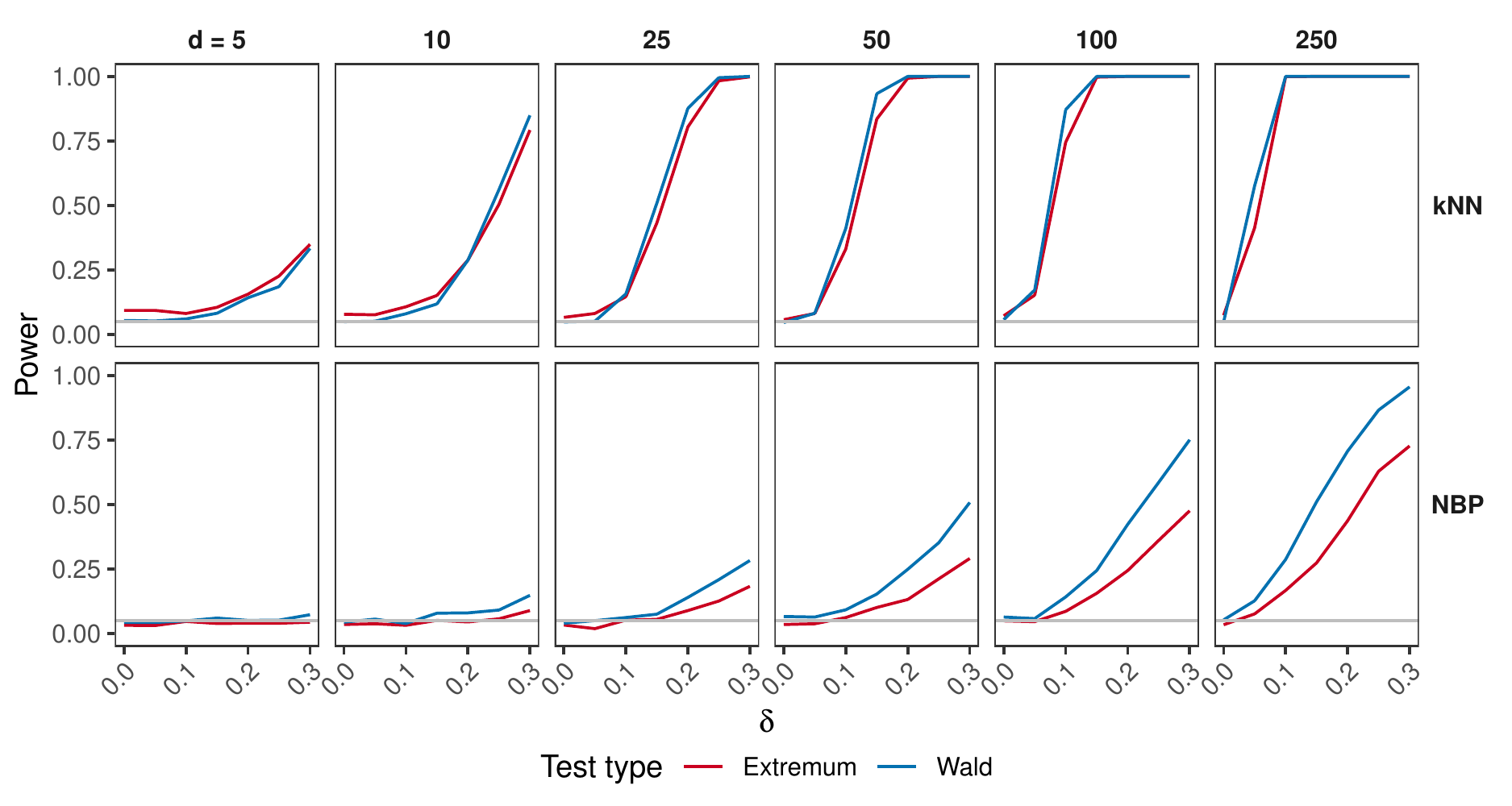}
			\caption{Correlation change with changing covariate dimension: $\rho_g = (g-1)\delta/(G-1)$ where $g$ is the group number. Comparison of extremum tests to Wald-like tests for the non-bipartite matching (NBP) and k-nearest neighbor (kNN) graphs.}
		\end{figure*}
		
		\clearpage
		\subsubsection{Increasing  number of treatment groups}\label{sec:extrem_graph_group}
		Now, we increase the group number, $G$ to see how this affects the power of extremum tests (max or min) as compared to Wald-like tests. As before, we first examine the path-based tests using the Mood runs test.
		
		\begin{figure*}[h]
			\includegraphics[width=\textwidth]{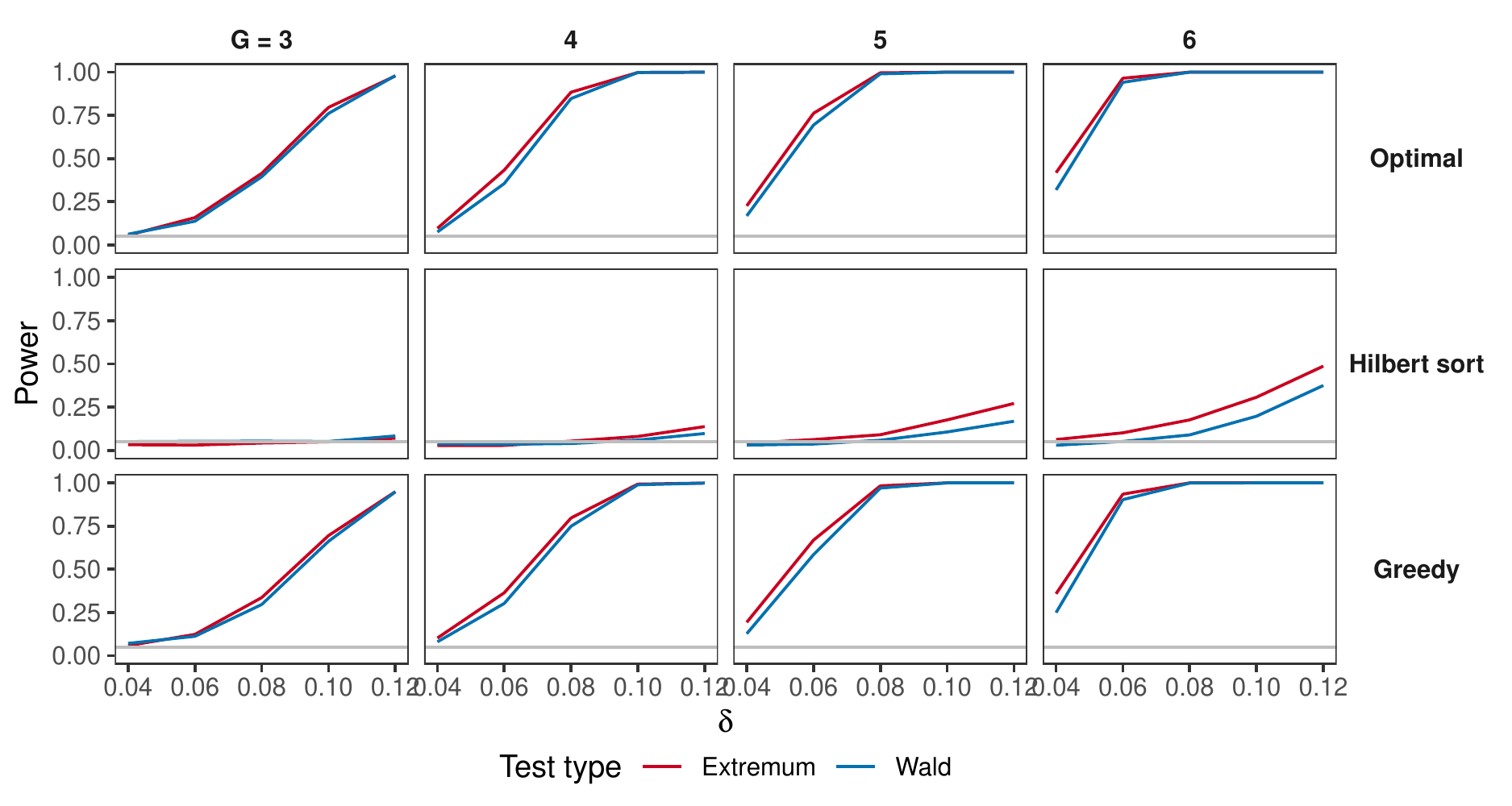}
			\caption{Location change with changing number of treatment groups: $\mu_g = (g-1)\delta$ where $g$ is the group number. Comparison of minimum statistic tests (extremum) to Wald-like tests for various paths using the Mood runs test.}
		\end{figure*}
		
		\begin{figure*}[h]
			\includegraphics[width=\textwidth]{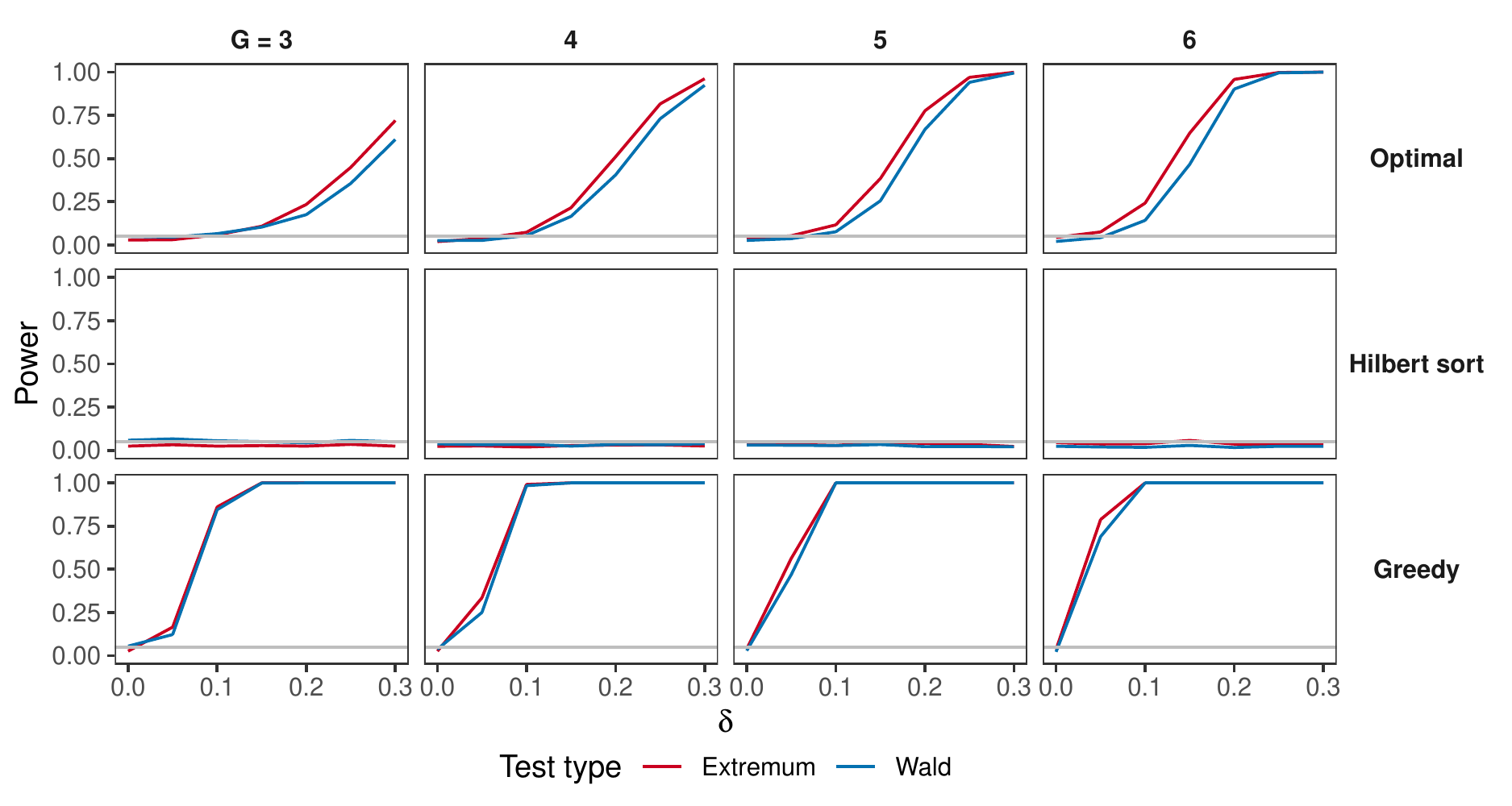}
			\caption{Scale change with changing number of treatment groups:  $\sigma_g ^2= 1 + (g-1)\delta$ where $g$ is the group number.  Comparison of minimum statistic tests (extremum) to Wald-like tests for various paths using the Mood runs test.}
		\end{figure*}
		
		\begin{figure*}[h]
			\includegraphics[width=\textwidth]{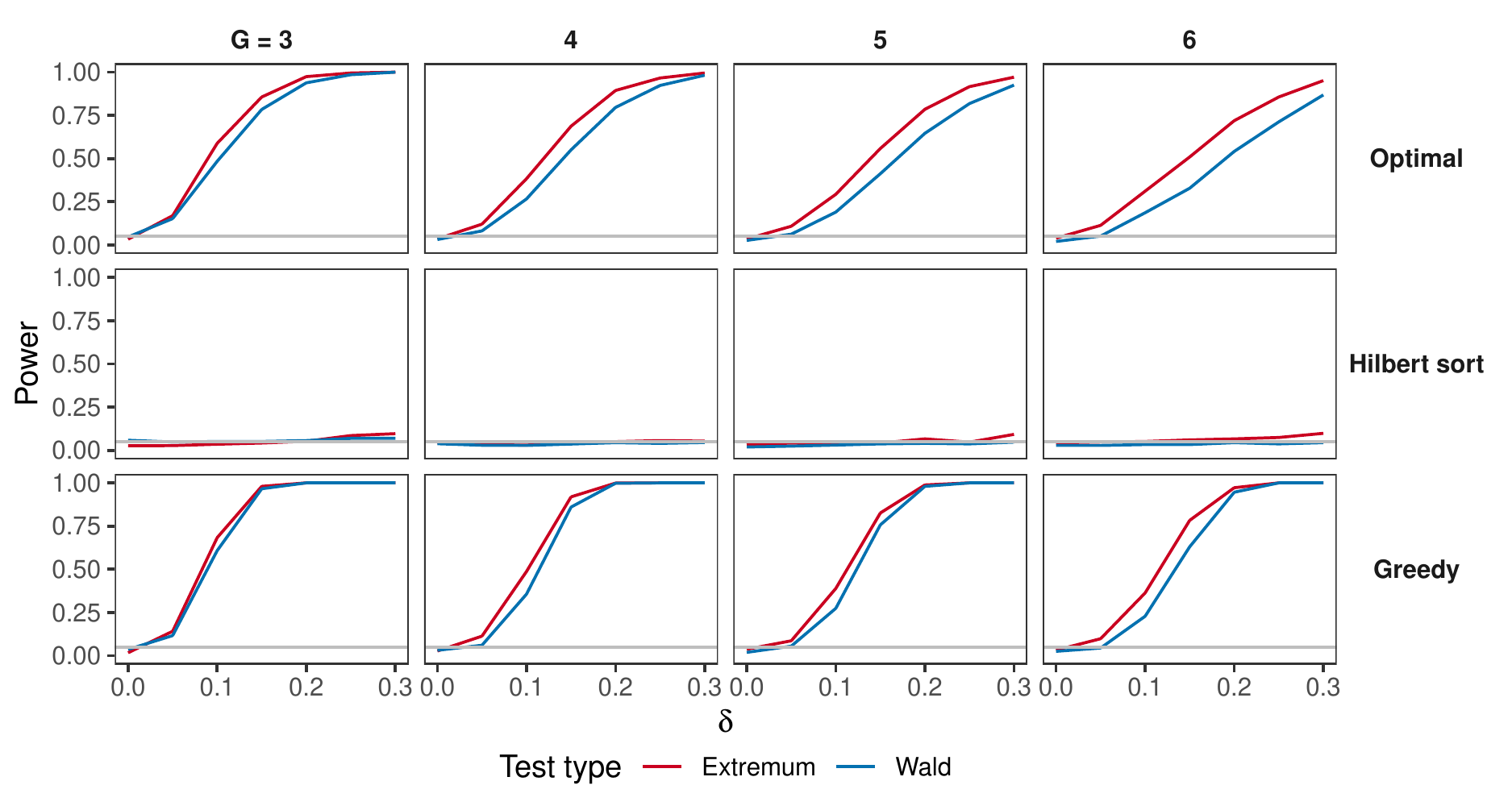}
			\caption{Correlation change with changing number of treatment groups:  $\rho_g = (g-1)\delta/(G-1)$ where $g$ is the group number.  Comparison of minimum statistic tests (extremum) to Wald-like tests for various paths using the Mood runs test.}
		\end{figure*}
		\clearpage
		For the crossmatch test on an non-bipartite (NBP) graph, the comparison is between a minimum-based extremum test and a Wald-like test, while for the kNN test on the kNN graph, the comparison is between a maximum-based extremum test and a Wald-like test.
		
		\begin{figure*}[h]
			\includegraphics[width=\textwidth]{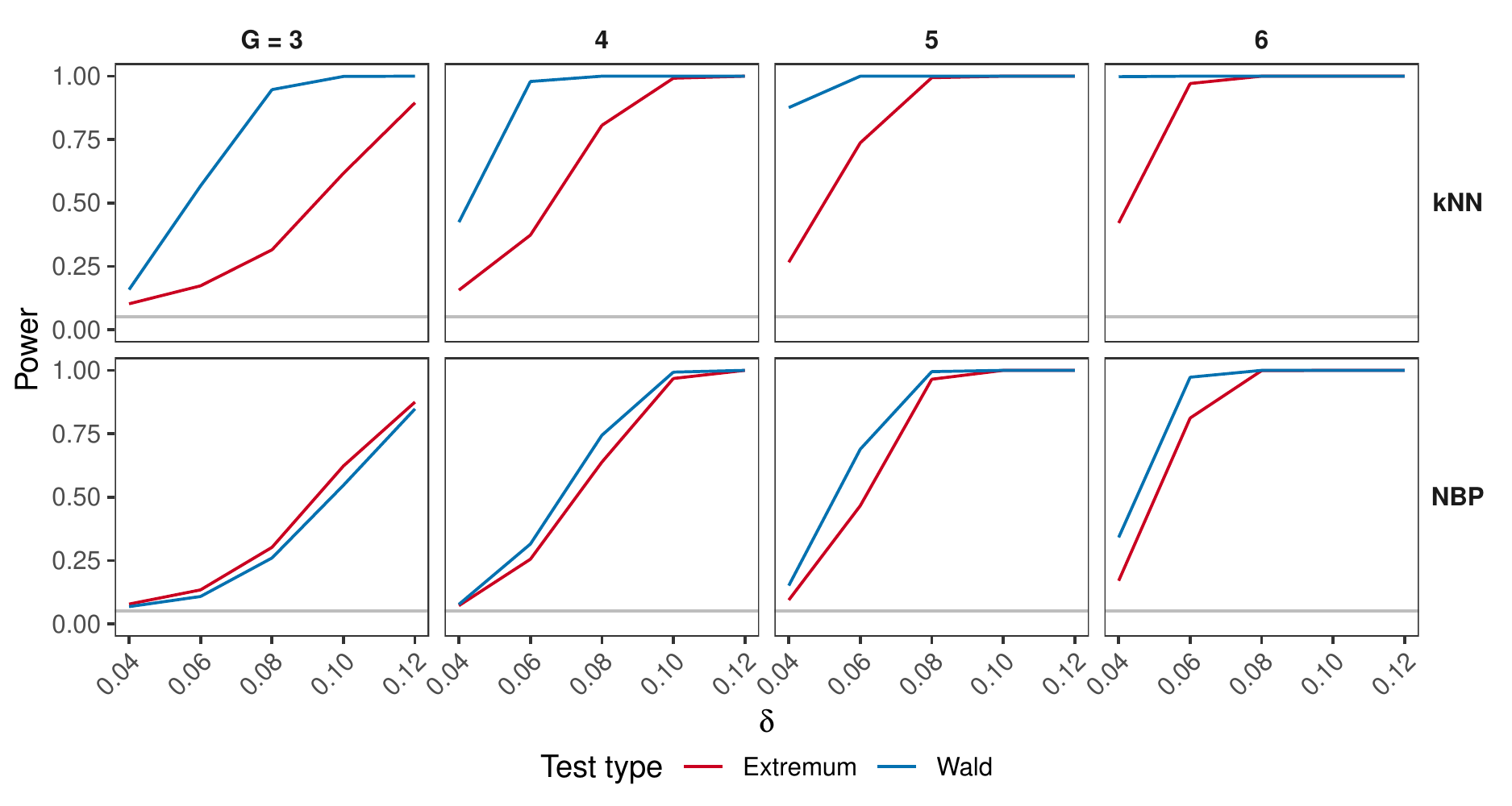}
			\caption{Location change with changing number of treatment groups:  $\mu_g = (g-1)\delta$ where $g$ is the group number. Comparison of extremum tests to Wald-like tests for the non-bipartite matching (NBP) and k-nearest neighbor (kNN) graphs.}
		\end{figure*}
		
		\begin{figure*}[h]
			\includegraphics[width=\textwidth]{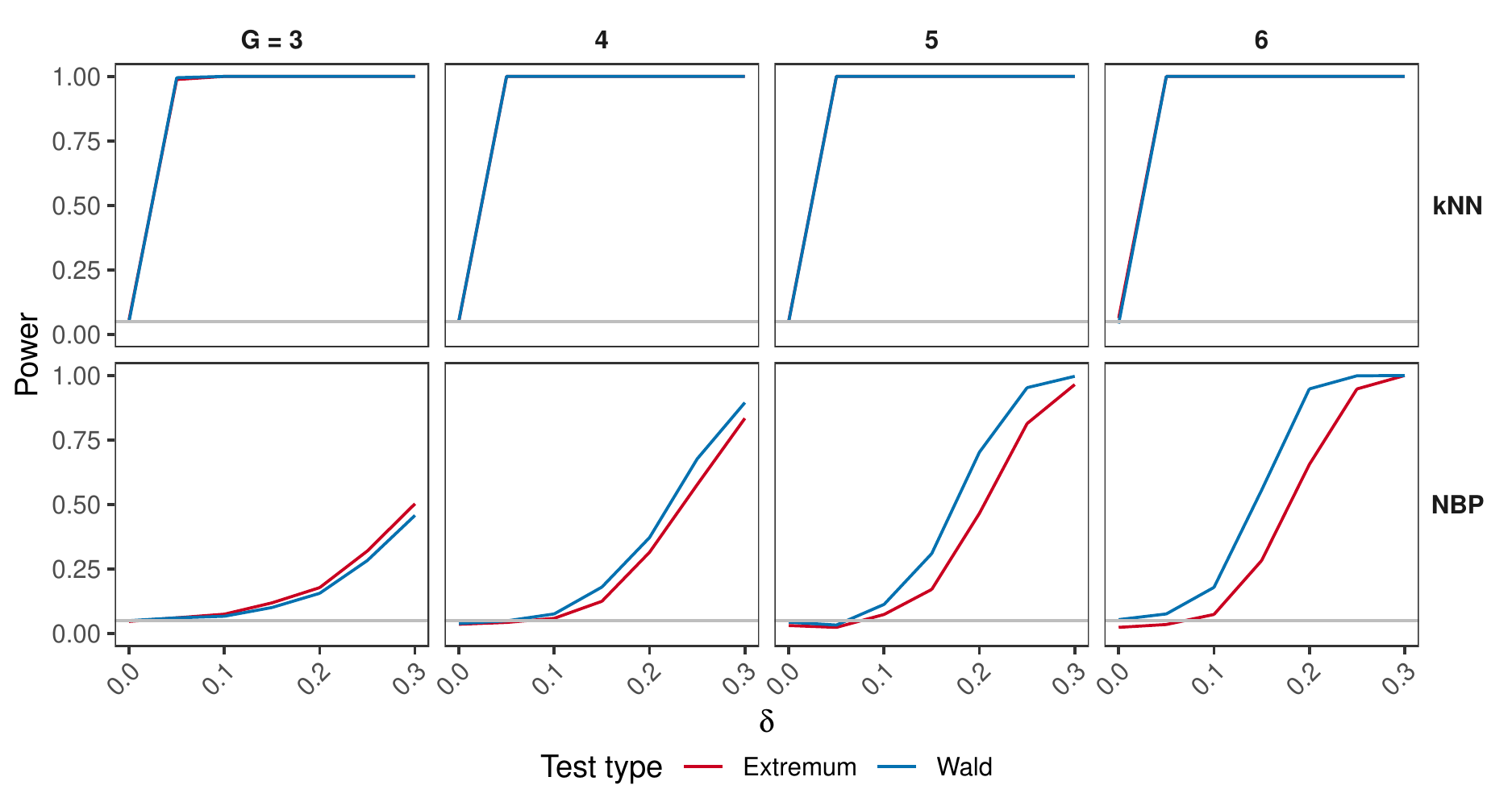}
			\caption{Scale change with changing number of treatment groups: $\sigma_g ^2= 1 + (g-1)\delta$ where $g$ is the group number. Comparison of extremum tests to Wald-like tests for the non-bipartite matching (NBP) and k-nearest neighbor (kNN) graphs.}
		\end{figure*}
		
		\begin{figure*}[h]
			\includegraphics[width=\textwidth]{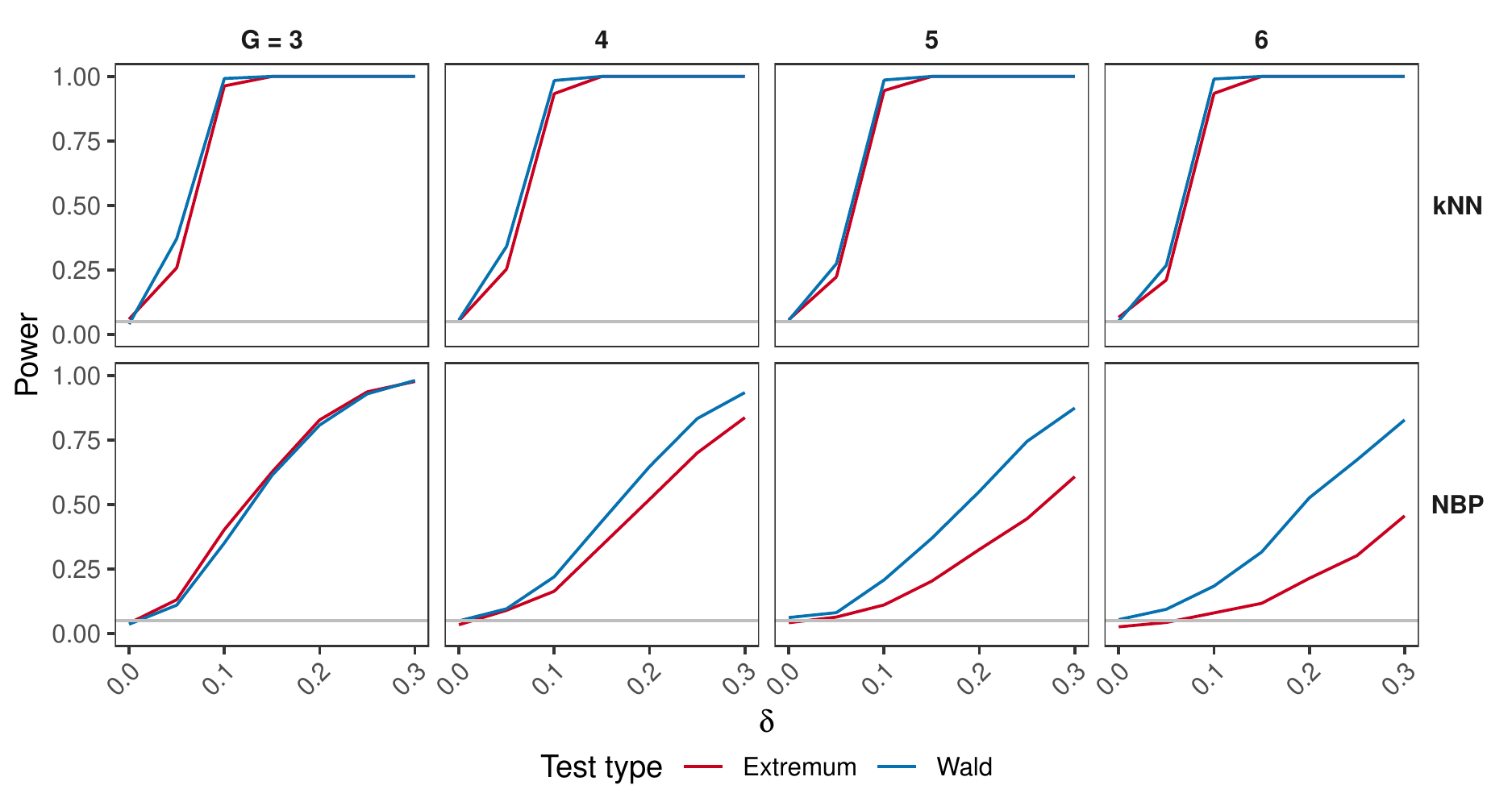}
			\caption{Correlation change with changing number of treatment groups: $\rho_g = (g-1)\delta/(G-1)$ where $g$ is the group number. Comparison of extremum tests to Wald-like tests for the non-bipartite matching (NBP) and k-nearest neighbor (kNN) graphs.}
		\end{figure*}
		
		\clearpage
		
		\subsection{Optimal versus greedy paths}\label{sec:path_res_graph}
		This section presents the detailed graphical results from Section \ref{sec:test_opt}. From these simulations, the optimal paths do slightly better in location changes but perform much worse in changes to the covariance of the data (scale and correlation changes).
		
		\subsubsection{Increasing covariate dimension}\label{sec:path_res_graph_dim}
		We increase covariate dimension, $d$ to see how this affects the power of the optimal, greedy, and Hilbert sort constructed Hamiltonian paths.
		
		\begin{figure*}[h]
			\includegraphics[width=\textwidth]{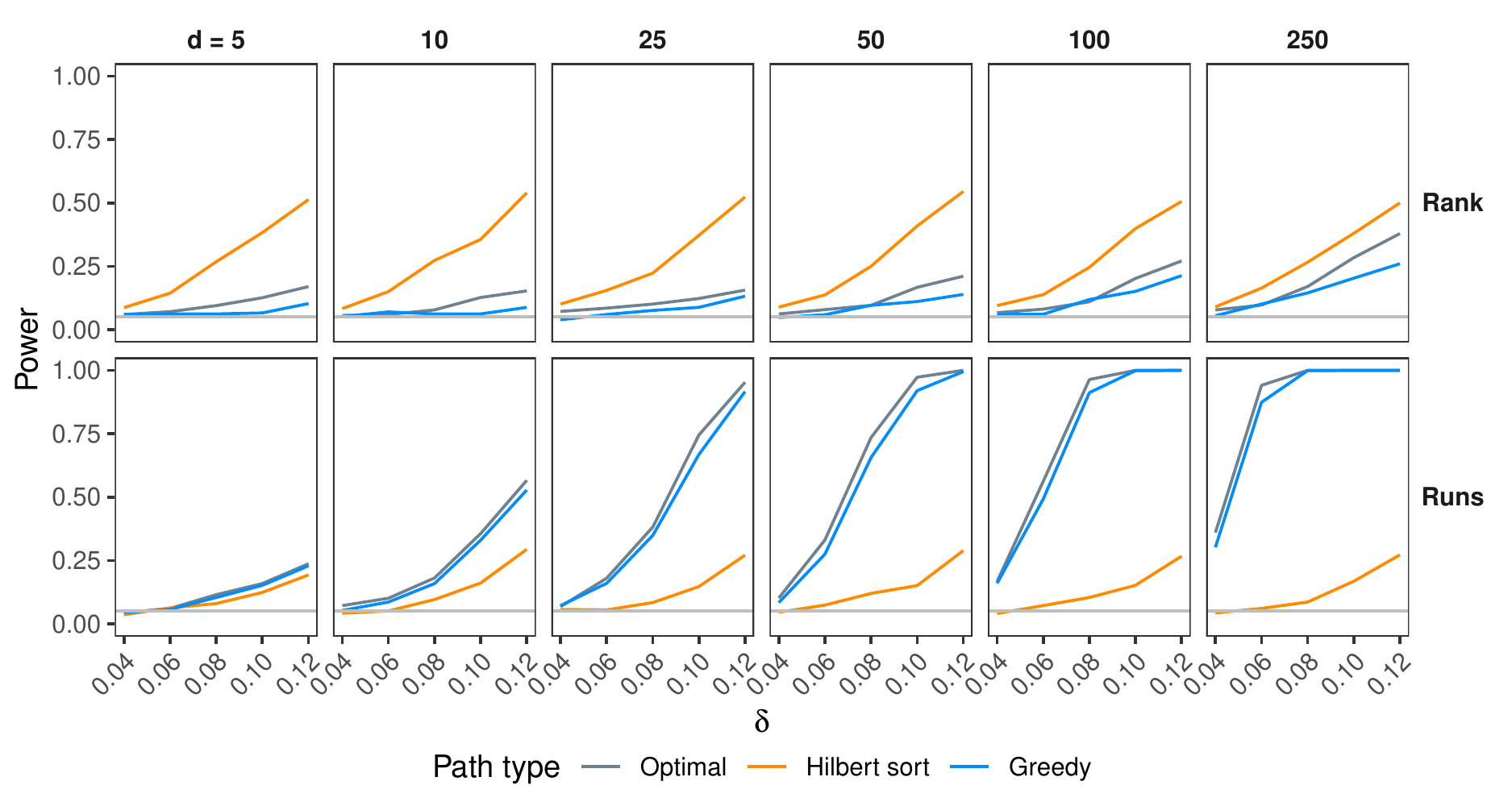}
			\caption{Location change with changing covariate dimension: $\mu_g = (g-1)\delta$ where $g$ is the group number. Rank refers to the Kruskal-Wallis test and Runs refers to the Mood runs test.}
		\end{figure*}
		
		\begin{figure*}[h]
			\includegraphics[width=\textwidth]{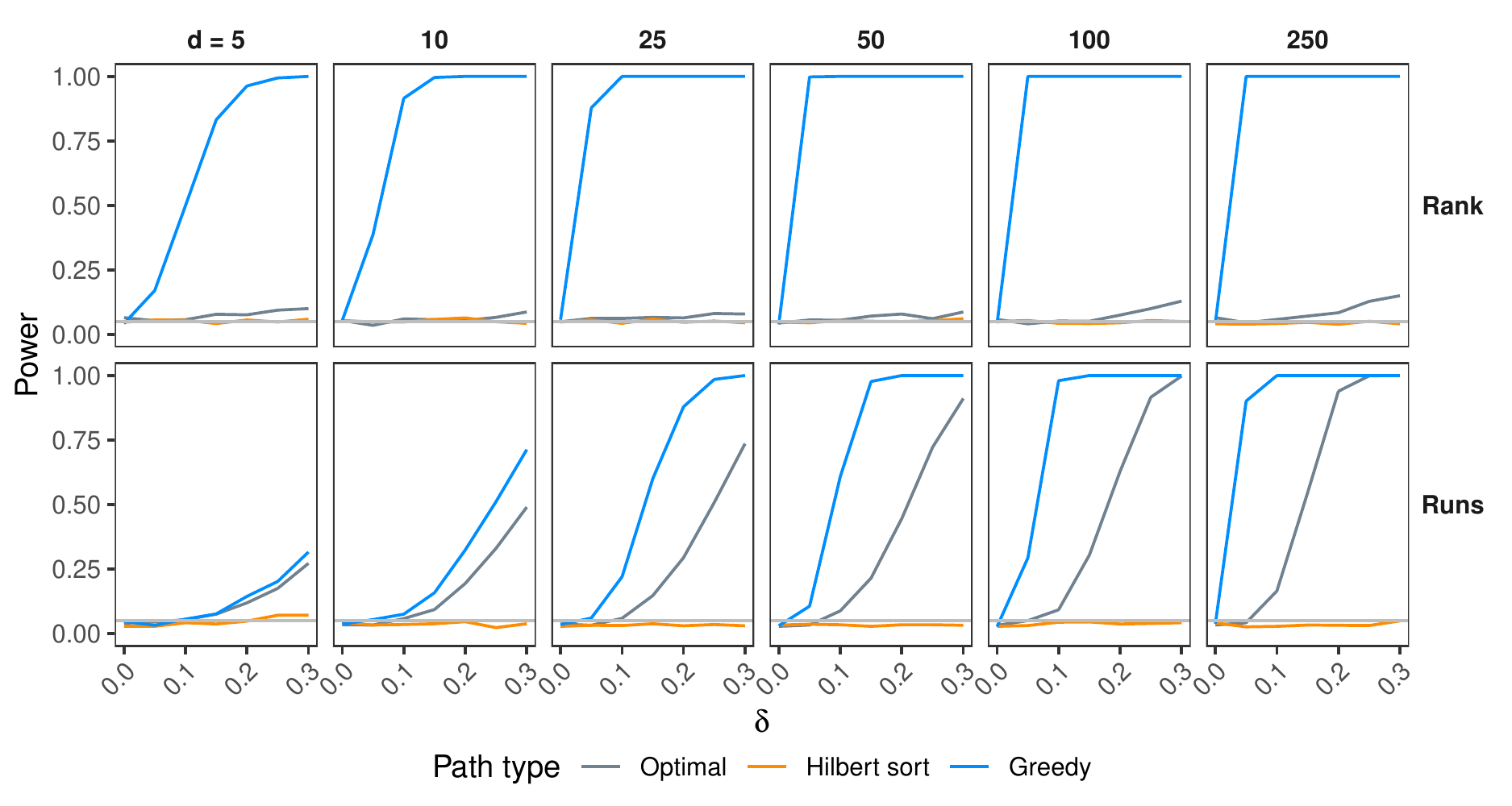}
			\caption{Scale change with changing covariate dimension: $\sigma_g ^2= 1 + (g-1)\delta$ where $g$ is the group number. Rank refers to the Kruskal-Wallis test and Runs refers to the Mood runs test.}
		\end{figure*}
		
		\begin{figure*}[h]
			\includegraphics[width=\textwidth]{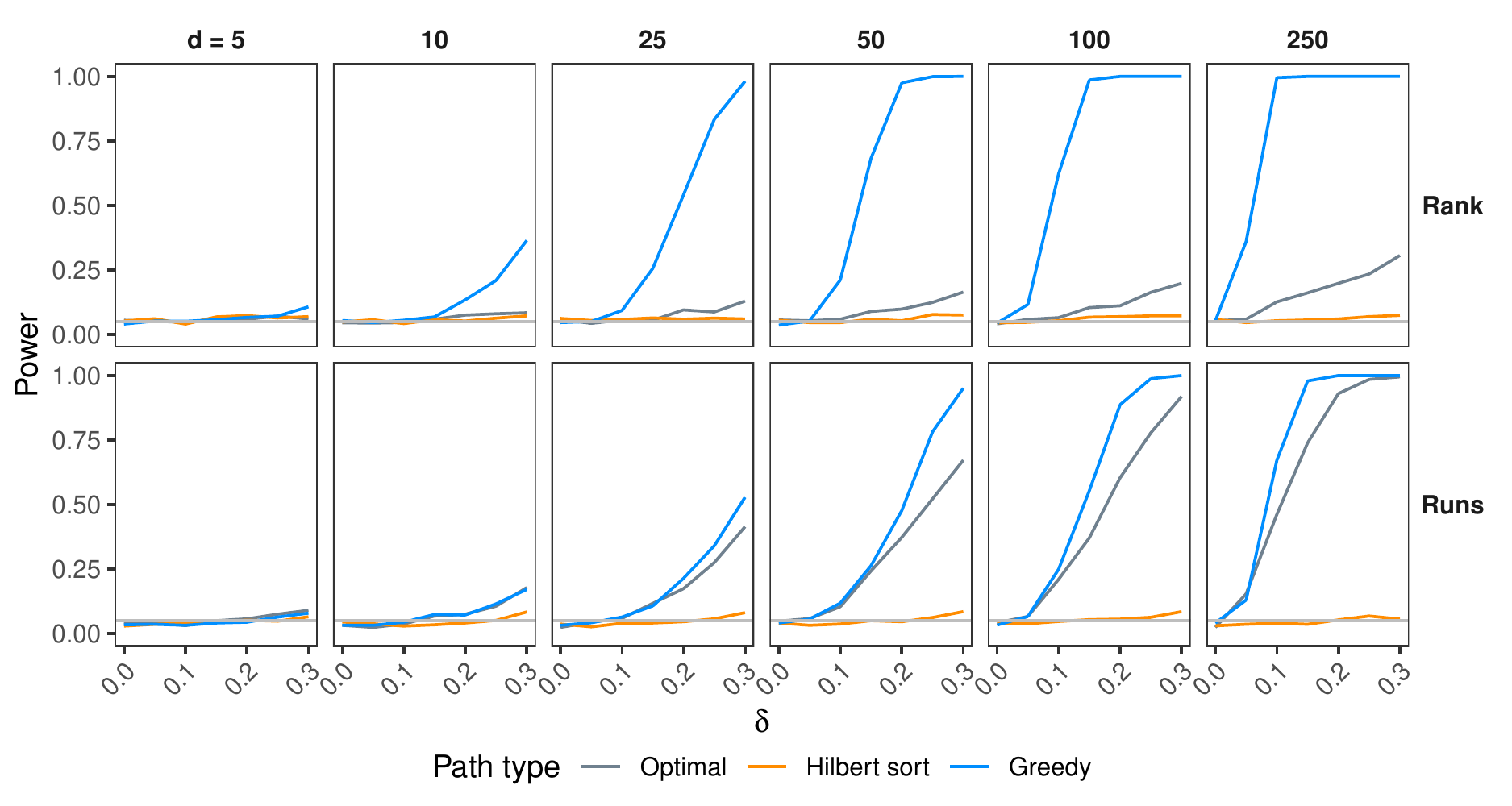}
			\caption{Correlation change with changing covariate dimension: $\rho_g = (g-1)\delta/(G-1)$ where $g$ is the group number. Rank refers to the Kruskal-Wallis test and Runs refers to the Mood runs test.}
		\end{figure*}
		
		\clearpage
		\subsubsection{Increasing number of treatment groups}\label{sec:path_res_graph_group}
		Similar to changes in covariate dimension, the optimal paths perform slightly better in location changes as we increase the group size but does much worse in changes to the scale and correlation.
		
		\begin{figure*}[h]
			\includegraphics[width=\textwidth]{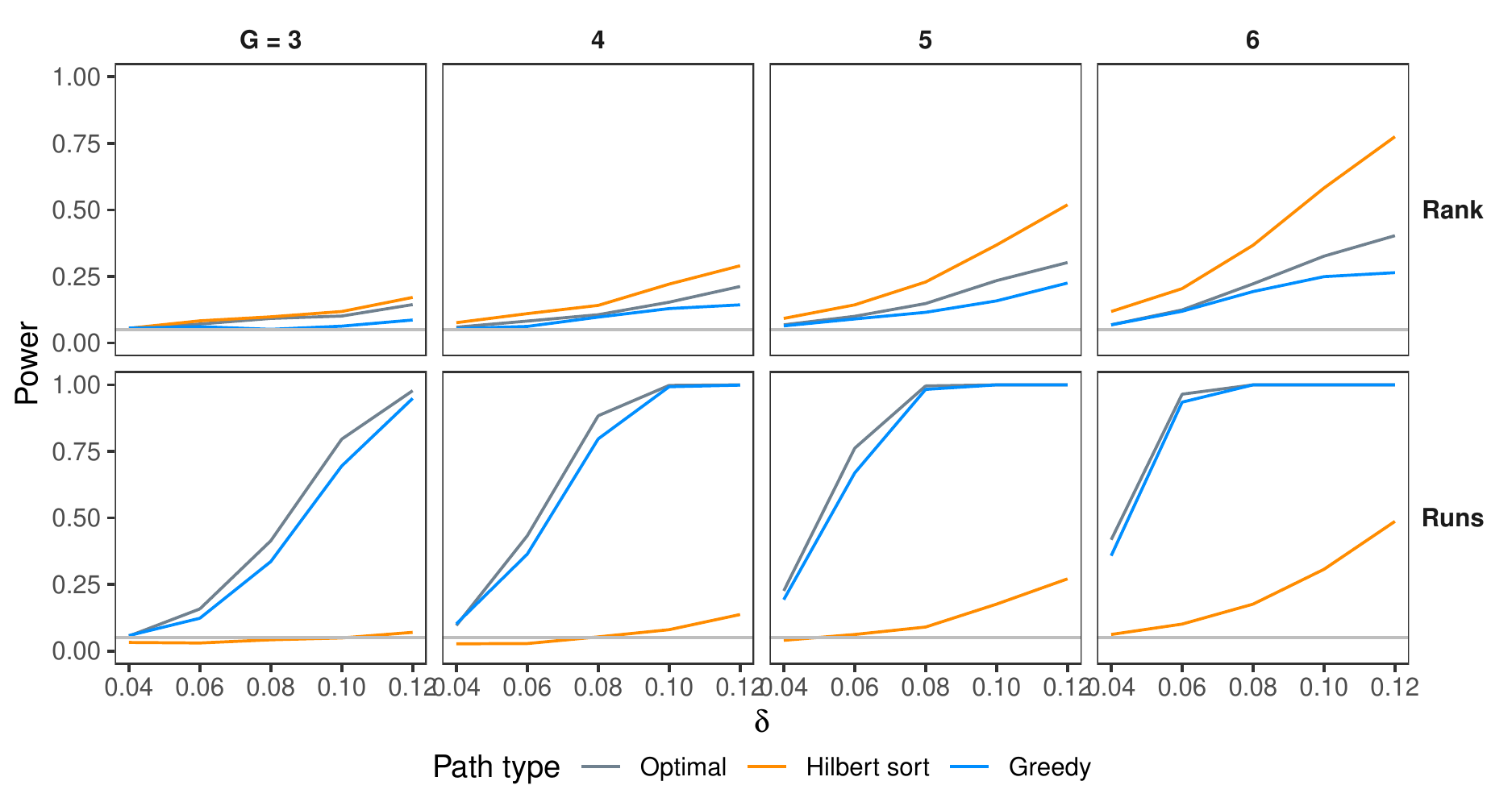}
			\caption{Location change with changing number of treatment groups: $\mu_g = (g-1)\delta$ where $g$ is the group number. Rank refers to the Kruskal-Wallis test and Runs refers to the Mood runs test.}
		\end{figure*}
		
		\begin{figure*}[h]
			\includegraphics[width=\textwidth]{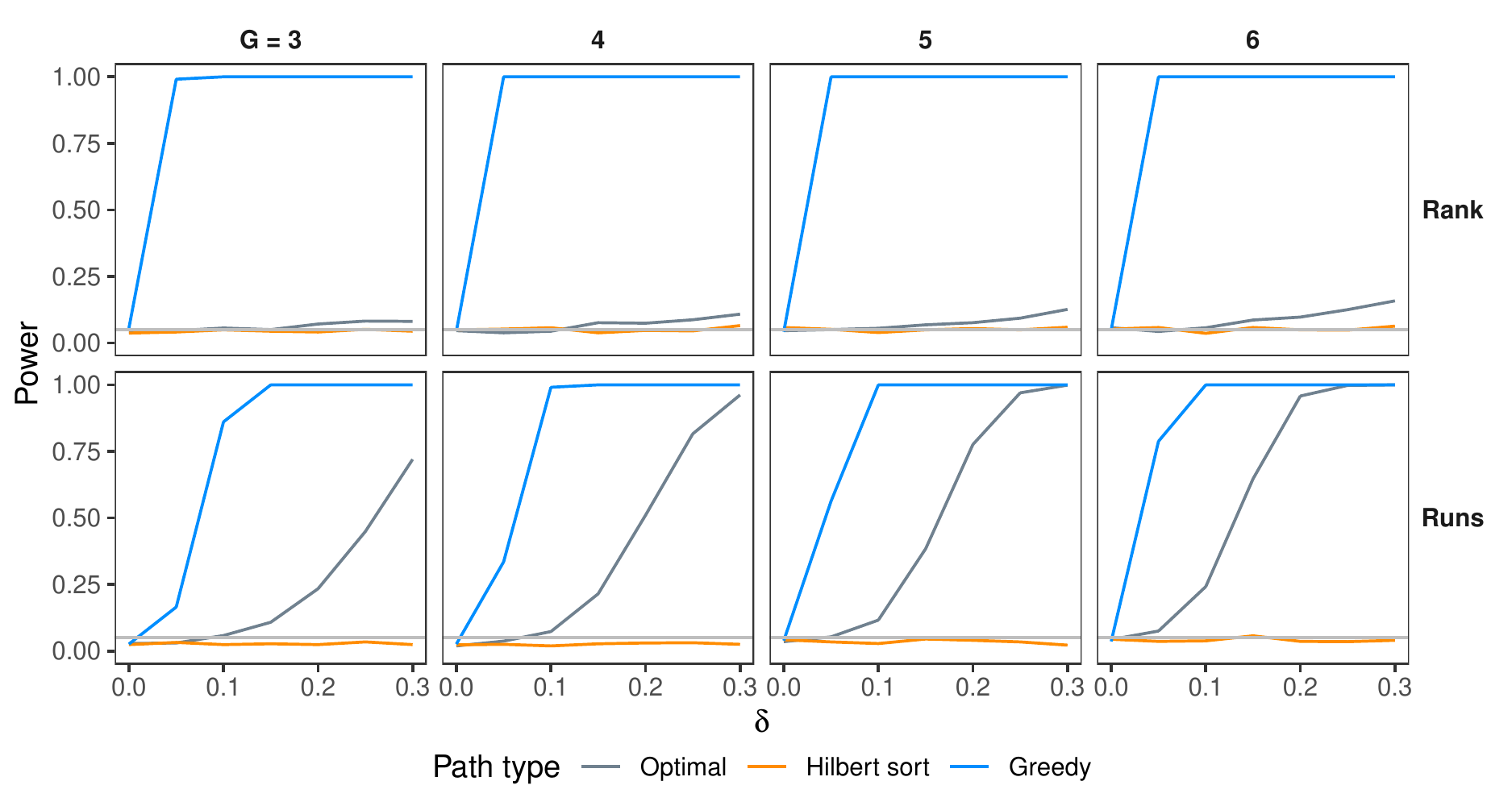}
			\caption{Scale change with changing number of treatment groups: $\sigma_g^2 = 1 + (g-1)\delta$ where $g$ is the group number. Rank refers to the Kruskal-Wallis test and Runs refers to the Mood runs test.}
		\end{figure*}
		
		\begin{figure*}[h]
			\includegraphics[width=\textwidth]{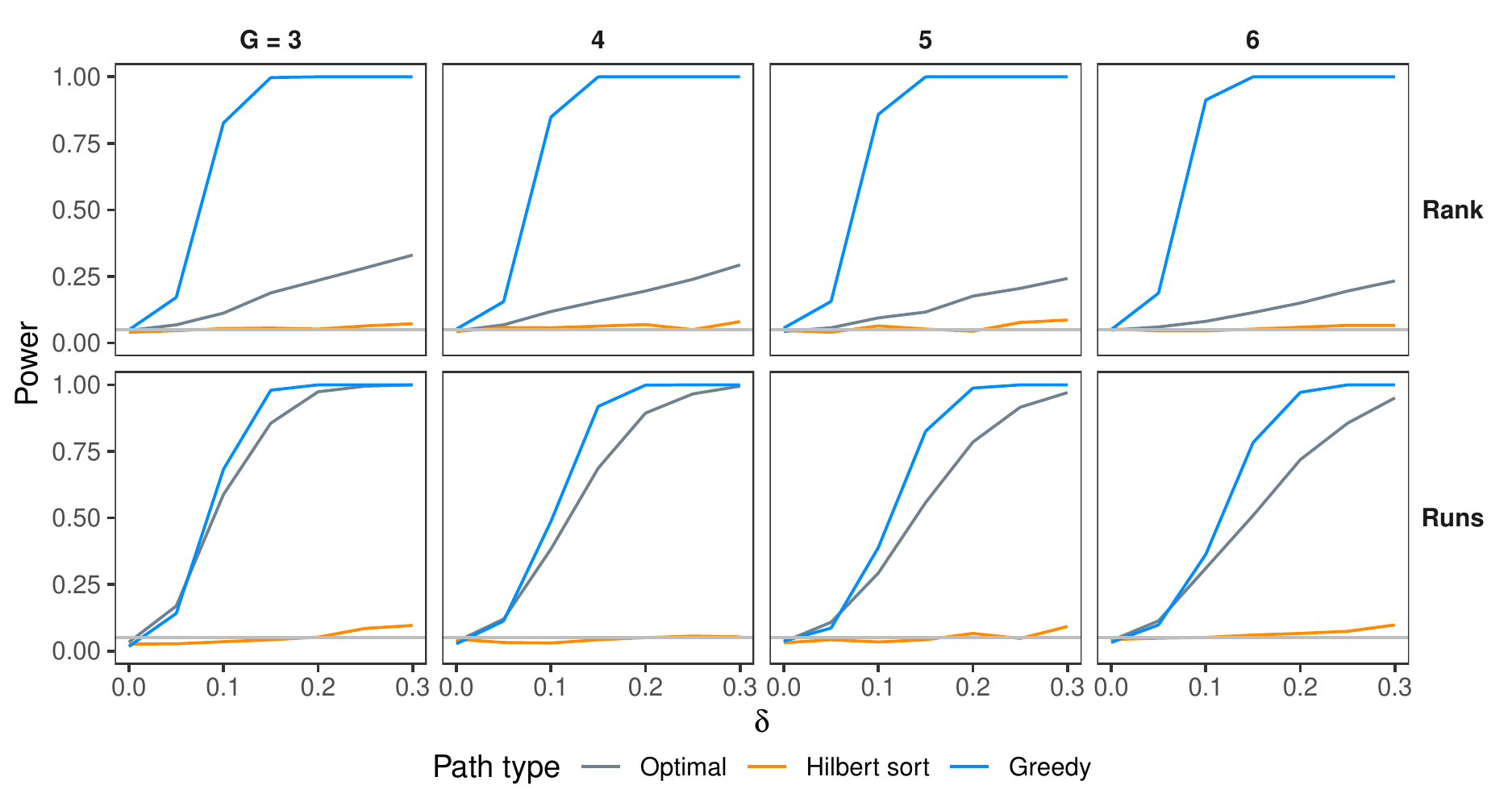}
			\caption{Correlation change with changing number of treatment groups: $\rho_g = (g-1)\delta/(G-1)$ where $g$ is the group number. Rank refers to the Kruskal-Wallis test and Runs refers to the Mood runs test.}
		\end{figure*}
		
		\clearpage
		\subsection{Evaluation of tests}\label{sec:test_res_graph}
		In this section we provide detailed graphical results from Section \ref{sec:test_sim}. In most settings, the $k$-nearest neighbor test outperforms other tests. Based on the results from the previous section, we only use the Hamiltonian path constructed via Kruskal's algorithm (greedy).

		\subsubsection{Increasing covariate dimension}\label{sec:test_res_graph_dim}
		We increase covariate dimension, $d$ to see how this affects the power of the tests under consideration.
		\begin{figure*}[h]
			\includegraphics[width = \textwidth]{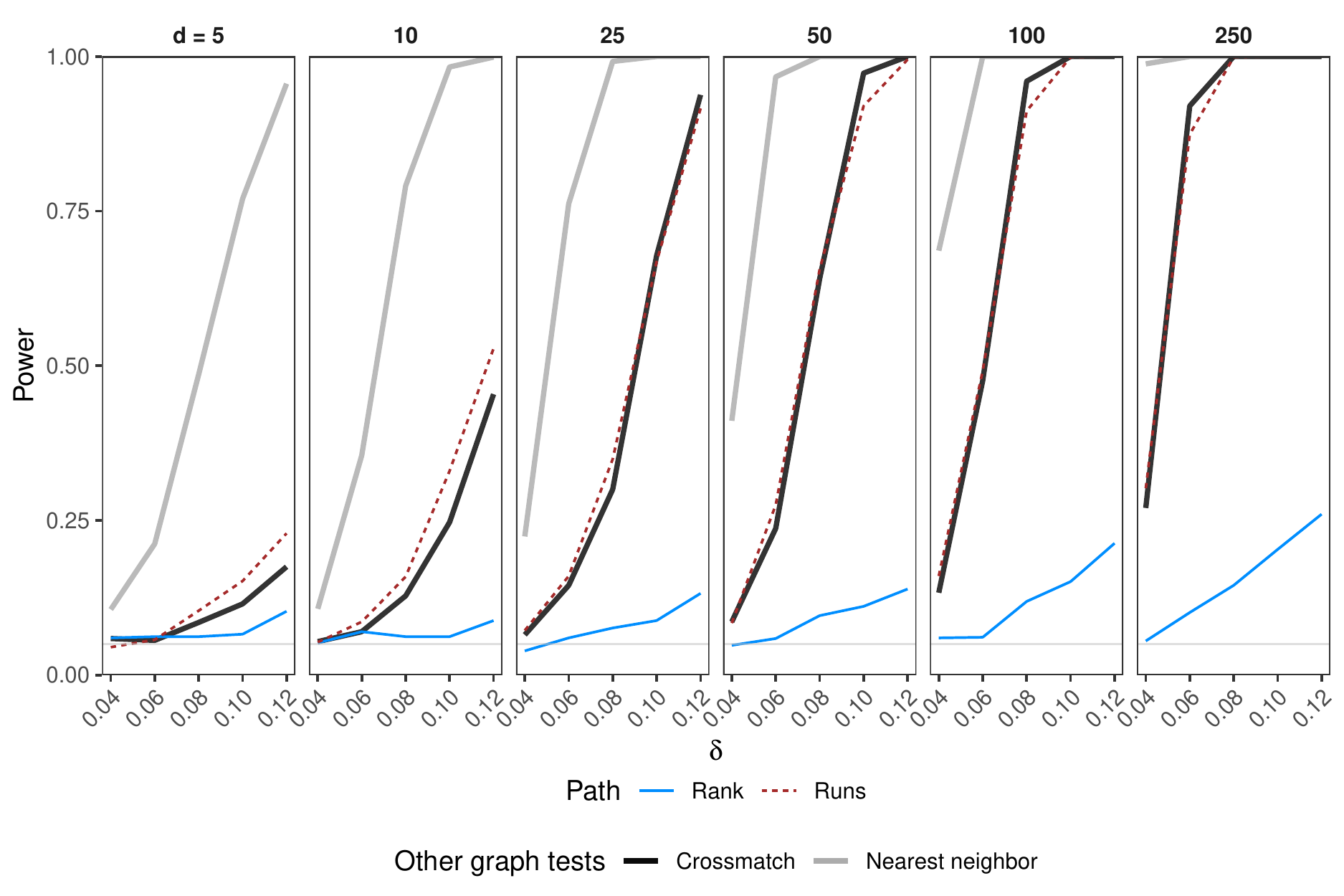}
			\caption{Location change with changing covariate dimension: $\mu_g = (g-1)\delta$ where $g$ is the group number. Rank refers to the Kruskal-Wallis test, Runs refers to the Mood runs test, Crossmatch is the multisample crossmatch, and Nearest neighbor is the multisample  $k$-nearest neighbor test presented in this paper.}
		\end{figure*}
		
		\begin{figure*}[h]
			\includegraphics[width = \textwidth]{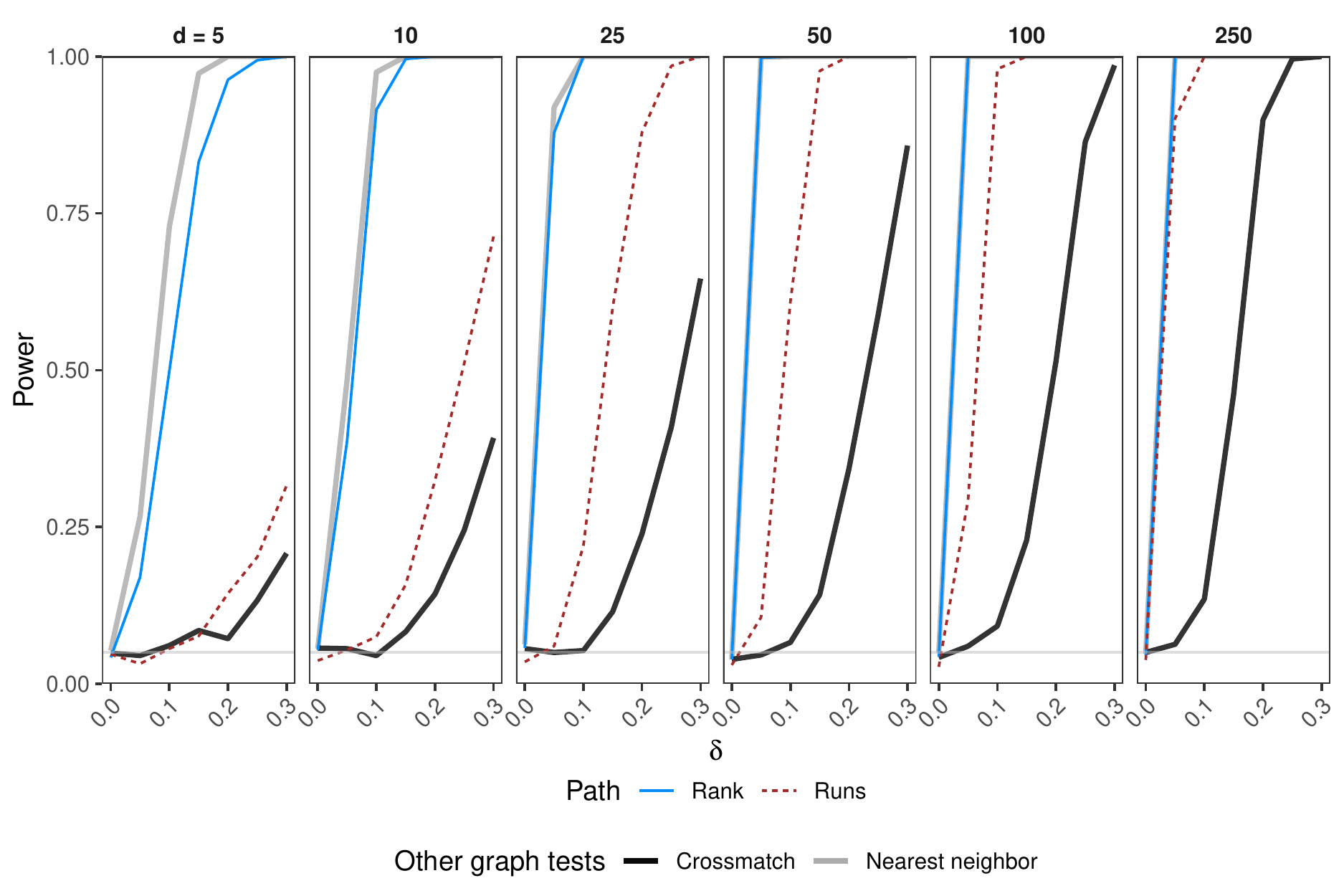}
			\caption{Scale change with changing covariate dimension: $\sigma_g ^2= 1 + (g-1)\delta$ where $g$ is the group number. Rank refers to the Kruskal-Wallis test, Runs refers to the Mood runs test, Crossmatch is the multisample crossmatch, and Nearest neighbor is the multisample  $k$-nearest neighbor test presented in this paper.}
		\end{figure*}
		
		\begin{figure*}[h]
			\includegraphics[width = \textwidth]{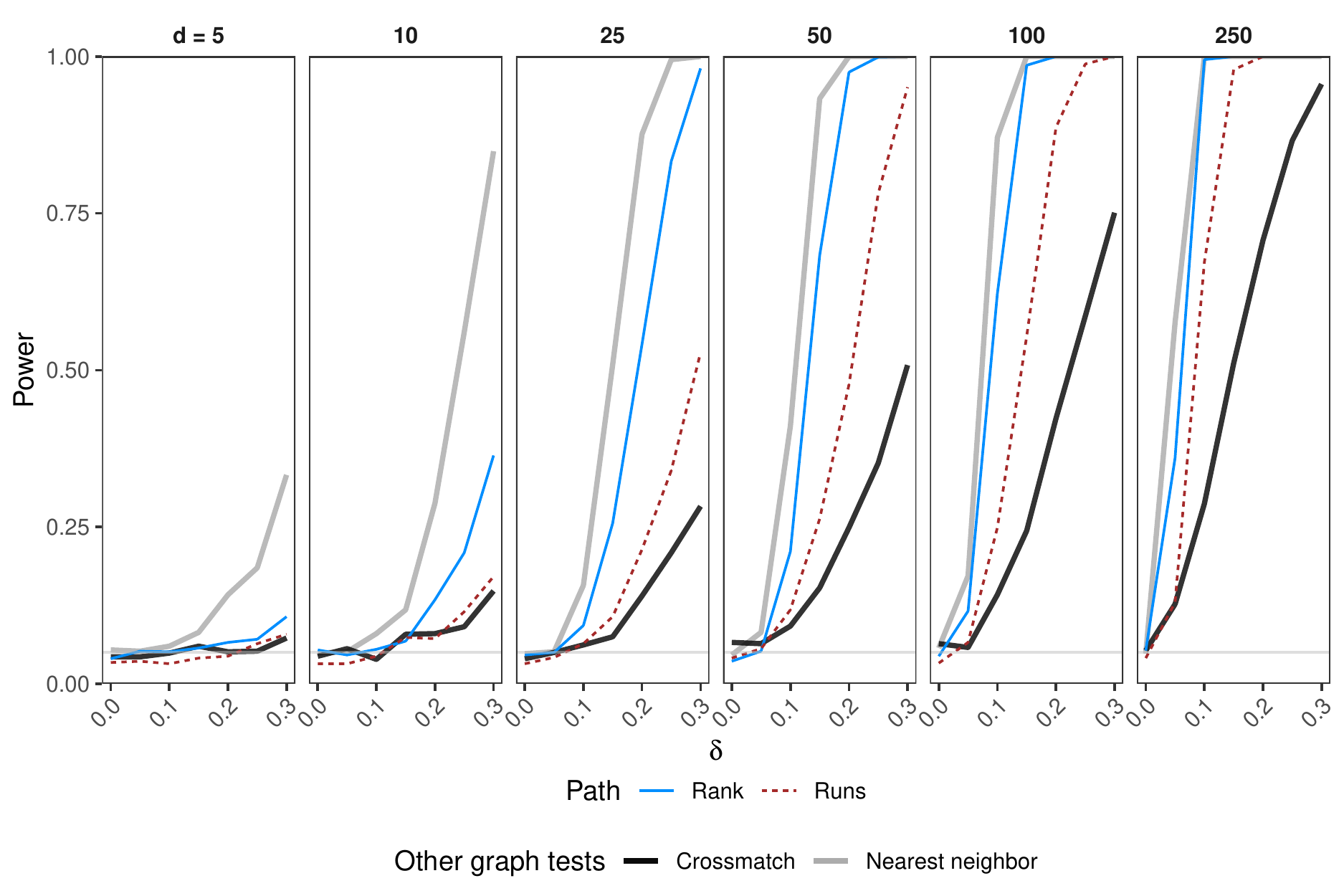}
			\caption{Correlation change with changing covariate dimension: $\rho_g = (g-1)\delta/(G-1)$ where $g$ is the group number. Rank refers to the Kruskal-Wallis test, Runs refers to the Mood runs test, Crossmatch is the multisample crossmatch, and Nearest neighbor is the multisample  $k$-nearest neighbor test presented in this paper.}
		\end{figure*}
		
		\clearpage
		\subsubsection{Increasing number of treatment groups}\label{sec:test_res_graph_group}
		We increase the number of treatment groups, $G$ to see how this affects the power of the tests under consideration.
		
		\begin{figure*}[h]
			\includegraphics[width = \textwidth]{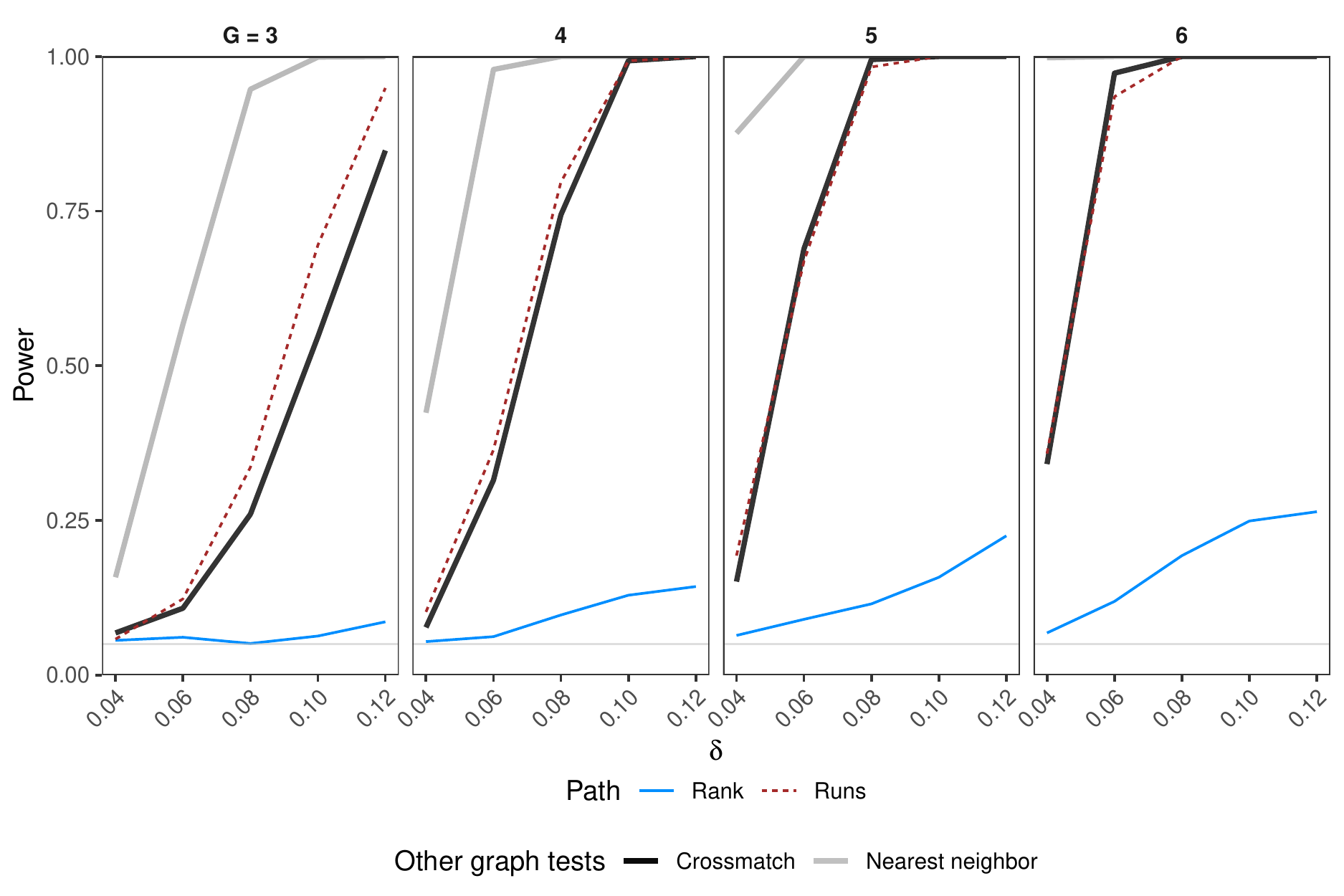}
			\caption{Location change with changing number of treatment groups: $\mu_g = (g-1)\delta$ where $g$ is the group number. Rank refers to the Kruskal-Wallis test, Runs refers to the Mood runs test, Crossmatch is the multisample crossmatch, and Nearest neighbor is the multisample  $k$-nearest neighbor test presented in this paper.}
		\end{figure*}
		
		\begin{figure*}[h]
			\includegraphics[width = \textwidth]{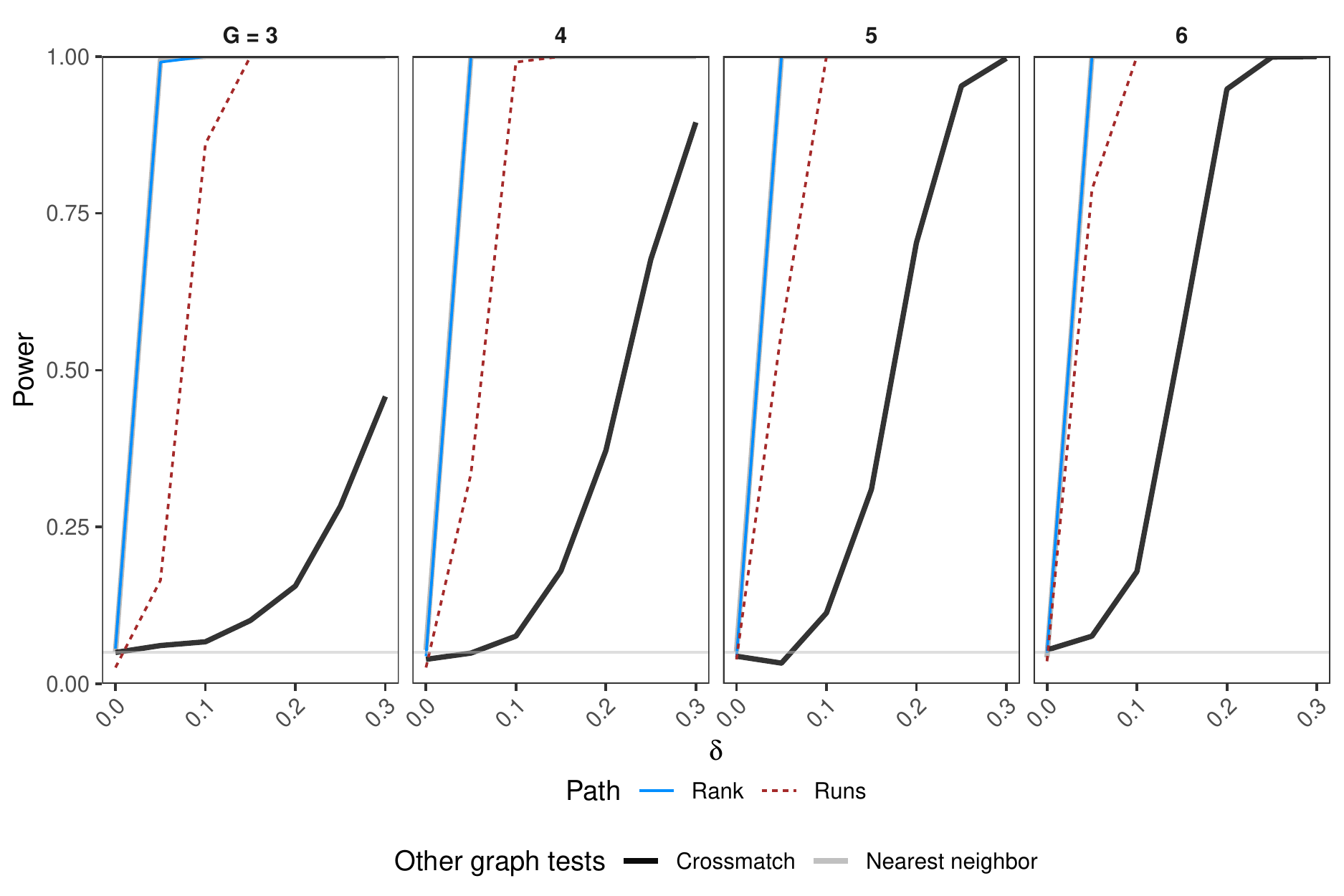}
			\caption{Scale change with changing number of treatment groups: $\sigma_g^2 = 1 + (g-1)\delta$ where $g$ is the group number. Rank refers to the Kruskal-Wallis test, Runs refers to the Mood runs test, Crossmatch is the multisample crossmatch, and Nearest neighbor is the multisample  $k$-nearest neighbor test presented in this paper.}
		\end{figure*}
		
		\begin{figure*}[h]
			\includegraphics[width = \textwidth]{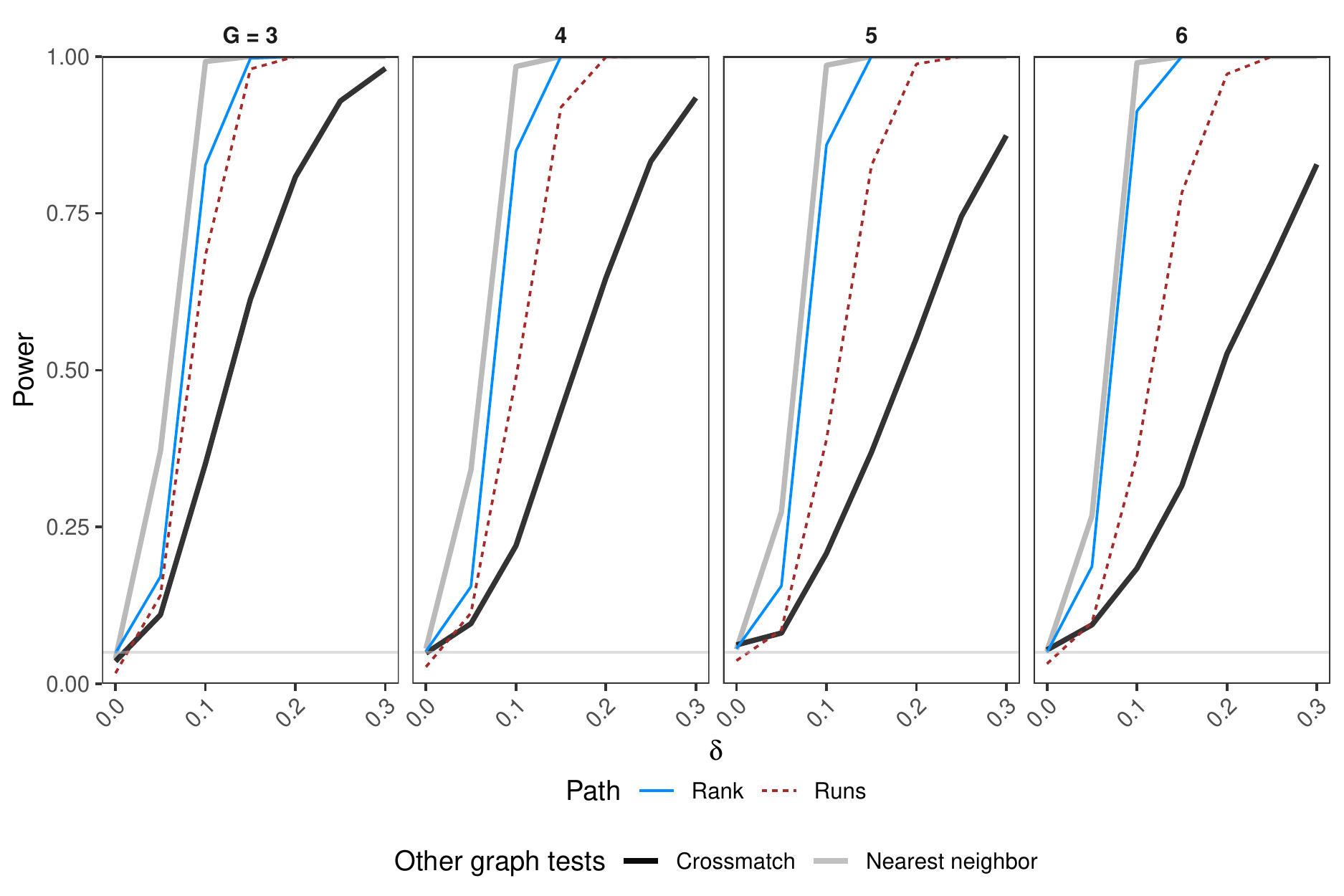}
			\caption{Correlation change with changing number of treatment groups: $\rho_g = (g-1)\delta/(G-1)$ where $g$ is the group number. Rank refers to the Kruskal-Wallis test, Runs refers to the Mood runs test, Crossmatch is the multisample crossmatch, and Nearest neighbor is the multisample  $k$-nearest neighbor test presented in this paper.}
		\end{figure*}
		
		\clearpage
		\subsection{Power of $k$-Nearest neighbor test with increasing k} \label{sec:knnpower}
\begin{figure*}[h]
			\includegraphics[width = \textwidth]{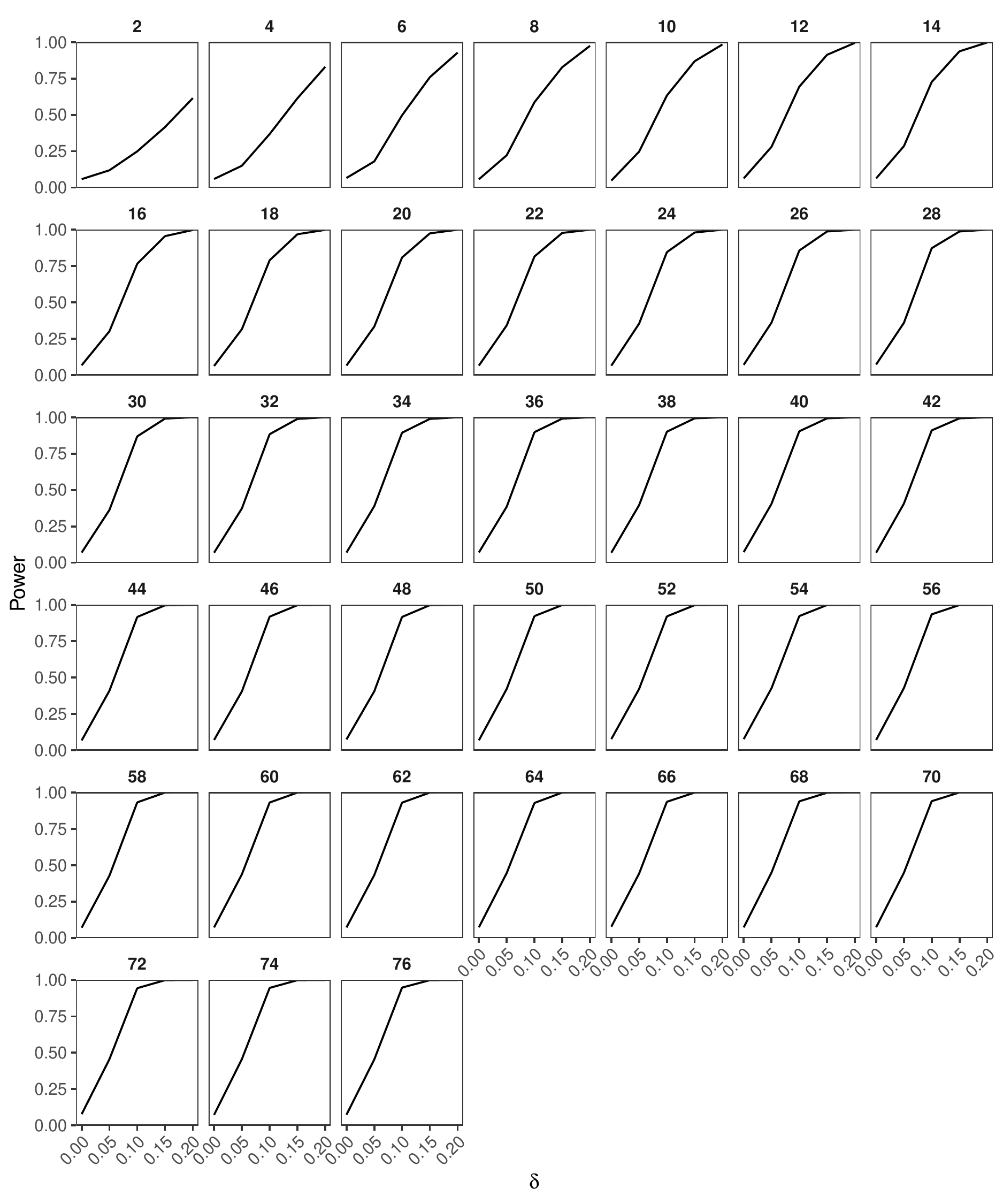}
			\caption{Power of $k$-nearest neighbor test with increasing $k$ in a scale change setting.}
			\label{fig:knn_k_power}
		\end{figure*}

		We also performed a quick check of the power of the $k$-nearest neighbor test as we increase $k$ in the scale change setting of Section \ref{sec:agarwal_desc} holding the number of covariates $d = 10$ and number of groups $G = 5$. We see in Figure \ref{fig:knn_k_power} that the power of the test does increase with $k$, verifying the results of \citet{Bhattacharya2019}. For a detailed description of the simulation setup in this setting, see Section \ref{sec:agarwal_desc}.
		
		\clearpage
		\printbibliography